\documentclass[11pt,a4paper]{article}
\usepackage{jheppub}

\usepackage{mathtools}
\usepackage{amsmath}
\usepackage{amssymb} 

\usepackage{tikz}
\usetikzlibrary{arrows.meta,positioning,calc}

\allowdisplaybreaks
\DeclarePairedDelimiter{\bra}{\langle}{\rvert}
\DeclarePairedDelimiter{\ket}{\lvert}{\rangle}

\usepackage{comment}

\usepackage{graphicx}
\usepackage{subcaption}

\newcommand{\dg}[1]{\scalebox{0.7}{#1}}
\newcommand{\inl}[1]{\raisebox{-0.2ex}{\scalebox{0.3}{#1}}}

\newcommand{\Rob}[1]{{\color{purple} #1}}
\newcommand{\comP}[1]{{\color{magenta}\textbf{P: #1}}}
\newcommand{\comE}[1]{{\color{green!50!black}\textbf{E:#1}}}
\newcommand{\gz}{{\color{lightgray}{0}}}
\newcommand{\I}{\mathrm{i}}
\newcommand{\E}{\mathrm{e}}
\newcommand{\C}{\mathbb{C}}

\newcommand{\lcyan}{cyan!20}
\newcommand{\lcyantwo}{cyan}

\newcommand{\redtwo}{red!60} %

\definecolor{lgreen}{RGB}{190,230,170}
\definecolor{dgreen}{RGB}{0,150,60}
\newcommand{\lgreen}{lgreen}
\newcommand{\dgreen}{dgreen}

\newcommand{\be}{\begin{equation}}
\newcommand{\ee}{\end{equation}}
\newcommand{\gLs}{(\gamma L)^2}
\newcommand{\doublet}[2]{\left(\begin{array}{c}#1\\#2\end{array}\right)}
\newcommand{\triplet}[3]{\left(\begin{array}{c}#1\\#2\\#3\end{array}\right)}
\newcommand{\twobytwo}[4]{\left(\begin{array}{cc} #1&#2\\#3&#4\end{array}\right)}
\newcommand{\threebythree}[9]{\left(\begin{array}{ccc} #1&#2&#3\\#4&#5&#6\\#7&#8&#9\\\end{array}\right)}
\newcommand{\iq}{q^{-1}}
\newcommand{\half}{\frac12}
\newcommand{\Rhat}{\hat{\mathrm{R}}}
\newcommand{\suq}{\mathrm{su}_q(3)}
\newcommand{\R}[4]{R^{#1\phantom{#2}#3}_{\phantom{#1}#2\phantom{#3}#4}}
\newcommand{\Rh}[4]{\hat{R}^{#1\phantom{#2}#3}_{\phantom{#1}#2\phantom{#3}#4}}
\newcommand{\Rt}[4]{\widetilde{R}^{#1\phantom{#2}#3}_{\phantom{#1}#2\phantom{#3}#4}}
\newcommand{\Urm}{\mathrm{U}}
\newcommand{\su}{\mathrm{su}}
\newcommand{\GL}{\mathrm{GL}}
\newcommand{\SU}{\mathrm{SU}}
\newcommand{\SO}{\mathrm{SO}}
\newcommand{\SL}{\mathrm{SL}}
\newcommand{\tq}[2]{\mathrm{t}^#1_{\phantom{#1}\!#2}}
\newcommand{\Tr}{\mathrm{Tr}}
\newcommand{\Acal}{\mathcal{A}}
\newcommand{\Fcal}{\mathcal{F}}
\newcommand{\Jcal}{\mathcal{J}}
\newcommand{\Rcal}{\mathcal{R}}
\newcommand{\Scal}{\mathcal{S}}
\newcommand{\Lcal}{\mathcal{L}}
\newcommand{\Wcal}{\mathcal{W}}
\newcommand{\mparagraph}[1]{\paragraph{#1}\mbox{}}
\newcommand{\Phibar}{\overline{\Phi}}
\newcommand{\Zset}{\mathbb{Z}}
\newcommand{\Ncal}{\mathcal{N}}
\newcommand{\Hcal}{\mathcal{H}}
\newcommand{\Gcal}{\mathcal{G}}
\newcommand{\Vcal}{\mathcal{V}}
\newcommand{\AdS}{\mathrm{AdS}}
\newcommand{\Srm}{\mathrm{S}}
\newcommand{\Cset}{{\,\,{{{^{_{\pmb{\mid}}}}\kern-.47em{\mathrm C}}}}}
\newcommand{\Supertwistor}{\Cset \mathrm{P}^{3|4}}
\newcommand{\qb}{{\bar{q}}}
\newcommand{\hb}{{\bar{h}}}
\newcommand{\zb}{{\bar{z}}}
\newcommand{\diff}{\mathrm{d}}
\newcommand{\p}{\partial}
\newcommand{\ra}{\rightarrow}
\newcommand{\gt}{\tilde{g}}
\newcommand{\bb}{\bar{b}}
\newcommand{\fb}{\bar{f}}
\newcommand{\astv}{\!\mathrel{\raisebox{.8pt}{{\scriptsize \hexstar}}}\!}

\newcommand{\Rcheck}{\check{R}}
\newcommand{\Qt}{\widetilde{Q}}
\newcommand{\tQ}{\widetilde{Q}}

\newcommand{\tphi}{\check{\phi}}
\newcommand{\phit}{\check{\phi}}
\newcommand{\pt}{\tilde{p}}
\newcommand{\bket}[1]{\bigg|#1\bigg\rangle}

\newcommand{\TXYXY}{\begin{tikzpicture}[rotate=-45,baseline=-0.1cm]\fill[\lcyan] (0,1)--(1,1)--(1,0)--(0,0)--(0,1);\fill[pink] (0,0.5) arc (-90:0:0.5);\fill[pink] (0,0.5)--(0.5,1)--(0,1)--(0,0.5);\fill[pink] (0.5,0) arc (180:90:0.5);\fill[pink](0.5,0)--(1,0)--(1,0.5)--(0.5,0); \draw[-,black!70!green] (0,1)--(1,1)--(1,0)--(0,0)--(0,1);\draw[-,black!70!green] (0,1)--(1,1)--(1,0)--(0,0)--(0,1);\draw[-] (1,0.1) arc (90:180:0.1); \draw[-,line width=0.5mm,black!60!red] (0,0.5) arc (-90:0:0.5); \draw[-,line width=0.5mm,black!60!red] (0.5,0) arc (180:90:0.5);
  \draw[-,line width=0.5mm,black!60!red] (0.3,0.8)--(0.3,0.6)--(0.5,0.6);
  \draw[-,line width=0.5mm,black!60!red] (0.7,0.2)--(0.7,0.4)--(0.5,0.4);
\end{tikzpicture}}

\newcommand{\TYXYX}{\begin{tikzpicture}[rotate=-45,baseline=-0.1cm]\fill[\lcyan] (0,1)--(1,1)--(1,0)--(0,0)--(0,1);\fill[pink] (0,0.5) arc (-90:0:0.5);\fill[pink] (0,0.5)--(0.5,1)--(0,1)--(0,0.5);\fill[pink] (0.5,0) arc (180:90:0.5);\fill[pink](0.5,0)--(1,0)--(1,0.5)--(0.5,0); \draw[-,black!70!green] (0,1)--(1,1)--(1,0)--(0,0)--(0,1);\draw[-,black!70!green] (0,1)--(1,1)--(1,0)--(0,0)--(0,1);\draw[-] (1,0.1) arc (90:180:0.1); \draw[-,line width=0.5mm,black!60!red] (0,0.5) arc (-90:0:0.5); \draw[-,line width=0.5mm,black!60!red] (0.5,0) arc (180:90:0.5);
  \draw[-,line width=0.5mm,black!60!red] (0.2,0.7)--(0.4,0.7)--(0.4,0.5);
    \draw[-,line width=0.5mm,black!60!red] (0.8,0.3)--(0.6,0.3)--(0.6,0.5);
\end{tikzpicture}}

\newcommand{\TYXXY}{\begin{tikzpicture}[rotate=-45,baseline=-0.1cm]\fill[\lcyan] (0,1)--(1,1)--(1,0)--(0,0)--(0,1);\fill[pink] (0,0.5) arc (-90:0:0.5);\fill[pink] (0,0.5)--(0.5,1)--(0,1)--(0,0.5);\fill[pink] (0.5,0) arc (180:90:0.5);\fill[pink](0.5,0)--(1,0)--(1,0.5)--(0.5,0); \draw[-,black!70!green] (0,1)--(1,1)--(1,0)--(0,0)--(0,1);\draw[-,black!70!green] (0,1)--(1,1)--(1,0)--(0,0)--(0,1);\draw[-] (1,0.1) arc (90:180:0.1); \draw[-,line width=0.5mm,black!60!red] (0,0.5) arc (-90:0:0.5); \draw[-,line width=0.5mm,black!60!red] (0.5,0) arc (180:90:0.5);
  \draw[-,line width=0.5mm,black!60!red] (0.3,0.8)--(0.3,0.6)--(0.5,0.6);
  \draw[-,line width=0.5mm,black!60!red] (0.8,0.3)--(0.6,0.3)--(0.6,0.5);
\end{tikzpicture}}

\newcommand{\TXYYX}{\begin{tikzpicture}[rotate=-45,baseline=-0.1cm]\fill[\lcyan] (0,1)--(1,1)--(1,0)--(0,0)--(0,1);\fill[pink] (0,0.5) arc (-90:0:0.5);\fill[pink] (0,0.5)--(0.5,1)--(0,1)--(0,0.5);\fill[pink] (0.5,0) arc (180:90:0.5);\fill[pink](0.5,0)--(1,0)--(1,0.5)--(0.5,0); \draw[-,black!70!green] (0,1)--(1,1)--(1,0)--(0,0)--(0,1);\draw[-,black!70!green] (0,1)--(1,1)--(1,0)--(0,0)--(0,1);\draw[-] (1,0.1) arc (90:180:0.1); \draw[-,line width=0.5mm,black!60!red] (0,0.5) arc (-90:0:0.5); \draw[-,line width=0.5mm,black!60!red] (0.5,0) arc (180:90:0.5);
  \draw[-,line width=0.5mm,black!60!red] (0.2,0.7)--(0.4,0.7)--(0.4,0.5);
  \draw[-,line width=0.5mm,black!60!red] (0.7,0.2)--(0.7,0.4)--(0.5,0.4);
\end{tikzpicture}}

\newcommand{\TpYXYX}{\begin{tikzpicture}[rotate=-45,baseline=-0.1cm]\fill[pink] (0,1)--(1,1)--(1,0)--(0,0)--(0,1);\fill[\lcyan] (0,0.5) arc (-90:0:0.5);\fill[\lcyan] (0,0.5)--(0.5,1)--(0,1)--(0,0.5);\fill[\lcyan] (0.5,0) arc (180:90:0.5);\fill[\lcyan](0.5,0)--(1,0)--(1,0.5)--(0.5,0); \draw[-,black!70!green] (0,1)--(1,1)--(1,0)--(0,0)--(0,1);\draw[-,black!70!green] (0,1)--(1,1)--(1,0)--(0,0)--(0,1);\draw[-] (1,0.1) arc (90:180:0.1); \draw[-,line width=0.5mm,black!60!red] (0,0.5) arc (-90:0:0.5); \draw[-,line width=0.5mm,black!60!red] (0.5,0) arc (180:90:0.5);
  \draw[-,line width=0.5mm,black!60!red] (0.2,0.7)--(0.4,0.7)--(0.4,0.5);
    \draw[-,line width=0.5mm,black!60!red] (0.8,0.3)--(0.6,0.3)--(0.6,0.5);
\end{tikzpicture}}

\newcommand{\TpYXXY}{\begin{tikzpicture}[rotate=-45,baseline=-0.1cm]\fill[pink] (0,1)--(1,1)--(1,0)--(0,0)--(0,1);\fill[\lcyan] (0,0.5) arc (-90:0:0.5);\fill[\lcyan] (0,0.5)--(0.5,1)--(0,1)--(0,0.5);\fill[\lcyan] (0.5,0) arc (180:90:0.5);\fill[\lcyan](0.5,0)--(1,0)--(1,0.5)--(0.5,0); \draw[-,black!70!green] (0,1)--(1,1)--(1,0)--(0,0)--(0,1);\draw[-,black!70!green] (0,1)--(1,1)--(1,0)--(0,0)--(0,1);\draw[-] (1,0.1) arc (90:180:0.1); \draw[-,line width=0.5mm,black!60!red] (0,0.5) arc (-90:0:0.5); \draw[-,line width=0.5mm,black!60!red] (0.5,0) arc (180:90:0.5);
  \draw[-,line width=0.5mm,black!60!red] (0.3,0.8)--(0.3,0.6)--(0.5,0.6);
  \draw[-,line width=0.5mm,black!60!red] (0.8,0.3)--(0.6,0.3)--(0.6,0.5);
\end{tikzpicture}}

\newcommand{\TpXYXY}{\begin{tikzpicture}[rotate=-45,baseline=-0.1cm]\fill[pink] (0,1)--(1,1)--(1,0)--(0,0)--(0,1);\fill[\lcyan] (0,0.5) arc (-90:0:0.5);\fill[\lcyan] (0,0.5)--(0.5,1)--(0,1)--(0,0.5);\fill[\lcyan] (0.5,0) arc (180:90:0.5);\fill[\lcyan](0.5,0)--(1,0)--(1,0.5)--(0.5,0); \draw[-,black!70!green] (0,1)--(1,1)--(1,0)--(0,0)--(0,1);\draw[-,black!70!green] (0,1)--(1,1)--(1,0)--(0,0)--(0,1);\draw[-] (1,0.1) arc (90:180:0.1); \draw[-,line width=0.5mm,black!60!red] (0,0.5) arc (-90:0:0.5); \draw[-,line width=0.5mm,black!60!red] (0.5,0) arc (180:90:0.5);
  \draw[-,line width=0.5mm,black!60!red] (0.3,0.8)--(0.3,0.6)--(0.5,0.6);
  \draw[-,line width=0.5mm,black!60!red] (0.7,0.2)--(0.7,0.4)--(0.5,0.4);
\end{tikzpicture}}

\newcommand{\TpXYYX}{\begin{tikzpicture}[rotate=-45,baseline=-0.1cm]\fill[pink] (0,1)--(1,1)--(1,0)--(0,0)--(0,1);\fill[\lcyan] (0,0.5) arc (-90:0:0.5);\fill[\lcyan] (0,0.5)--(0.5,1)--(0,1)--(0,0.5);\fill[\lcyan] (0.5,0) arc (180:90:0.5);\fill[\lcyan](0.5,0)--(1,0)--(1,0.5)--(0.5,0); \draw[-,black!70!green] (0,1)--(1,1)--(1,0)--(0,0)--(0,1);\draw[-,black!70!green] (0,1)--(1,1)--(1,0)--(0,0)--(0,1);\draw[-] (1,0.1) arc (90:180:0.1); \draw[-,line width=0.5mm,black!60!red] (0,0.5) arc (-90:0:0.5); \draw[-,line width=0.5mm,black!60!red] (0.5,0) arc (180:90:0.5);
  \draw[-,line width=0.5mm,black!60!red] (0.2,0.7)--(0.4,0.7)--(0.4,0.5);
  \draw[-,line width=0.5mm,black!60!red] (0.7,0.2)--(0.7,0.4)--(0.5,0.4);
\end{tikzpicture}}

\newcommand{\TZZZZ}{\begin{tikzpicture}[rotate=45,baseline=0.6cm]\fill[\lcyan] (0,1)--(1,1)--(1,0)--(0,0)--(0,1);\draw[-,black!70!green] (0,1)--(1,1)--(1,0)--(0,0)--(0,1);\draw[-] (0,0.1) arc (90:0:0.1); \end{tikzpicture}\hspace{-0.004cm}}

\newcommand{\TXZXZ}{\begin{tikzpicture}[rotate=45,baseline=0.6cm]\fill[pink] (0,1)--(1,1)--(1,0)--(0,0)--(0,1);\draw[-] (0,0.1) arc (90:0:0.1); \fill[\lcyan] (0,0.5) arc (-90:0:0.5);\fill[\lcyan] (0,0.5)--(0,1)--(0.5,1)--(0,0.5);\draw[-,black!70!green] (0,1)--(1,1)--(1,0)--(0,0)--(0,1);\draw[-,line width=0.5mm,black!60!red] (0,0.5) arc (-90:0:0.5);  \draw[-,line width=0.5mm,black!60!red] (0.2,0.7)--(0.4,0.7)--(0.4,0.5);
\end{tikzpicture}\hspace{-0.01cm}}
\newcommand{\TZXZX}{\begin{tikzpicture}[rotate=45,baseline=0.6cm]\fill[\lcyan] (0,1)--(1,1)--(1,0)--(0,0)--(0,1);  \fill[pink] (0.5,0) arc (180:90:0.5);\fill[pink](0.5,0)--(1,0)--(1,0.5)--(0.5,0);\draw[-,line width=0.5mm,black!60!red] (0.5,0) arc (180:90:0.5);\draw[-,black!70!green] (0,1)--(1,1)--(1,0)--(0,0)--(0,1);\draw[-] (0,0.1) arc (90:0:0.1);  \draw[-,line width=0.5mm,black!60!red] (0.7,0.2)--(0.7,0.4)--(0.5,0.4);\end{tikzpicture}\hspace{-0.04cm}}
\newcommand{\TXZZX}{\begin{tikzpicture}[rotate=45,baseline=0.6cm]\fill[\lcyan] (0,1)--(1,1)--(1,0)--(0,0)--(0,1);\fill[pink] (0,0)--(1,0)--(1,0.5)--(0,0.5)--(0,0);\draw[-,black!70!green] (0,1)--(1,1)--(1,0)--(0,0)--(0,1);\draw[-] (0,0.1) arc (90:0:0.1); \draw[-,line width=0.5mm,black!60!red] (0,0.5)--(1,0.5);\draw[-,line width=0.5mm,black!60!red] (0.4,0.4)--(0.5,0.5)--(0.4,0.6);\end{tikzpicture}\hspace{-0.04cm}}
\newcommand{\TZXXZ}{\begin{tikzpicture}[rotate=45,baseline=0.6cm]\fill[\lcyan] (0,1)--(1,1)--(1,0)--(0,0)--(0,1);\draw[-,black!70!green] (0,1)--(1,1)--(1,0)--(0,0)--(0,1);\draw[-] (0,0.1) arc (90:0:0.1);   \fill[pink] (0.5,0)--(0.5,1)--(1,1)--(1,0)--(0.5,0);\draw[-,line width=0.5mm,black!60!red] (0.5,0.01)--(0.5,1);\draw[-,line width=0.5mm,black!60!red] (0.4,0.4)--(0.5,0.5)--(0.6,0.4);
        \end{tikzpicture}}

\newcommand{\TpXXXX}{\begin{tikzpicture}[rotate=45,baseline=0.6cm]\fill[\lcyan] (0,1)--(1,1)--(1,0)--(0,0)--(0,1);\fill[pink] (0,0.5) arc (-90:0:0.5);\fill[pink] (0,0.5)--(0.5,1)--(0,1)--(0,0.5);\fill[pink] (0.5,0) arc (180:90:0.5);\fill[pink](0.5,0)--(1,0)--(1,0.5)--(0.5,0); \draw[-,black!70!green] (0,1)--(1,1)--(1,0)--(0,0)--(0,1);\draw[-,black!70!green] (0,1)--(1,1)--(1,0)--(0,0)--(0,1);\draw[-] (0,0.1) arc (90:0:0.1); \draw[-,line width=0.5mm,black!60!red] (0,0.5) arc (-90:0:0.5); \draw[-,line width=0.5mm,black!60!red] (0.5,0) arc (180:90:0.5);\draw[-,line width=0.5mm,black!60!red] (0.2,0.7)--(0.4,0.7)--(0.4,0.5);
  \draw[-,line width=0.5mm,black!60!red] (0.7,0.2)--(0.7,0.4)--(0.5,0.4);\end{tikzpicture}\hspace{-0.04cm}}

\newcommand{\TpZZZZ}{\begin{tikzpicture}[rotate=45,baseline=0.6cm]\fill[pink] (0,1)--(1,1)--(1,0)--(0,0)--(0,1);\draw[-,black!70!green] (0,1)--(1,1)--(1,0)--(0,0)--(0,1);\draw[-] (0,0.1) arc (90:0:0.1); \end{tikzpicture}\hspace{-0.004cm}}

\newcommand{\TpXZXZ}{\begin{tikzpicture}[rotate=45,baseline=0.6cm]\fill[\lcyan] (0,1)--(1,1)--(1,0)--(0,0)--(0,1);\draw[-] (0,0.1) arc (90:0:0.1); \fill[pink] (0,0.5) arc (-90:0:0.5);\fill[pink] (0,0.5)--(0,1)--(0.5,1)--(0,0.5);\draw[-,black!70!green] (0,1)--(1,1)--(1,0)--(0,0)--(0,1);\draw[-,line width=0.5mm,black!60!red] (0,0.5) arc (-90:0:0.5);\draw[-,line width=0.5mm,black!60!red] (0.2,0.7)--(0.4,0.7)--(0.4,0.5);   \end{tikzpicture}\hspace{-0.01cm}}
\newcommand{\TpXZZX}{\begin{tikzpicture}[rotate=45,baseline=0.6cm]\fill[pink] (0,1)--(1,1)--(1,0)--(0,0)--(0,1);\fill[\lcyan] (0,0)--(1,0)--(1,0.5)--(0,0.5)--(0,0);\draw[-,black!70!green] (0,1)--(1,1)--(1,0)--(0,0)--(0,1);\draw[-] (0,0.1) arc (90:0:0.1); \draw[-,line width=0.5mm,black!60!red] (0,0.5)--(1,0.5);\draw[-,line width=0.5mm,black!60!red] (0.4,0.4)--(0.5,0.5)--(0.4,0.6);\end{tikzpicture}\hspace{-0.04cm}}
\newcommand{\TpZXXZ}{\begin{tikzpicture}[rotate=45,baseline=0.6cm]\fill[pink] (0,1)--(1,1)--(1,0)--(0,0)--(0,1);\draw[-,black!70!green] (0,1)--(1,1)--(1,0)--(0,0)--(0,1);\draw[-] (0,0.1) arc (90:0:0.1);   \fill[\lcyan] (0.5,0)--(0.5,1)--(1,1)--(1,0)--(0.5,0);\draw[-,line width=0.5mm,black!60!red] (0.5,0.01)--(0.5,1);\draw[-,line width=0.5mm,black!60!red] (0.4,0.4)--(0.5,0.5)--(0.6,0.4);
        \end{tikzpicture}}
 \newcommand{\TpZXZX}{\begin{tikzpicture}[rotate=45,baseline=0.6cm]\fill[pink] (0,1)--(1,1)--(1,0)--(0,0)--(0,1);  \fill[\lcyan] (0.5,0) arc (180:90:0.5);\fill[\lcyan](0.5,0)--(1,0)--(1,0.5)--(0.5,0);\draw[-,line width=0.5mm,black!60!red] (0.5,0) arc (180:90:0.5);\draw[-,black!70!green] (0,1)--(1,1)--(1,0)--(0,0)--(0,1);\draw[-] (0,0.1) arc (90:0:0.1);\draw[-,line width=0.5mm,black!60!red] (0.7,0.2)--(0.7,0.4)--(0.5,0.4);\end{tikzpicture}\hspace{-0.04cm}}
\newcommand{\TXXXX}{\begin{tikzpicture}[rotate=45,baseline=0.6cm]\fill[pink] (0,1)--(1,1)--(1,0)--(0,0)--(0,1);\fill[\lcyan] (0,0.5) arc (-90:0:0.5);\fill[\lcyan] (0,0.5)--(0.5,1)--(0,1)--(0,0.5);\fill[\lcyan] (0.5,0) arc (180:90:0.5);\fill[\lcyan](0.5,0)--(1,0)--(1,0.5)--(0.5,0); \draw[-,black!70!green] (0,1)--(1,1)--(1,0)--(0,0)--(0,1);\draw[-,black!70!green] (0,1)--(1,1)--(1,0)--(0,0)--(0,1);\draw[-] (0,0.1) arc (90:0:0.1); \draw[-,line width=0.5mm,black!60!red] (0,0.5) arc (-90:0:0.5); \draw[-,line width=0.5mm,black!60!red] (0.5,0) arc (180:90:0.5);
\draw[-,line width=0.5mm,black!60!red] (0.2,0.7)--(0.4,0.7)--(0.4,0.5);
  \draw[-,line width=0.5mm,black!60!red] (0.7,0.2)--(0.7,0.4)--(0.5,0.4);
\end{tikzpicture}\hspace{-0.04cm}}

\newcommand{\TYYYY}{\begin{tikzpicture}[rotate=45,baseline=0.6cm]\fill[pink] (0,1)--(1,1)--(1,0)--(0,0)--(0,1);\fill[\lcyan] (0,0.5) arc (-90:0:0.5);\fill[\lcyan] (0,0.5)--(0.5,1)--(0,1)--(0,0.5);\fill[\lcyan] (0.5,0) arc (180:90:0.5);\fill[\lcyan](0.5,0)--(1,0)--(1,0.5)--(0.5,0); \draw[-,black!70!green] (0,1)--(1,1)--(1,0)--(0,0)--(0,1);\draw[-,black!70!green] (0,1)--(1,1)--(1,0)--(0,0)--(0,1);\draw[-] (0,0.1) arc (90:0:0.1); \draw[-,line width=0.5mm,black!60!red] (0,0.5) arc (-90:0:0.5); \draw[-,line width=0.5mm,black!60!red] (0.5,0) arc (180:90:0.5);
\draw[-,line width=0.5mm,black!60!red] (0.3,0.8)--(0.3,0.6)--(0.5,0.6);
  \draw[-,line width=0.5mm,black!60!red] (0.8,0.3)--(0.6,0.3)--(0.6,0.5);
\end{tikzpicture}\hspace{-0.04cm}}

\newcommand{\TpYYYY}{\begin{tikzpicture}[rotate=45,baseline=0.6cm]\fill[\lcyan] (0,1)--(1,1)--(1,0)--(0,0)--(0,1);\fill[pink] (0,0.5) arc (-90:0:0.5);\fill[pink] (0,0.5)--(0.5,1)--(0,1)--(0,0.5);\fill[pink] (0.5,0) arc (180:90:0.5);\fill[pink](0.5,0)--(1,0)--(1,0.5)--(0.5,0); \draw[-,black!70!green] (0,1)--(1,1)--(1,0)--(0,0)--(0,1);\draw[-,black!70!green] (0,1)--(1,1)--(1,0)--(0,0)--(0,1);\draw[-] (0,0.1) arc (90:0:0.1); \draw[-,line width=0.5mm,black!60!red] (0,0.5) arc (-90:0:0.5); \draw[-,line width=0.5mm,black!60!red] (0.5,0) arc (180:90:0.5);
\draw[-,line width=0.5mm,black!60!red] (0.3,0.8)--(0.3,0.6)--(0.5,0.6);
  \draw[-,line width=0.5mm,black!60!red] (0.8,0.3)--(0.6,0.3)--(0.6,0.5);
\end{tikzpicture}\hspace{-0.04cm}}

\newcommand{\TXYXYid}{\begin{tikzpicture}[rotate=45,baseline=0.6cm]\fill[pink] (0,1)--(1,1)--(1,0)--(0,0)--(0,1);\fill[\lcyan] (0,0.5) arc (-90:0:0.5);\fill[\lcyan] (0,0.5)--(0.5,1)--(0,1)--(0,0.5);\fill[\lcyan] (0.5,0) arc (180:90:0.5);\fill[\lcyan](0.5,0)--(1,0)--(1,0.5)--(0.5,0); \draw[-,black!70!green] (0,1)--(1,1)--(1,0)--(0,0)--(0,1);\draw[-,black!70!green] (0,1)--(1,1)--(1,0)--(0,0)--(0,1);\draw[-] (0,0.1) arc (90:0:0.1); \draw[-,line width=0.5mm,black!60!red] (0,0.5) arc (-90:0:0.5); \draw[-,line width=0.5mm,black!60!red] (0.5,0) arc (180:90:0.5);
\draw[-,line width=0.5mm,black!60!red] (0.2,0.7)--(0.4,0.7)--(0.4,0.5);
  \draw[-,line width=0.5mm,black!60!red] (0.8,0.3)--(0.6,0.3)--(0.6,0.5);
\end{tikzpicture}\hspace{-0.04cm}}

\newcommand{\TYXYXid}{\begin{tikzpicture}[rotate=45,baseline=0.6cm]\fill[pink] (0,1)--(1,1)--(1,0)--(0,0)--(0,1);\fill[\lcyan] (0,0.5) arc (-90:0:0.5);\fill[\lcyan] (0,0.5)--(0.5,1)--(0,1)--(0,0.5);\fill[\lcyan] (0.5,0) arc (180:90:0.5);\fill[\lcyan](0.5,0)--(1,0)--(1,0.5)--(0.5,0); \draw[-,black!70!green] (0,1)--(1,1)--(1,0)--(0,0)--(0,1);\draw[-,black!70!green] (0,1)--(1,1)--(1,0)--(0,0)--(0,1);\draw[-] (0,0.1) arc (90:0:0.1); \draw[-,line width=0.5mm,black!60!red] (0,0.5) arc (-90:0:0.5); \draw[-,line width=0.5mm,black!60!red] (0.5,0) arc (180:90:0.5);
   \draw[-,line width=0.5mm,black!60!red] (0.3,0.8)--(0.3,0.6)--(0.5,0.6);
   \draw[-,line width=0.5mm,black!60!red] (0.7,0.2)--(0.7,0.4)--(0.5,0.4);
\end{tikzpicture}\hspace{-0.04cm}}

\newcommand{\TpXYXYid}{\begin{tikzpicture}[rotate=45,baseline=0.6cm]\fill[\lcyan] (0,1)--(1,1)--(1,0)--(0,0)--(0,1);\fill[pink] (0,0.5) arc (-90:0:0.5);\fill[pink] (0,0.5)--(0.5,1)--(0,1)--(0,0.5);\fill[pink] (0.5,0) arc (180:90:0.5);\fill[pink](0.5,0)--(1,0)--(1,0.5)--(0.5,0); \draw[-,black!70!green] (0,1)--(1,1)--(1,0)--(0,0)--(0,1);\draw[-,black!70!green] (0,1)--(1,1)--(1,0)--(0,0)--(0,1);\draw[-] (0,0.1) arc (90:0:0.1); \draw[-,line width=0.5mm,black!60!red] (0,0.5) arc (-90:0:0.5); \draw[-,line width=0.5mm,black!60!red] (0.5,0) arc (180:90:0.5);
\draw[-,line width=0.5mm,black!60!red] (0.2,0.7)--(0.4,0.7)--(0.4,0.5);
  \draw[-,line width=0.5mm,black!60!red] (0.8,0.3)--(0.6,0.3)--(0.6,0.5);
\end{tikzpicture}\hspace{-0.04cm}}

\newcommand{\TpYXYXid}{\begin{tikzpicture}[rotate=45,baseline=0.6cm]\fill[\lcyan] (0,1)--(1,1)--(1,0)--(0,0)--(0,1);\fill[pink] (0,0.5) arc (-90:0:0.5);\fill[pink] (0,0.5)--(0.5,1)--(0,1)--(0,0.5);\fill[pink] (0.5,0) arc (180:90:0.5);\fill[pink](0.5,0)--(1,0)--(1,0.5)--(0.5,0); \draw[-,black!70!green] (0,1)--(1,1)--(1,0)--(0,0)--(0,1);\draw[-,black!70!green] (0,1)--(1,1)--(1,0)--(0,0)--(0,1);\draw[-] (0,0.1) arc (90:0:0.1); \draw[-,line width=0.5mm,black!60!red] (0,0.5) arc (-90:0:0.5); \draw[-,line width=0.5mm,black!60!red] (0.5,0) arc (180:90:0.5);
   \draw[-,line width=0.5mm,black!60!red] (0.3,0.8)--(0.3,0.6)--(0.5,0.6);
   \draw[-,line width=0.5mm,black!60!red] (0.7,0.2)--(0.7,0.4)--(0.5,0.4);
\end{tikzpicture}\hspace{-0.04cm}}

\title{
Dynamical Temperley-Lieb algebras from Quiver CFTs}

\author[a]{Rob Klabbers,}
\author[b]{Elli Pomoni,}
\author[b]{Ana L. Retore,}
\author[b]{Paul Ryan,}
\author[c,d]{Konstantinos Zoubos}

\affiliation[a]{Humboldt-Universit{\"a}t zu Berlin, Zum Großen Windkanal 2, 12489 Berlin, Germany}

\affiliation[b]{Deutsches Elektronen-Synchrotron DESY, Notkestr. 85, 22607 Hamburg, Germany}

\affiliation[c]{Department of Physics, University of Pretoria, Private Bag X20, Hatfield 0028, South Africa}

\affiliation[d]{National Institute for Theoretical and Computational Sciences (NITheCS), Gauteng, South Africa}

\emailAdd{rob.klabbers@hu-berlin.de}
\emailAdd{elli.pomoni@desy.de}
\emailAdd{ana.retore@desy.de}
\emailAdd{paul.ryan@desy.de}
\emailAdd{kzoubos@up.ac.za}

\abstract{

\bigskip

The one-loop dilatation operator of a superconformal planar gauge theory can  be written as a sum of local generators. The algebra these generators satisfy is a powerful organising principle for the spectrum. 
For SU(2)-like bosonic subsectors we show that this algebra can be read off directly from  Feynman diagrams, which we identify as a natural graphical language for it. For $\mathcal{N}=4$ super Yang-Mills this simply reproduces the 
Temperley-Lieb (TL) algebra of the XXX spin chain. 
After orbifolding and marginally deforming to  $\mathcal{N}=2$ superconformal quiver theories, the same recipe yields dynamical generalisations of the TL algebra. In these theories, the faces of the diagrams are coloured according to the quiver nodes, 
and separated by domain walls corresponding to bifundamental fields. 
There are two inequivalent SU(2)-like subsectors, which lead to two different TL algebras. One sector is built from the two bifundamental scalars and (in the simplest case) gives a dynamical TL algebra 
 with a colour-dependent loop fugacity in its quadratic relation. The other sector is built from one bifundamental and one adjoint scalar, and (always) gives a novel algebra, in which the cubic TL relation becomes dynamical.  
 In this latter case, we show that each generator of this new algebra is related to a standard XXZ TL generator by a colour-dependent twist. This twist is local and not an algebra homomorphism, and can be written as a dynamical Drinfeld twist.
 In both sectors, the TL structure provides a natural graphical approach to
the spectral problem, as well as clarifying the structure of the irreducible representations of these chains. Finally, we give a preliminary description of the loop models associated to these algebras.

}

\begin{document}
\maketitle

\section{Introduction}\label{sec:intro}

The discovery of integrable structures in the spectral problem of planar $\Ncal=4$ super Yang-Mills (SYM) theory \cite{Minahan:2002ve} ushered in a new era in the quest for exact results in gauge theory, see e.g. \cite{Beisert:2010jr,Bombardelli:2016rwb} for reviews. A key factor underlying these developments has been the mapping of gauge theory observables to spin-chain operators, such as the dilatation operator to a spin-chain Hamiltonian. At one loop, the action of the dilatation operator is nearest neighbour, and in specific subsectors reduces to that of the Heisenberg XXX Hamiltonian, a one-dimensional quantum integrable system that has been extensively studied since Bethe provided a solution via what is now known as the coordinate Bethe ansatz (see e.g. \cite{BaxterBook,Gomez96} for introductions). 

Given the success of the integrability programme for $\Ncal=4$ SYM, it is very natural 
to consider gauge theories with less symmetry and understand to which extent integrable structures survive. There are two possible philosophies to moving away from the $\Ncal=4$ maximally supersymmetric point: one can deform within the space of known integrable structures, which might however not immediately correspond to a well-understood or physically relevant gauge-theoretic deformation, or one can study the spectral problem of concrete gauge theories with less symmetry, with the distinct possibility, however, of venturing quickly into uncharted territory with regards to integrability. In this paper we follow the second approach. Before turning to it,
however, some comments on the first are in order, as it provides
the context for the algebraic structures that we will encounter.

Within the first approach, a very natural and well-studied
deformation of the isotropic XXX model is the anisotropic XXZ spin
chain, in which the $\SU(2)$ symmetry is broken to $\Urm(1)$ by a
parameter $q$. Together with the associated six-vertex model
\cite{Lieb:1967zz}, the XXZ model is one of the cornerstones of the
theory of quantum integrable systems: it remains solvable by Bethe
ansatz methods \cite{BaxterBook,Faddeev:1994nk,Gomez96,Arutyunovbook},
and is the prototypical model with trigonometric, rather than
rational, dependence on the spectral parameter.

A key structural feature of the XXZ model is that its bulk Hamiltonian is
a sum of generators of the Temperley-Lieb (TL) algebra
\cite{Temperley:1971iq}, a diagrammatic algebra which provides an
important organising principle for its spectrum \cite{Pasquier:1989kd}.
The TL algebra can be thought of as (a quotient
of) a $q$-deformation of the group algebra of the symmetric group: at $q=1$, the TL
generators reduce to combinations of the identity and the
permutations of neighbouring sites, out of which the XXX Hamiltonian
is built. In this sense, the TL algebra plays for XXZ the role that
the symmetric group plays for XXX.

At the same time, the XXZ model illustrates well the limitations of
the first approach from the perspective of gauge theory: despite its importance, it has not been realised
in the planar limit of a four-dimensional gauge theory. In particular, it should
not be confused with the integrable deformations of $\Ncal=4$ SYM
obtained by Drinfeld twists, such as the $\beta$ and $\gamma$
deformations (see e.g.\ \cite{Zoubos:2010kh,vanTongeren:2013gva} for
reviews): their integrable structure is captured by a twist of a Yangian, so they remain rational rather than trigonometric models
\cite{Beisert:2005if,Ihry:2008gm,Garus:2017bgl}. 

As already mentioned, in this paper we will instead follow the second approach: rather than deforming the integrable structure, we start from concrete gauge theories with less supersymmetry and ask what algebraic structure their dilatation operator carries. As we will
see, this route nevertheless leads us to structures closely related
to the XXZ model and its TL algebra.

Guided by the gauge theory, a line of investigation has focused on
the $\Ncal=2$ SCFTs obtained from $\Ncal=4$ SYM by orbifolding and then marginally deforming \cite{Gadde:2009dj,Gadde:2010zi,Liendo:2011xb,Pomoni:2011jj,Pomoni:2013poa,Mitev:2014yba,Mitev:2015oty,Pomoni:2021pbj,Bozkurt:2024tpz,Bertle:2024djm,Bozkurt:2025exl,Bath:2026gcu}, see \cite{Pomoni:2019oib} for a review. For the $\Zset_2$ orbifold case, the spin chains arising in the planar limit were shown not to be integrable by standard criteria \cite{Gadde:2009dj,Gadde:2010zi}. In particular, the $S$-matrices for magnon scattering do not satisfy the Yang-Baxter equation (apart from the orbifold point itself, where integrability is inherited from $\Ncal=4$ SYM \cite{Beisert:2005he}). %
The work \cite{Pomoni:2021pbj} considered their algebraic
structure in more detail and found a reformulation as a dynamical
spin chain, a structure typically encountered in relation to elliptic quantum
groups \cite{Felder:1994be}. For the latter, the usual Yang-Baxter
equation is replaced by the dynamical Yang-Baxter equation (see e.g. \cite{EtingofLatourBook,KonnoBook} for introductions), in which
the $R$-matrix also depends on a dynamical parameter that is shifted
according to the states on which it acts. Therefore, in the quiver
CFTs one should not expect the standard Yang-Baxter equation to
be the appropriate criterion for integrability.
This direction was further explored in 
\cite{Bertle:2024djm}, which showed invariance of the gauge-theoretic Lagrangian under a groupoid-type quantum symmetry constructed from the $F$-term and $D$-term relations of the theory. The algebraic structure found in \cite{Bertle:2024djm} has a natural categorical description, which closely parallels recent constructions of quantum groups for restricted SOS models \cite{Felder:2020tct,AnicFelder26}.  Other recent work studied the three- and four-magnon coordinate Bethe ansatz \cite{Bozkurt:2024tpz,Bozkurt:2025exl}, as well as the spin chains related to the more general class of ADE orbifolds \cite{Bath:2026gcu}. These results have made it clear that the $\Ncal=2$ quivers
do have plenty of additional structure which calls for a deeper understanding. 

In this work we make further progress towards uncovering the algebraic structure of the $\Ncal=2$ quiver CFTs. Our particular focus will be on constructing the analogue, in the cyclic quiver theories, of the Temperley-Lieb algebra appearing in the XXZ model. To do this, we take a step back and consider how the TL algebra at $q=1$ emerges in planar $\Ncal=4$ SYM. Given the connection to the XXX model, this is of course expected. However, as we will see, this connection can be made very explicit in the formalism of $\Ncal=1$ superspace, where the TL elements can be directly obtained from the gauge-theoretic Feynman diagrams which are part of the computation of the planar one-loop dilatation operator.  

As the $\Ncal=1$ superspace description generalises straightforwardly to any
$\Ncal=2$ quiver SCFT (see \cite{Pomoni:2019oib} for a review and
\cite{Pomoni:2011jj,Bath:2026gcu} for a detailed construction), we can then retrace
the same steps and use the gauge theory dilatation operator to define
analogues of the TL algebra for the bosonic $\SU(2)$-like subsectors of
the corresponding spin chains. 
In $\Ncal=2$ quiver SCFTs there are two
inequivalent such sectors \cite{Pomoni:2021pbj,Bertle:2024djm}: the
XY \emph{dense} subsector, which is built from the two bifundamental
scalars $X$ and $Y$ and enjoys an unbroken $\SU(2)$ symmetry, and the
XZ \emph{dilute} subsector, which is built from one bifundamental
scalar $X$ and the adjoint scalar $Z$, and in which the $\SU(2)$
symmetry is naively broken. The terminology is borrowed from the loop
model representation of the corresponding algebras, which we discuss
in Section~\ref{sec:loop}. There, each bifundamental field is
represented by a line acting as a domain wall, which separates regions
coloured by the two gauge nodes it connects, while the adjoint field
$Z$, which does not change the colour, carries no such line. Hence in
the XY sector there is a domain wall at every site of the chain,
while in the XZ sector the domain walls only appear at the sites
occupied by $X$ fields.
 As expected, the
resulting algebras are \emph{dynamical} Temperley-Lieb algebras, in
the sense that their defining relations depend on the spin-chain
configuration, which is characterised by a colouring of the regions
of the diagrams according to the quiver nodes. 

Dynamical Temperley-Lieb
algebras have recently been studied in \cite{Ren:2023mtn}, 
which extends previous implementations of site-dependent Temperley-Lieb algebras in the context of SOS models \cite{Pasquier:1986jc, Pearce:1990ila,Owczarek:1987ue} and brings them into the dynamical framework, with the goal of implementing a version of Baxterisation \cite{Jones:1990hq} 
for such models. 
 As we will show, for the $\Zset_2$ quiver, the XY sector gives
an algebra of the type studied in \cite{Ren:2023mtn}, where the loop fugacity appearing in the quadratic TL relations becomes colour-dependent. On the other hand, the XZ sector leads to a novel generalisation, in which it is instead the \emph{cubic}
Temperley-Lieb relation that becomes dynamical.

Although we will leave a formal study of these new Temperley-Lieb
algebras for future work, we will show that they do provide a graphical approach
to the spectral problem of the $\Ncal=2$ theories. To see this, we
study (first for the $\Zset_2$ and then the $\Zset_K$ quivers) finite-length open chains, and show that their spectrum is naturally organised in
terms of link states, the diagrammatic basis on which the
TL generators act. In this basis the Hamiltonian breaks
into smaller blocks, and new insights into the spectrum emerge, such
as the origin of certain energy splittings that one encounters as
one deforms away from the orbifold point.

To get a further handle on these algebras, we then show how they
can be related to more standard structures, by removing their
dynamical dependence through transformations which act differently (i.e.\ locally)
on each generator. In the XY sector, the colour dependence of each
generator only enters through an overall coupling, so a simple
colour-dependent rescaling is enough to reduce the relations to those
of the Temperley-Lieb algebra at $q=1$, as in $\Ncal=4$ SYM. The XZ
sector requires more work. There, each generator is locally an XXZ
Temperley-Lieb generator, with a value of $q$ given by the ratio of
the gauge couplings of the neighbouring colours, which therefore
varies along the chain. Using a twist to map every generator to the
value of $q$ of the leftmost colour makes manifest a direct sum
structure $\mathrm{TL}(q_1)\oplus\dots\oplus\mathrm{TL}(q_K)$, with one
copy of the standard XXZ algebra for each colour. Alternatively, all
generators can be twisted to the $q=1$ TL algebra. Our twist has the
same form as the coassociator of the dynamical twists which turn
$U_q(\mathfrak{sl}(2))$ into a quasi-Hopf algebra
\cite{Babelon:1995rz,Arnaudon:1997hm}, but is used in the opposite
direction, to twist the dynamics away. Since each generator is
transformed separately, neither the rescaling nor the twist preserves
the spectrum of the Hamiltonian. We do, however, expect this connection to be a key step towards understanding whether these TL algebras can be Baxterised. 

\begin{figure}[h]
\centering
$\mathbb{D}\big|_{\text{one-loop}} \;\sim\; H=-\sum_i e_i$,\\[2pt]
where $e_i$ are generators of an algebra $\mathcal{A}$\\[0.5\baselineskip]
\hspace{-0.5cm}with\\[0.5\baselineskip]
\begin{tikzpicture}[
    every node/.style={align=center},
    arr/.style={-{Stealth[length=3mm]}, thick, shorten <=0.5\baselineskip, shorten >=0.5\baselineskip},
    darr/.style={-{Implies}, double, double distance=1.5pt, semithick, shorten <=0.5\baselineskip, shorten >=0.5\baselineskip},
  ]
  \node (n4) {$\mathcal{N}=4$ SYM};
  \node (tl) [base right=0pt of n4] {$\Rightarrow\; \mathcal{A}=\mathrm{TL}$ with $q=1$};
  \node (n2) [below=2.6cm of n4] {$\mathcal{N}=2$ SCFT};
  \draw[arr] (n4) -- (n2)
     node[midway, right=4pt] {$\mathbb{Z}_K$ orbifolding\\ $+$ marginal deformations};
  \node (left)  [anchor=north] at ([shift={(-4.6,-2.2)}]n2.south)
        {$\mathcal{A}=$ Dynamical TL  of\\ the type described in \cite{Ren:2023mtn},  \\
        see \eqref{eq:XYTLrescaled} and \eqref{eq:XY_dTL_relations_K2}};
  \node (mid)   [anchor=north] at ([shift={(0,-3.2)}]n2.south)
        {$\mathcal{A}=$ Novel dynamical TL,  \\
        see \eqref{eq:ZK-XY-rels} and \eqref{eq:XY-ZK-dyn}};
  \node (right) [anchor=north] at ([shift={(4.6,-2.2)}]n2.south)
        {$\mathcal{A}=$ Novel dynamical TL,  \\
        see \eqref{eq:ZK-XZ-rels} and \eqref{eq:XZ-ZK-dyn}};
  \draw[darr] ([xshift=-6pt]n2.south) -- (left.north)
     node[midway, above left=2pt, align=right] {$XY$\\ $K=2$};
  \draw[darr] (n2.south) -- (mid.north)
     node[midway, right=3pt, align=left] {$XY$\\ $K\ge 3$};
  \draw[darr] ([xshift=6pt]n2.south) -- (right.north)
     node[midway, above right=2pt, align=left] {$XZ$\\ sector};
\end{tikzpicture}
\caption{In this paper, we describe a connection between quiver gauge theories and TL algebras and their generalisations. This figure summarises which TL algebra describes the one-loop Hamiltonian for each type of quiver gauge theory and sector thereof.  }
\label{fig:main-results}
\end{figure}
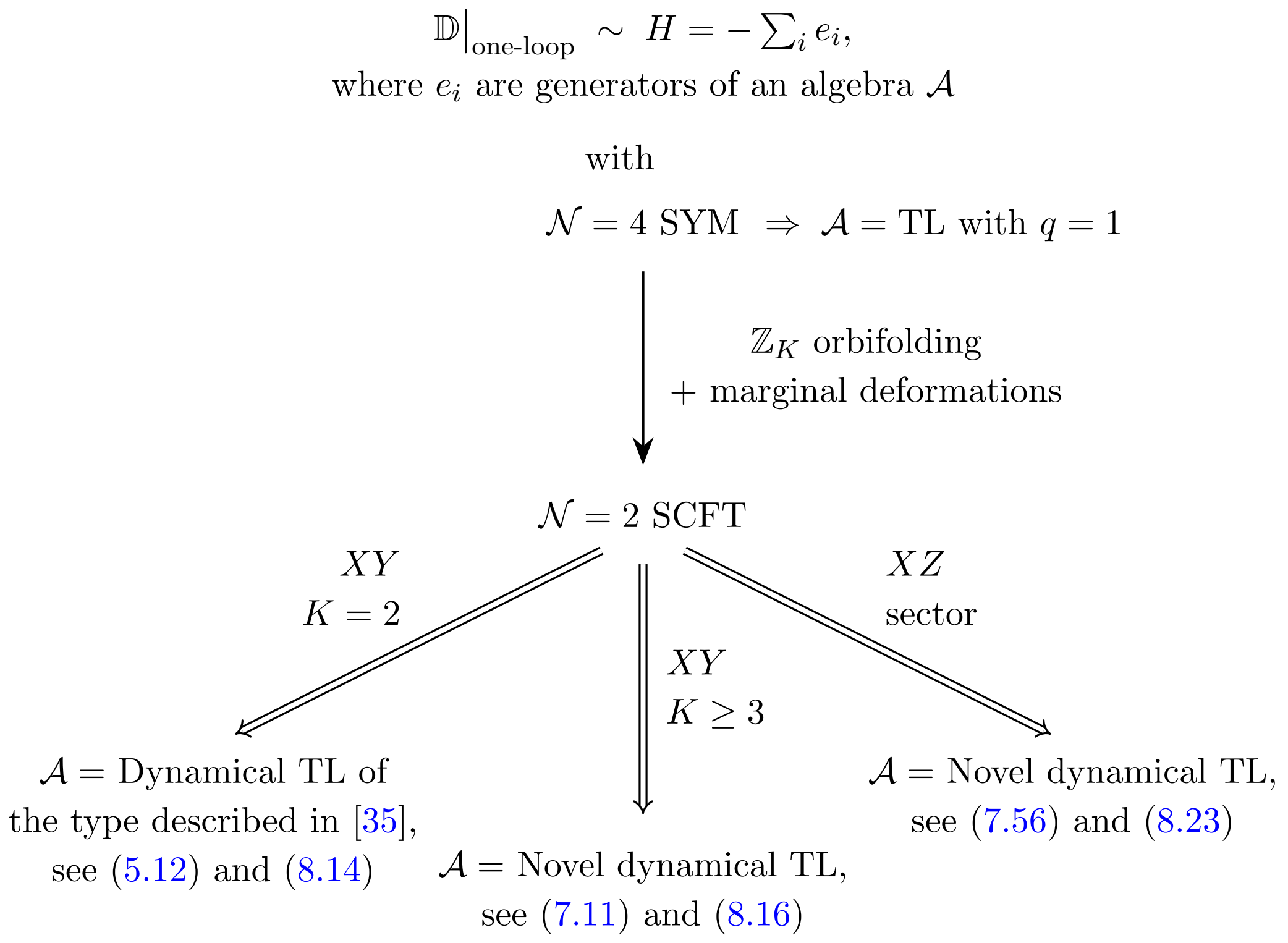

Our main results are summarised in Figure~\ref{fig:main-results}. The structure of this paper is as follows: In the next section, we will review the standard Temperley-Lieb algebra and its relation to the XXZ spin chain. In Section~\ref{sec:N=4SYM}, we will show how the TL algebra at $q=1$ arises from the superspace description of the $\Ncal=4$ SYM one-loop dilatation operator. We then turn to the $\Ncal=2$ orbifolds of $\Ncal=4$ SYM and their marginal deformations. After a brief review of these theories in Section~\ref{sec:orbifold}, we will turn to the construction of the dynamical Temperley-Lieb algebra in the (``dense'') XY sector (in Section~\ref{sec:XYsector}) and the (``dilute'') XZ sector (in Section~\ref{sec:XZsector}) of the $\Zset_2$ orbifold theory. In both cases we will illustrate how one works with this new algebra by explicitly considering some examples for small lengths. In Section~\ref{sec:Zk} we will consider the extension to $\Zset_K$ orbifolds. In Section~\ref{sec:twist} we will see how each generator of the dynamical TL algebra of the XZ sector is related, generator by generator, to a standard XXZ (or XXX) TL generator by colour-dependent twists reminiscent of dynamical Drinfeld twists. Finally, in Section~\ref{sec:loop} we formulate our algebras in the language of loop models. We conclude with several open questions and a few technical appendices.

\section{XXZ spin chains and the Temperley-Lieb algebra}\label{sec:XXZ}

In this section we briefly review some well-known aspects of the XXZ spin chain model, as it was in this context that the Temperley-Lieb algebra was originally constructed \cite{Temperley:1971iq}. For more details on the XXZ model, its integrable structure, and its quantum group symmetry, we refer to textbooks such as \cite{Gomez96}. In particular, we showcase how the description of the XXZ Hamiltonian as elements of the TL algebra is very powerful in determining the spectrum: on each magnon sector, the action of the Hamiltonian is fully captured by elementary manipulations on diagrams and can subsequently be diagonalised straightforwardly. Moreover, we will see how this description naturally captures the type of (anti-)symmetrisation encoded in the Hamiltonian, and provides a natural -- link states -- basis to construct the irreducible representations that make up the spectrum. In later sections, we will see how from Feynman diagrammatics we can obtain the same results for the one-loop (bulk) Hamiltonians of $\mathcal{N}=4$ and $\mathcal{N}=2$ quiver gauge theories. 

\paragraph{XXZ and Temperley-Lieb.}
The XXZ model is a nearest-neighbour spin chain 
whose Hamiltonian is a deformation of that of the XXX model in the sense that the coefficient of one of the three Pauli matrix terms is taken to be different from the other two. Denoting by $q\in \C^\times$ the deformation parameter, we set this Hamiltonian to be 
\begin{equation}
	\label{eq:xxz_ham}
	H_{\rm XXZ} = -\frac{1}{2} \left(\sum_{i=1}^{L-1} \left( \sigma_i^x \sigma_{i+1}^x + \sigma_i^y \sigma_{i+1}^y \right) + \sum_{i=1}^{L-1} \frac{q+q^{-1}}{2}\sigma_i^z \sigma_{i+1}^z + \frac{q-q^{-1}}{2} (\sigma_L^z - \sigma_1^z)\right)\, ,
\end{equation}
with $\{\sigma_i^x, \sigma_i^y, \sigma_i^z\}$ the Pauli matrices forming a local representation of $\mathfrak{sl}(2)$ on the $i$-th tensor leg of the spin chain Hilbert space $V^{\otimes L}$ with $V \cong \C^2 = \C \ket{\uparrow} \oplus \C \ket{\downarrow}$. The overall choice of sign is such that the Hamiltonian is ferromagnetic, as this is the case of relevance to planar gauge theory. We will focus exclusively on the case where $q$ is not a root of unity. 
The boundary term $-(q-q^{-1})(\sigma_L^z-\sigma_1^z)/4$ is chosen such that the Hamiltonian has $U_q(\mathfrak{sl}(2))$ symmetry \cite{Gomez96}, as will be discussed in more detail below. It is convenient to add a constant to express the Hamiltonian in the form 
\be
\label{eq:TL_ham}
H=H_{\rm XXZ}+(L-1)(q+q^{-1})/4=\sum_{i=1}^{L-1} \Hcal_{i,i+1}\;\; \text{where}\;\;\Hcal_{i,i+1}=\left(\begin{array}{cccc}0&0&0&0\\0&q&-1&0\\0&-1&q^{-1}&0\\0&0&0&0\end{array}\right)\;,
\ee
and such that the lowest energy (that of the vacuum state) is zero. The Temperley-Lieb generators $e_i$ are now simply defined as 
\be\label{eq:TL_ham_gens}
	e_i=-\Hcal_{i,i+1}\;,
\ee
or in other words 
\begin{equation}
	\label{eq:TL_rep}
	e_i  = I\otimes \ldots \otimes E \otimes \ldots \otimes I\, , \quad E= -\begin{pmatrix}
		0 & 0& 0 & 0 \\
		0 & q & -1 & 0 \\
		0 & -1& q^{-1} & 0 \\
		0 & 0& 0 & 0
	\end{pmatrix}\, , 
\end{equation}
with $E$ acting on the $i$th and $(i+1)$th spaces as a $q$-antisymmetriser. 
In these conventions, and defining $\beta = -(q+q^{-1})$, the $e_i$ satisfy the relations\footnote{The minus sign in the quadratic TL-algebra relation is conventional, and can be scaled away by sending $e_i \mapsto - e_i$. As we will see in the next sections, though, the minus sign is natural from the supersymmetric gauge theory point of view, and hence we will keep it.} 
\be \label{TLrels}
e_i^2=\beta e_i\;\;,\;\; e_ie_{i\pm1}e_i=e_i \;\;,\;\; e_ie_j=e_je_i \;,\;|i-j|>1\;.
\ee
The (unital, associative, complex) algebra generated by the $L-1$ generators $e_i$ subject to these relations is the Temperley-Lieb algebra, which we denote by $\mathrm{TL}(q)$, labelling it by $q$ rather than by the loop fugacity $\beta$. We suppress the dependence on $L$, which can be inferred from context. 

\paragraph{Temperley-Lieb as a diagrammatic algebra.}
The generators $e_i$ admit a natural graphical description as 
\be \label{TLXXZ}
e_i=\;\;
\, , 
\ee
where each dot is a tensor leg, continuing strands are untouched by the action of $e_i$ and connected horizontal dots indicate $q$-antisymmetrisation. In this notation, composition can be performed graphically by stacking one operator on top of another. Focusing just on sites $i,i+1$, we have:
\be
e_i^2=\;\;%
=
\beta~ e_i\, ,
\ee
where the removal of a loop has produced a factor of $\beta=-(q+q^{-1})$. For this reason, $\beta$ is known as the loop fugacity. We can similarly depict the cubic relation graphically:
\be
e_1e_2e_1=\;\;%
=e_1\, .
\ee
In the second step we have straightened the strand, which can then be seen to represent the identity at site 3. The other cubic relation in \eqref{TLrels} can be obtained similarly. 

For completeness one should also include the identity element, which simply extends all strands from the bottom to the top of a diagram:
\be \label{TLidentity}
I=\;\;%
\, . 
\ee

\paragraph{Quantum Schur-Weyl duality.} For generic values of $q$ (that is, $q$ not a root of unity) $\mathrm{TL}(q)$ is semisimple, making its representation theory straightforward. For simplicity, in this section we restrict to the case $L=2N$. 

As its name indicates, the XXZ model does not have the (isotropic) global $\mathfrak{sl}(2)$-symmetry, and only the $S^z$-part of the symmetries of the XXX model remain unbroken. Nevertheless, the particular boundary conditions in the Hamiltonian \eqref{eq:xxz_ham} are such that it commutes with an action of the quantum group $U_q(\mathfrak{sl}(2))$ on $V^{\otimes L}$ generated by
\begin{equation}
	\label{eq:uqsl2}
	\begin{aligned}
		q^{S^z} &= \prod_{i=1}^L q^{\sigma^z_i/2} = q^{\sigma^z/2} \otimes q^{\sigma^z/2} \otimes \ldots \otimes q^{\sigma^z/2}\, , \\
		S_q^\pm &= \sum_{i=1}^L \underbrace{q^{-\sigma^z/2} \otimes\ldots \otimes q^{-\sigma^z_{}/2}}_{(i-1)-\text{fold}} \otimes \, \sigma^\pm \otimes \underbrace{ q^{\sigma^z_{}/2}  \ldots \otimes q^{\sigma^z/2}}_{(L-i)-\text{fold}}\, , 
	\end{aligned}
\end{equation}
satisfying 
\begin{equation}
	\label{eq:uqsl2_relations}
	q^{S^z} S_q^\pm q^{-S^z} = q^{\pm1} S_q^\pm \, , \quad [S_q^+,S_q^-] = \frac{q^{2S^z}-q^{-2S^z}}{q-q^{-1}}\,. 
\end{equation}
In other words, the sum of generators in \eqref{eq:TL_ham} is $U_q(\mathfrak{sl}(2))$-invariant. Even more, the action of the entire algebra $\mathrm{TL}(q)$ (and not just the generators) commutes with this $U_q(\mathfrak{sl}(2))$-action in such a way that one cannot extend either algebra without breaking this commutativity; they are maximally commutative. Hence they can be diagonalised simultaneously on the Hilbert space, yielding a decomposition\begin{equation}
	\label{eq:schur_weyl}
	V^{\otimes L} \cong \bigoplus_j V_j \otimes T_j\, , 
\end{equation}
with the $j$ indexing all appearing irreducible representations, $V_j$ is an irreducible representation of $U_q(\mathfrak{sl}(2))$ and $T_j$ one of $\mathrm{TL}(q)$. This is an incarnation of quantum Schur-Weyl duality \cite{chariQuantumAffineAlgebras1996}.

As we will show in more detail below, $H$ preserves each of the $T_j$, and can be diagonalised on each separately. Since $H$ further commutes with $U_q(\mathfrak{sl}(2))$, \eqref{eq:schur_weyl} implies that the full spectrum of $H$ can be determined with the help of combinatorics and diagrammatic computations  \cite{Read:2007ti}: its eigenvalues follow from diagrammatics on the $T_j$, and their multiplicities are measured by the dimension of each $V_j$. Since the latter are the standard irreducible representations of $U_q(\mathfrak{sl}(2))$, these dimensions are known too. 

\paragraph{The link state basis.}
To see how the above works in practice, let us now turn to the representation space $T_j \subset V^{\otimes L}$ of the Temperley-Lieb algebra. As the algebra element $e_i$ is a $q$-antisymmetriser, we expect that its representations can be constructed by suitably symmetrising and antisymmetrising the vector spaces at each site. 
Therefore, we consider the following type of diagrams 
\be \label{eq:links_unlinks}

\ee
consisting of $2(N-j)$ linked and $2j$ unlinked spins, with the links being non-intersecting.\footnote{The restriction $L=2N$ makes sure both sets are of even cardinality, simplifying the counting.} Since the links represent $q$-antisymmetrisation, a pair of two linked spins form a $q$-singlet $\ket{\uparrow \downarrow} - q^{-1}\ket{\downarrow \uparrow}$. The unlinked lines represent such spins that $q$-antisymmetrising any neighbouring pair (ignoring any pairs of linked spins in between) annihilates them. So, for two sites we have:
\be
=\left\{\begin{array}{c}\ket{\downarrow\downarrow}\\\ket{\uparrow\downarrow} +q \ket{\downarrow\uparrow}\\
\ket{\uparrow\uparrow}\end{array}\right.\, .
\ee
The unlinked state can represent any state in the $q$-deformed triplet. As another example, for length $4$ the diagram
\be
%
\ee
 can represent any of the following:
\begin{equation}
	\ket{\uparrow \uparrow \downarrow \uparrow} - q^{-1} \ket{\uparrow \downarrow \uparrow  \uparrow}\, , \quad \ket{\downarrow \uparrow \downarrow \downarrow} - q^{-1} \ket{\downarrow \downarrow \uparrow  \downarrow}\, , \quad (\ket{\uparrow \uparrow \downarrow \downarrow}+ q\ket{\downarrow \uparrow \downarrow \uparrow} )  - q^{-1} (\ket{\uparrow \downarrow \uparrow  \downarrow}+ q\ket{\downarrow \downarrow \uparrow  \uparrow})\, .
\end{equation}
Each state is $q$-antisymmetric in spaces $2$ and $3$, whilst each is also $q$-symmetric in spaces $1$ and $4$ (as they are adjacent after ignoring the linked pair). For fixed $j$, these diagrams form a basis of the irreducible module $T_j$, and each diagram stands for a whole spin-$j$ $U_q(\mathfrak{sl}(2))$ multiplet $V_j$, one state for each way of filling in the unlinked sites. The diagram states are linearly independent but in general not orthogonal. Since at generic $q$ the dimensions are the same as at $q=1$, $\dim T_j$ equals the number of standard Young tableaux of shape $(N+j,N-j)$, which is also the number of non-crossing link diagrams with $2j$ unlinked sites. As we will soon show in an example, the basis obtained by (anti-)symmetrisation as prescribed by Young tableaux contains crossing diagrams, and hence is nevertheless different to the link basis. 

These diagrams can be obtained by cutting in half diagrams depicting certain elements of $\mathrm{TL}(q)$, and can alternatively be described by sequences of nested parentheses and dots, with the parentheses obeying usual typographical nesting rules. For example, the diagram \eqref{eq:links_unlinks} would be $(())\circ \circ ()$ (see \cite{Read:2007ti}). 

Applying any of the $e_i$ to a diagram of $T_j$ can only yield the following options: if sites $(i,i+1)$ were unlinked before, $e_i$ annihilates the diagram. In all other cases $e_i$ creates a link between the sites at $i$ and $i+1$, and potentially moves a linked pair one step in either direction. Hence the resulting diagram is also in $T_j$, and $\mathrm{TL}(q)$ leaves $T_j$ invariant. In particular, this means $H$ also leaves $T_j$ invariant, and we can diagonalise $H$ on each $T_j$ separately. 

\paragraph{Example: The $L=4$ Hamiltonian.}To see how this works, consider length $4$ again. $T_2$ is the module containing no antisymmetrised indices. Therefore it is spanned by a single diagram, consisting of only unlinked spins. It corresponds to 5 possible magnon sectors, of $S^z$-weights $-2,\ldots 2$ 
\be 
\label{eq:link_state_l4}
v_1=\begin{tikzpicture}[scale=0.78,baseline={([yshift=-0.5ex]current bounding box.center)}]
  \draw[-,thick] (-0.25,0)--(3.25,0);
    \draw[-,thick,\lcyantwo] (0,0)--(0,-0.75); \draw[-,thick,\lcyantwo] (1,0)--(1,-0.75);
  \draw[-,thick,\lcyantwo] (2,0)--(2,-0.75); \draw[-,thick,\lcyantwo] (3,0)--(3,-0.75);
  \filldraw[red,thick] (0,0) circle (1pt); \filldraw[red,thick] (1,0) circle (1pt);
  \filldraw[red,thick] (2,0) circle (1pt); \filldraw[red,thick] (3,0) circle (1pt); 
\end{tikzpicture}\;=\;\left\{\begin{array}{l}\ket{\downarrow\downarrow\downarrow\downarrow}\\
\ket{\uparrow\downarrow\downarrow\downarrow}+q\ket{\downarrow\uparrow\downarrow\downarrow}+q^{2}\ket{\downarrow\downarrow\uparrow\downarrow}+q^{3}\ket{\downarrow\downarrow\downarrow\uparrow}\\
\ket{\uparrow\uparrow\downarrow\downarrow}+q\ket{\uparrow\downarrow\uparrow\downarrow}+q^{2}\ket{\uparrow\downarrow\downarrow\uparrow}+q^{2}\ket{\downarrow\uparrow\uparrow\downarrow}+q^{3}\ket{\downarrow\uparrow\downarrow\uparrow}+q^{4}\ket{\downarrow\downarrow\uparrow\uparrow}\\
\ket{\uparrow\uparrow\uparrow\downarrow}+q\ket{\uparrow\uparrow\downarrow\uparrow}+q^{2}\ket{\uparrow\downarrow\uparrow\uparrow}+q^{3}\ket{\downarrow\uparrow\uparrow\uparrow}\;\\
\ket{\uparrow\uparrow\uparrow\uparrow}
\end{array}\right.
\ee
and depicts only unlinked spins. Let us stress that one should not confuse this diagram with the identity element \eqref{TLidentity} in $\mathrm{TL}(q)$. Acting with $H$ on the identity element simply gives back $H$. When acting on the link state \eqref{eq:link_state_l4} $H$ acts in an explicit $\mathrm{TL}(q)$-representation, and follows the annihilation rules described above. Hence $H$ annihilates this state by the above rules for $T_j$.

The next module, $T_1$, contains one antisymmetrisation and is spanned by the three diagrams\footnote{We use the standard convention that e.g. $|\downarrow\uparrow\rangle \otimes |\downarrow\downarrow \rangle = |\downarrow\uparrow\downarrow\downarrow\rangle$.}
\be\label{T1}\begin{split}
v_2&=
=\;\left\{\begin{array}{c}(\ket{\uparrow\downarrow}-q^{-1}\ket{\downarrow\uparrow})\otimes\ket{\downarrow\downarrow}\\
(\ket{\uparrow\downarrow}-q^{-1}\ket{\downarrow\uparrow})\otimes(\ket{\uparrow\downarrow}+q\ket{\downarrow\uparrow})\\
(\ket{\uparrow\downarrow}-q^{-1}\ket{\downarrow\uparrow})\otimes\ket{\uparrow\uparrow}\end{array}%
\right.\;.
\end{split}\ee
We see that to each diagram belong three vectors with distinct magnon number, corresponding to their $q^{S^z}$-eigenvalue. Specifying this eigenvalue to be $q^{-1},1$ or $q^{+1}$ picks out the top, middle or bottom states of each diagram respectively. 
In each resulting magnon sector, the Hamiltonian in the basis $\{v_2,v_3,v_4\}$ is 
\begin{equation}
H=	%
\, . 
\end{equation}
This can easily be determined graphically by stacking the TL-operators vertically on top of a state. For example, $e_1$ acts on $v_2$ as follows 
\begin{align} \label{TLaction1}
e_1\left(%
\, \;,
\end{align}
where the factor came from removing the closed loop. Acting with $e_2$ we find
\begin{align}
e_2\left(%
\;,
\end{align}
where there is no factor as all we did was straighten the line. The third generator $e_3$ on this same state annihilates it, because it is a  $q$-antisymmetriser acting on a $q$-symmetrised state.

\begin{figure}[t!]
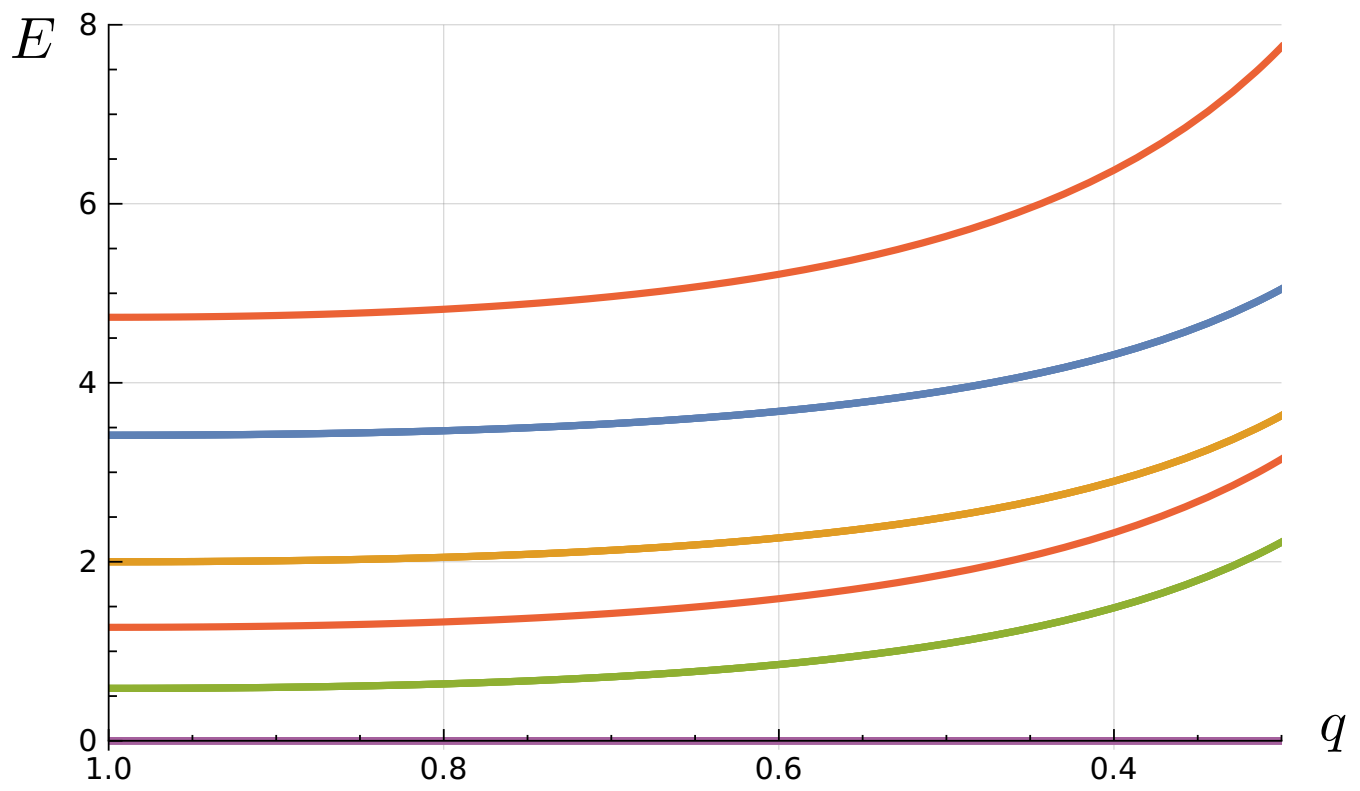

  \begin{center}
    %
  \caption{The spectrum of the XXZ Hamiltonian \eqref{eq:TL_ham} for $L=4$. The two singlet states are denoted in red, the spin-1 states in green, orange and blue, and the spin-2 (BPS) states in purple.} \label{GraphL4XXZ}
  \end{center}
\end{figure}

Finally, $T_0$ is the module containing two pairs of antisymmetrised indices, and is spanned by two diagrams
\be\label{T2}
\begin{split}v_5&=%
=\ket{\uparrow}\otimes(\ket{\uparrow\downarrow}-q^{-1}\ket{\downarrow\uparrow})\otimes\ket{\downarrow}-q^{-1}\ket{\downarrow}\otimes(\ket{\uparrow\downarrow}-q^{-1}\ket{\downarrow\uparrow})\otimes\ket{\uparrow}\;.
\end{split}
\ee
The Hamiltonian in this basis is
\be
H= %
\, .\label{eq:HXXZ-2links}
\ee
Graphically, we can check that \begin{align}
e_1\left(%
\,,\\
\intertext{
and acting with $e_3$ gives exactly the same. On the other hand, acting with $e_2$ gives}
e_2\left(%
\,,
\end{align}
where there is no factor as no closed loop is formed and we only straightened the line, so that $e_2\,v_5=v_6$. Altogether, we obtain exactly the first line in \eqref{eq:HXXZ-2links}.
 Counting all the states, we have found that the appearing $V_j$ have dimensions $5$, $3$ and $1$, thus in total we have found 
\begin{equation}
	5 \times 1+ 3 \times 3+ 1 \times 2   = 16 = 2^4
\end{equation}
states, as expected for $L=4$. Diagonalising the small matrices yields the full spectrum:
\begin{equation}
\begin{tabular}{c|c}
    Energy & Multiplicity \\
    \hline
   $ 0 $   &  $5$ \\
    $q+q^{-1}+\sqrt{2}$ & 3 \\
   $ q+q^{-1} $ & 3 \\
    $q+q^{-1}-\sqrt{2}$ & 3 \\
    $\half(3(q+q^{-1})+\sqrt{8+(q+q^{-1})^2})$ & 1 \\
 $\half(3(q+q^{-1})-\sqrt{8+(q+q^{-1})^2})$ & 1 
\end{tabular}
\end{equation}
Figure~\ref{GraphL4XXZ} shows the XXZ spectrum as one deforms away from the XXX point $q=1$ towards lower values of $q$. 

\paragraph{Relating link states to Young tableaux.} For readers more familiar with Young tableaux as the organising principle of $\mathfrak{sl}(2)$ representations, we note that the link states are simply a different choice of basis, which has the advantage that it can be represented graphically without crossings. In Appendix~\ref{app:young} we make this explicit for the $L=4$ link states \eqref{eq:link_state_l4}, \eqref{T1} and \eqref{T2}. 

\paragraph{Baxterisation and generalisations.} The above discussion has only scratched the surface of the Temperley-Lieb algebra and its applications. Referring to e.g. \cite{Pasquier:1989kd,Ridout:2012gg} for further aspects, we will simply state the well-known fact that the TL generator can be promoted to an $R$-matrix satisfying the Yang-Baxter equation by introducing a spectral parameter, a process known as Baxterisation \cite{Jones:1990hq}. Specifically, for the XXZ case, one can define the permuted $R$ matrix $\check{R}=PR$ as 
\be
\check{R}_{i,i+1}(u)=1+\frac{\sin{u}}{\sin(u-\eta)}\, e_i\;,
\ee
with the fugacity related to $\eta$ as $\beta=-2\cos\eta$ (i.e. $q=e^{i\eta}$). The fact that this $R$-matrix satisfies the Yang-Baxter equation implies the integrability of the XXZ spin chain (which also holds true for other boundary conditions than those in \eqref{eq:xxz_ham}, whose algebraic description requires the one- and two-boundary extensions of the TL algebra \cite{Martin:1993jka,deGier:2009twoboundaryTL}) and allows for the computation of the spectrum via the Algebraic Bethe Ansatz approach, see \cite{Faddeev:1994nk, Gomez96} for pedagogical treatments. 

Any representation of $\mathrm{TL}(q)$ defines a spin chain with Hamiltonian $-\sum_i e_i$, and many such chains beyond XXZ can be solved by Bethe ansatz \cite{GhiottoMalvezzi00}. For the representation theory of the algebra, see \cite{Westbury95,Ridout:2012gg}. The diagrammatic calculus above is that of Kauffman's description \cite{Kauffman87} of the Jones polynomial \cite{Jones85}, later axiomatised as planar algebras \cite{jones1999planaralgebrasi}.

The Temperley-Lieb algebra admits many generalisations \cite{Martin91}, of which two are relevant for us. In the \emph{dilute} TL algebra \cite{BehrendPearce97,Belletete:2013zla} boundary points may remain free of strands, while in \emph{coloured} versions the colours are carried either by the strands \cite{GrimmPearce93,Grimm:2003am,DiFrancesco:1998hb,Babichenko:2007vz} or by the regions between them \cite{Howat_2026}. The latter algebras are closely related to face models \cite{Pasquier:1986jc,Pasquier:1987xj, Pearce:1990ila}, where the heights at the corners of the faces play the role of the colour. Recently, such models have also been approached from the point of view of fusion categories \cite{Blakeney:2025ext}. Both the dilute and coloured-region generalisations will reappear in the $\Ncal=2$ quiver theories we will consider. Most relevant for us are the \emph{dynamical} TL algebras of \cite{Ren:2023mtn}, whose structure constants depend on a label shifted from site to site. As we will review in Section \ref{sec:twist}, the possible values of this label can be mapped to the colours of the regions, while the rules for shifting the label (or not) can also produce the dilute property. Therefore the formalism is powerful enough to describe both our desired extensions.  

\section{The $q=1$ Temperley-Lieb algebra from $\mathcal{N}=4$ SYM}\label{sec:N=4SYM}

  The above discussion of the standard Temperley-Lieb algebra has as input the symmetry structure  of the XXZ model, with the XXX model recovered as the limiting case $q=1$. Since the XXX model arises from the one-loop planar dilatation operator of $\Ncal=4$ SYM, one can 
  wonder whether the Temperley-Lieb algebra at $q=1$ can also be naturally described in the gauge theory. 
   In the following sections we will see that answering this question
  leads us to 
  a
  generalisation of the Temperley-Lieb algebra relevant to the $\Ncal=2$ cyclic quiver theories.

  \paragraph{$\Ncal=4$ SYM in superspace.}   The most direct route to the TL algebra is to use the superspace formulation of the gauge theory (for introductions, see \cite{Gates:1983nr,Wess:1992cp}, as well as the lecture notes \cite{Pomoni:2019oib} which focus on the current $\Ncal=2$ context). Its power 
  is particularly evident in holomorphic sectors, where at one loop there is only one type of chiral diagram which is divergent and therefore contributes to the planar dilatation operator (for a review, see \cite{Sieg:2010jt}). Although the more commonly used component formalism is of course equivalent, the intermediate steps of the computation involve more diagrams and therefore slightly obscure the TL structure.   
  Even more importantly, in superspace all the vertices are cubic, which will allow us to directly match the structure of the dilatation operator to that of the Temperley-Lieb generator.  
\begin{figure}[h]\begin{center}\begin{tikzpicture}[scale=0.8]
    \filldraw[\lcyantwo,thick] (0,0) circle (10pt);
    \draw[->,thick,\lcyantwo] (-0.5,0.3) arc (220:-30:0.6);
    \draw[->,thick,\lcyantwo] (0.5,0.2) arc (130:-150:0.6);
    \draw[->,thick,\lcyantwo] (0.1,-0.6) arc (-10:-260:0.6);
    \node[scale=0.9] at (0,0) {$1$};
    \node[scale=0.7] at (0,1.6) {$X$};
    \node[scale=0.7] at (1.7,-0.9) {$Y$};
         \node[scale=0.7] at (-1.4,-1.2) {$Z$};
  \end{tikzpicture}\end{center}
  \caption{The quiver diagram representing $\Ncal=4$ SYM. The node indicates the single $\SU(N)$ gauge group. The three holomorphic scalars are depicted as arrows coming back to the node, implying that they are in the adjoint representation of $\SU(N)$.} \label{Fig:N4quiver}
  \end{figure}
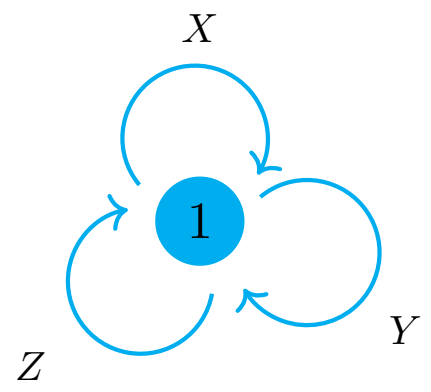
  
  Recall that the field content of $\Ncal=4$ SYM is an $\SU(N)$ gauge field, four fermions and six scalar fields, all in the adjoint representation of the gauge group. In $\Ncal=1$ superspace language, the gauge field and one of the fermions are combined into a vector superfield $V$, while three holomorphic combinations of the scalars and fermions combine into three chiral superfields, denoted $\Phi^i=X,Y,Z$. This field content is illustrated by the quiver in Figure~\ref{Fig:N4quiver}. The chiral-field part of the action of $\Ncal=4$ SYM in $\Ncal=1$ superspace notation is
  \be
  \label{eq:N4_SYM_action}
  \mathcal{S}=\int\diff^4x \diff^4\theta \Tr \left(e^{-g V}\overline{\Phi}^ie^{g V}\Phi^i\right)
  +\left(\int\diff^4x \diff^2\theta\Wcal+ \text{h.c.}\right)~.
  \ee
We suppress the vector kinetic and ghost terms %
, which we will not need.  The $\Ncal=4$ SYM superpotential is
\be \label{N4superpotential}
 \Wcal=ig \left(\Tr(XYZ)-\Tr(XZY)\right)~,
 \ee
 with $g$ the Yang-Mills coupling constant. As all the fields are in the adjoint representation, it is convenient to use double-line notation  \cite{tHooft:1973alw}. The holomorphic vertices encoded in $\Wcal$ 
 are
 \be
 \begin{tikzpicture}[scale=0.7,baseline=-0.8cm]
   \draw[-,thick,violet] (0,0)--(1,-1)--(1,-2);
   \draw[-,thick,violet] (1.2,-2)--(1.2,-1)--(2.2,0);
   \draw[-,thick,violet] (0.15,0.15)--(1.1,-0.8)--(2.05,0.15);
\node at (-0.3,0.3) {$X$};\node at (1.15,-2.3) {$Y$};\node at (2.4,0.3) {$Z$};
 \end{tikzpicture}=g \;\quad \;
  \begin{tikzpicture}[scale=0.7,baseline=-0.8cm]
   \draw[-,thick,violet] (0,0)--(1,-1)--(1,-2);
   \draw[-,thick,violet] (1.2,-2)--(1.2,-1)--(2.2,0);
   \draw[-,thick,violet] (0.15,0.15)--(1.1,-0.8)--(2.05,0.15);
\node at (-0.3,0.3) {$Z$};\node at (1.15,-2.3) {$Y$};\node at (2.4,0.3) {$X$};
 \end{tikzpicture}=-g~,
 \ee
and the antiholomorphic ones 
 \be
 \begin{tikzpicture}[scale=0.7,baseline=0.8cm]
   \draw[-,thick,violet] (0,0)--(1,1)--(1,2);
   \draw[-,thick,violet] (1.2,2)--(1.2,1)--(2.2,0);
   \draw[-,thick,violet] (0.15,-0.15)--(1.1,0.8)--(2.05,-0.15);
\node at (-0.3,-0.3) {$\bar{Z}$};\node at (1.15,2.3) {$\bar{Y}$};\node at (2.4,-0.3) {$\bar{X}$};
 \end{tikzpicture}=g \;\quad \;
  \begin{tikzpicture}[scale=0.7,baseline=0.8cm]
   \draw[-,thick,violet] (0,0)--(1,1)--(1,2);
   \draw[-,thick,violet] (1.2,2)--(1.2,1)--(2.2,0);
   \draw[-,thick,violet] (0.15,-0.15)--(1.1,0.8)--(2.05,-0.15);
\node at (-0.3,-0.3) {$\bar{X}$};\node at (1.15,2.3) {$\bar{Y}$};\node at (2.4,-0.3) {$\bar{Z}$};
 \end{tikzpicture}=-g~.
 \ee
The ordering of the fields in the traces in \eqref{N4superpotential} maps directly to the cyclic ordering of the vertices.  We will use an anticlockwise convention for reading off the vertices. 
That is, if the ordering is $XYZ$ or $\bar{X}\bar{Y}\bar{Z}$ (read anticlockwise) we have a plus sign, while if the ordering is $XZY$ or $\bar{X}\bar{Z}\bar{Y}$ we get a minus.

The above vertices are expressed in the standard convention  where all the fields are taken to be outgoing. In the following, it will be convenient to flip the direction of some of the fields, so that they are incoming. These fields will then be replaced by their conjugates. The advantage of this convention is that the quantum numbers at the top of a vertex will equal those at the bottom.\footnote{Specifically, consider the $\SU(3)$ weights, which using the standard $\lambda^3, \lambda^8$ matrices can be assigned as $X:(1/2,\sqrt{3}/6), Y:(-1/2,\sqrt{3}/6), Z:(0,-\sqrt{3}/3)$, and therefore $\bar{Y}$ and $XZ$ have the same weights.} Our notation for such vertices is as follows:
\be \label{N4verticeshol}
 \begin{tikzpicture}[scale=0.8,baseline=1.7cm]
\draw[->,\lcyantwo,thick](-2,1.6)--(-2,2.6);\node at (-2.3,2.1) {$t$};
   \draw[violet,thick](0.91,1.3)--(0.9,2);
   \draw[violet,thick](1.09,1.3)--(1.09,2);
   \draw[violet,thick](1.8,3) arc (0:-180:0.8);
   \draw[violet,thick](2,3) arc (0:-85:1);
  \draw[violet,thick](0,3) arc (-180:-95:1);
   \node at(0.1,3.3) {$X$};\node at (1.9,3.3) {$Z$};
\node at (1,1){$\bar{Y}$};
 \end{tikzpicture}=g\;,\quad
  \begin{tikzpicture}[scale=0.8,baseline=1.7cm]

   \draw[violet,thick](0.91,1.3)--(0.9,2);
   \draw[violet,thick](1.09,1.3)--(1.09,2);
   \draw[violet,thick](1.8,3) arc (0:-180:0.8);
   \draw[violet,thick](2,3) arc (0:-85:1);
  \draw[violet,thick](0,3) arc (-180:-95:1);
   \node at(0.1,3.3) {$Z$};\node at (1.9,3.3) {$X$};
\node at (1,1){$\bar{Y}$};
\end{tikzpicture}=-g\;.\;\quad 
 \ee
where we have flipped the direction of the outgoing $Y$ field to make it an incoming $\bar{Y}$. For the antiholomorphic vertices we flip the direction of both fields at the bottom, converting them to incoming holomorphic fields:
 \be
 \begin{tikzpicture}[scale=0.8,baseline=1cm]
\draw[->,\lcyantwo,thick](-2,0.6)--(-2,1.6);\node at (-2.3,1.1) {$t$};
   \draw[violet,thick](1.8,0) arc (0:180:0.8);
   \draw[violet,thick](2,0) arc (0:85:1);
   \draw[violet,thick](0,0) arc (180:95:1);
   \draw[violet,thick](0.91,0.99)--(0.9,1.7);
   \draw[violet,thick](1.09,0.99)--(1.09,1.7);
   \node at  (0.1,-0.3) {$Z$};\node at (1.9,-0.3) {$X$};
    \node at (1.1,2) {$\bar{Y}$};
 \end{tikzpicture}=g\;,\quad
 \begin{tikzpicture}[scale=0.8,baseline=1cm]
   \draw[violet,thick](1.8,0) arc (0:180:0.8);
   \draw[violet,thick](2,0) arc (0:85:1);
   \draw[violet,thick](0,0) arc (180:95:1);
   \draw[violet,thick](0.91,0.99)--(0.9,1.7);
   \draw[violet,thick](1.09,0.99)--(1.09,1.7);
   \node at  (0.1,-0.3) {$X$};\node at (1.9,-0.3) {$Z$};
    \node at (1.1,2) {$\bar{Y}$};
 \end{tikzpicture}=-g\;.\;
\ee
To emphasise which fields are incoming and which are outgoing we
have indicated the time direction 
in the preceding diagrams, 
but we will suppress it in the following. 

\paragraph{The one-loop dilatation operator.} We are interested in the planar spectral problem of the theory, for which we wish to compute the two-point function 
\be
\langle \mathcal{O}(x)\bar{\mathcal{O}}(y)\rangle=\frac{1}{|x-y|^{2\Delta}}
\ee
to determine the scaling dimension $\Delta$.
Let us focus on an $\SU(2)$ sector, for instance the one made up of only holomorphic $X$ and $Z$ fields. Then the operators above will be traces of the type $\mathcal{O}(x)=\Tr(X(x)Z(x)\cdots Z(x))$ consisting of products of $L$ operators all inserted at the same spacetime point $x$, and can be depicted in double-line notation as
\be \label{Ox}
\mathcal{O}_\Delta(x)=\,
\,\, . 
\ee
At tree level, the scaling dimension of this operator is just $\Delta=L$. Divergent loop corrections lead to $g$-dependent corrections to $\Delta$, with $\Delta-L$ known as the anomalous dimension. In $\Ncal=1$ superspace, a typical one-loop contribution will look like 
\be
%
\, , 
\ee
with all lines indicating index loops and the top line constituting the conjugate operator $\bar{\mathcal{O}}$. We note that this diagram has one more index loop compared to the tree-level diagram where all the fields in $\mathcal{O}$ are Wick contracted with those in $\bar{\mathcal{O}}$ with propagators. The one-loop Feynman integral (taken to be at the bottom loop) is divergent in dimensional regularisation, and therefore contributes to the anomalous dimension. In the 't Hooft limit where $N$ is large but $\lambda=g^2N$ is held fixed, non-planar diagrams are suppressed. Furthermore, other superspace one-loop diagrams (e.g. those involving gauge fields) are finite \cite{Sieg:2010jt}. Therefore, this is the only type of diagram that contributes to the anomalous dimension. Following the effective vertex method \cite{Beisert:2002bb,Beisert:2003tq}, instead of computing the two-point function directly we can consider the dilatation operator, which measures the scaling dimension of $\mathcal{O}_\Delta$, as 
an operator
 \be \label{Hsuperspace}
 \Hcal_{i,i+1}=(-1)\times \;\;\begin{tikzpicture}[scale=0.6,baseline=0.8cm]
\draw[violet,thick](1.8,0) arc (0:180:0.8);
  \draw[violet,thick](2,0) arc (0:85:1); \draw[violet,thick](0,0) arc
  (180:95:1); \draw[violet,thick](0.91,0.99)--(0.9,2);
  \draw[violet,thick](1.09,0.99)--(1.09,2); \draw[violet,thick](1.8,3) arc
  (0:-180:0.8); \draw[violet,thick](2,3) arc (0:-85:1);
  \draw[violet,thick](0,3) arc (-180:-95:1);
  \draw[violet,thick](-1.7,0)--(-1.7,3); \draw[violet,thick](-1.9,0)--(-1.9,3);
  \draw[violet,thick] (3.7,0)--(3.7,3);  \draw[violet,thick] (3.9,0)--(3.9,3);
  \draw[violet,thick,dashed] (-2.3,1.5)--(-2.9,1.5);
  \draw[violet,thick,dashed] (4.3,1.5)--(4.9,1.5);
  \node at (-1.6,-0.4)[scale=0.9] {$i-1$};  \node at (0.1,-0.4)[scale=0.9] {$i$};\node at (1.9,-0.4)[scale=0.9] {$i+1$};\node at (3.8,-0.4)[scale=0.9] {$i+2$}; 
 \end{tikzpicture} \, , 
\ee
which acts on $\mathcal{O}_\Delta(x)$ by Wick contraction.\footnote{In standard conventions, the fields at the bottom of $\Hcal$ are antiholomorphic and contract with the holomorphic fields in $\mathcal{O}$. In our preferred ``bottom incoming"- ``top outgoing" conventions, holomorphic fields propagate to holomorphic so the bottom fields of $\Hcal$ will also be holomorphic.} The factor of $(-1)$ enters because in our graphical approach we do not include a factor of $(-1)$ from the loop integration (see \cite{Sieg:2010jt}), which is crucial for the ferromagnetic behaviour of the spin chain. Omitting it leads to simpler diagrammatics but it needs to be restored in the end to obtain the correct Hamiltonian. 
The full one-loop dilatation operator
is then simply
\be
H=\sum_{i=1}^{L-1} I_1\otimes I_2\cdots \otimes \Hcal_{i,i+1}\cdots \otimes I_{L-1}\otimes I_L\, ,
\ee
and can be interpreted as a nearest-neighbour spin-chain Hamiltonian \cite{Minahan:2002ve}. For single-trace operators the chain is periodic, and the sum should also include the term $\Hcal_{L,1}$ coupling the last and first sites. Throughout this paper we work with open chains, for which this term is absent.

Before proceeding to clarify the link to the Temperley-Lieb algebra, let us briefly review how this superspace picture translates to the component computation of the dilatation operator that might be more familiar to some readers. Recall that the superspace chiral superfields are expansions in the fermionic coordinate $\theta^\alpha$. Specifically, we can write $\Phi^i=\phi^i+\sqrt{2}\theta^\alpha\psi_\alpha^i+\theta^2 F^i+\cdots$, where the fermions $\psi^i_\alpha$ are the superpartners of the scalar fields $\phi^i$, while $F^i$ are scalar auxiliary fields which are required by off-shell supersymmetry. The $\cdots$ contain derivatives of $\phi^i$ and $\psi^i$. Expanding the kinetic terms in \eqref{eq:N4_SYM_action} leads to the standard kinetic terms for the scalars and fermions coupled to the gauge field, however they do not lead to a kinetic term for $F^i$, just a contribution of the form $\bar{F}^iF^i$. Having no kinetic term, $F^i$ is an auxiliary field and can be integrated out algebraically. 

We wish to work with operators made up only of the scalar components of the superfields $X$ and $Z$. Therefore, we project the incoming and outgoing fields to their lowest components, $X|_0=\phi_X$ and $Z|_0=\phi_Z$. However, the Feynman rules for this process include an integration over superspace, $\int\diff^4\theta=\int \diff^2\theta\diff^2\bar{\theta}$. As this integral needs to be saturated for the answer not to vanish, it picks out the auxiliary-field components of the $Y$ and $\bar{Y}$, reducing the $Y\bar{Y}$ propagator to that of $F_Y\bar{F}_Y$. As mentioned, these fields do not propagate, so the vertical propagator collapses to a point. Integrating out the $F_Y$'s produces a quartic potential term for the scalars. Pictorially, the above procedure can be depicted as follows:\footnote{Here we temporarily revert to the standard outgoing convention for all fields, which is more appropriate to obtain a term in the Lagrangian.} 
\be
\int\diff^4\theta~
 \begin{tikzpicture}[scale=0.8,baseline=1.1cm]
\draw[violet,thick](1.8,0) arc (0:180:0.8);
  \draw[violet,thick](2,0) arc (0:85:1); \draw[violet,thick](0,0) arc
  (180:95:1); \draw[violet,thick](0.91,0.99)--(0.9,2);
  \draw[violet,thick](1.09,0.99)--(1.09,2); \draw[violet,thick](1.8,3) arc
  (0:-180:0.8); \draw[violet,thick](2,3) arc (0:-85:1);
  \draw[violet,thick](0,3) arc (-180:-95:1);
 \node at (0.1,-0.4)[scale=0.7] {$\bar{X}|_0$};\node at (1.9,-0.4)[scale=0.7] {$\bar{Z}|_0$}; \node at
  (0.1,3.3)[scale=0.7] {$X|_0$};\node at (1.9,3.3)[scale=0.7] {$Z|_0$};
   \node at (1.3,1.25)[scale=0.6] {$\bar{Y}$};\node at (1.3,1.8)[scale=0.6] {$Y$};
 \end{tikzpicture}=
 \begin{tikzpicture}[scale=0.8,baseline=1.1cm]
\draw[violet,thick](1.8,0) arc (0:180:0.8);
  \draw[violet,thick](2,0) arc (0:85:1); \draw[violet,thick](0,0) arc
  (180:95:1); \draw[violet,thick](0.91,0.99)--(0.9,2);
  \draw[violet,thick](1.09,0.99)--(1.09,2); \draw[violet,thick](1.8,3) arc
  (0:-180:0.8); \draw[violet,thick](2,3) arc (0:-85:1);
  \draw[violet,thick](0,3) arc (-180:-95:1);
  \node at (0.1,-0.4)[scale=0.7] {$\bar{\phi}_X$};\node at (1.9,-0.4)[scale=0.7] {$\bar{\phi}_Z$};
  \node at(0.1,3.3)[scale=0.7] {$\phi_X$};\node at (1.9,3.3)[scale=0.7] {$\phi_Z$};
   \node at (1.35,1.25)[scale=0.6] {$\bar{F}_{Y}$};\node at (1.35,1.8)[scale=0.6] {$F_Y$};
 \end{tikzpicture}\;\;\rightarrow
\quad \begin{tikzpicture}[scale=0.8,baseline=0.7cm]
  \draw[violet,thick](0,0)--(1,1)--(0,2);
  \draw[violet,thick](0.15,-0.15)--(1.15,0.85)--(2.15,-0.15);
  \draw[violet,thick](2.3,0)--(1.3,1)--(2.3,2);
  \draw[violet,thick](0.15,2.15)--(1.15,1.15)--(2.15,2.15);
 \node at (-0.25,-0.3)[scale=0.7] {$\bar{\phi}_X$};\node at (2.45,-0.3)[scale=0.7] {$\bar{\phi}_Z$};
  \node at (-0.25,2.3)[scale=0.7] {$\phi_X$};\node at (2.45,2.3)[scale=0.7] {$\phi_Z$};
\end{tikzpicture}
 \ee
We have chosen one of the four possible combinations of $X$ and $Z$ fields. There are three other combinations in \eqref{Hsuperspace}, and including them brings us to the familiar component form of the scalar potential, namely $\Vcal=-g^2\Tr([\phi_X,\phi_Z][\bar{\phi}_X,\bar{\phi}_Z])$, which is the usual starting point for the computation of the dilatation operator in the $\SU(2)$ sector \cite{Minahan:2002ve, Beisert:2003tq}, see \cite{Minahan:2010js, deLeeuw:2015hxa} for reviews. We see that the vertical propagator can be removed from the notation.  However, perhaps one of the main takeaways of this work is that it can also be very useful to stay in superspace and therefore retain the $\bar{Y}$ propagator, as it will play an important role in the quiver case. 

\paragraph{The $\Ncal=4$ SYM Temperley-Lieb algebra.}

Let us now notice that, acting on each pair of neighbouring sites, the Hamiltonian \eqref{Hsuperspace} will receive contributions from both vertices in \eqref{N4verticeshol}, which differ by a sign. Therefore the Hamiltonian is an antisymmetriser, giving zero when acting on symmetric combinations and multiplying antisymmetric ones by two.\footnote{One typically writes $\Hcal_{i,i+1}=I_{i,i+1}-P_{i,i+1}$, with $P$ the permutation operation.}
This observation makes the link to the Temperley-Lieb algebra immediate. Let us define an operator summarising  the above one-loop contribution to the dilatation operator for any two neighbouring sites:
\be \label{TLN=4}
e=
\, .
\ee
\noindent 
Its action on the full chain is then: 
\be 
e_i=I_1\otimes I_2\otimes\cdots \otimes I_{i-1}\otimes e\otimes I_{i+2}\otimes \cdots \otimes I_{L}.\label{eq:e_i}
\ee
This is thought of as an operator acting on states of the form \eqref{Ox} by Wick contracting the antiholomorphic fields at the bottom with the holomorphic fields of the state. 
The vertical dashed line indicates the $\bar{Y}$ propagator. 

Let us now check that the algebra of the elements \eqref{TLN=4}
satisfies the defining relations of $\mathrm{TL}(q=1)$. Concatenating two elements $e_i$ by placing one on top of each other, we obtain graphically:
\be
e^2=

\ee
Let us focus on the internal loop, which from \eqref{TLN=4} corresponds to an incoming $\bar{Y}$ field splitting and rejoining. It evaluates to 
\be
%
\, . 
  \ee
As mentioned, although this is an actual loop integral in the gauge theory, we do not include its overall (divergent) value but only the contribution which comes directly from the superspace vertices.\footnote{In dimensional regularisation, the additional factor that we suppress is equal to $-1/((4\pi)^2\epsilon)$ \cite{Sieg:2010jt}. It is perhaps worth clarifying that, despite appearances, the loop does \emph{not} correspond to the renormalisation of the $\bar{Y}$ propagator, which would give zero in superspace. It should be understood as a consecutive action of the dilatation operator on the new composite operator arising after the first action.}  What is important in our present context is that a loop in  Temperley-Lieb language can be replaced  by a $\bar{Y}$ strand, times a factor of $-2g^2$. We emphasise that the minus sign arises because one of the vertices in each diagram is cyclic, while the other anticyclic. We conclude that $e_i$ as defined in \eqref{eq:e_i} satisfies the quadratic relation
\be
e_i^2=-2g^2 e_i\;.
\ee
Now let us consider the cubic relation. In the gauge language, the action of $e_1e_2e_1$ on a state with the field at site $3$ being an $X$ can be represented as follows:\footnote{We again emphasise that this diagram represents three consecutive actions of the one-loop Hamiltonian and not a higher-loop Hamiltonian. Also, as above, the focus here is purely on the contributions from the superpotential and we suppress the divergent loop factors.}
\be
  \scalebox{0.7}{
}\, .
\ee
For a nonzero answer, the fields at sites 1 and 2 should be either $XZ$ or $ZX$, and similarly for the outgoing ones. However, regardless of these choices, the choice of the third field completely determines the fields within the dashed area. This area contains two holomorphic and two antiholomorphic vertices, so the overall power of the coupling one obtains is $g^4$. The fact that the outgoing field at site 3 also has to be an $X$ allows us to rewrite the diagram as a product of a single Temperley-Lieb generator times an identity strand, resulting in the standard Temperley-Lieb cubic relation (up to a factor of $g^4$):

\be
\, \, .
\ee
 The case with a $Z$ field on site 3, as well as the $e_2e_1e_2$ relation can be understood similarly and produce the same result:
 \be
%
\, \, .
\ee

We can therefore summarise the action of the generators \eqref{eq:e_i} as
\be
e_i^2=-2g^2 e_i \;\;,\;\; e_ie_{i\pm 1} e_i=g^4 e_i \;\;, \;e_ie_j=e_je_i \;\text{for} \; |i-j|>1\, .
\ee
We can absorb the powers of $g$ by rescaling $e_i\ra g^2\tilde{e}_i$ so that we obtain the standard TL relations \eqref{TLrels} with $q=1$:
\be
\tilde{e}_i^2=-2 \tilde{e}_i \;\;,\;\; \tilde{e}_i\tilde{e}_{i\pm 1} \tilde{e}_i=\tilde{e}_i \;\;, \tilde{e}_i\tilde{e}_j=\tilde{e}_j\tilde{e}_i \;\text{for} \; |i-j|>1\, .
\ee
We conclude that our gauge-theoretic operator is, up to a factor of $g^2$, the Temperley-Lieb algebra generator \eqref{TLXXZ} evaluated at $q=1$:
\be \label{N4TLequivalence}
\begin{tikzpicture}[baseline=0.6cm]
\fill[\lcyan] (-0.25,0) rectangle (1.25,1.5);
 \draw[-,thick] (-0.25,0)--(1.25,0);
\draw[-,thick] (-0.25,1.5)--(1.25,1.5);
\draw[-,black,dashed,thick] (1,1.5) arc (0:-180:0.5);
\draw[-,black,dashed,thick] (1,0) arc (0:180:0.5);
\draw[-,gray,dashed,thick] (0.5,0.5)--(0.5,1);
  \filldraw[red,thick] (0,1.5) circle (1pt); \filldraw[red,thick] (1,1.5) circle (1pt);
\filldraw[teal,thick] (0,0) circle (1pt); \filldraw[teal,thick] (1,0) circle (1pt);
\end{tikzpicture}\;=g^2\;\begin{tikzpicture}[baseline=0.6cm]
 \draw[-,thick] (-0.25,0)--(1.25,0);
\draw[-,thick] (-0.25,1.5)--(1.25,1.5);
\draw[-,\lcyantwo,thick] (1,1.5) arc (0:-180:0.5);
\draw[-,\lcyantwo,thick] (1,0) arc (0:180:0.5);
  \filldraw[red,thick] (0,1.5) circle (1pt); \filldraw[red,thick] (1,1.5) circle (1pt);
\filldraw[teal,thick] (0,0) circle (1pt); \filldraw[teal,thick] (1,0) circle (1pt);
\end{tikzpicture} \;\qquad\text{(at $q=1$),}
\ee
Having established this equivalence, we could of course now simply drop the vertical line from the formalism. However, we will soon see that, in the quiver theories, keeping the vertical line is crucial, as there it (as well as all the other lines) can separate differently-coloured regions. Similarly, although in this case the power of $g^2$ can easily be absorbed, in the following we will encounter cases where it becomes dynamical, i.e. it will depend on the aforementioned colours and cannot be absorbed globally. Therefore, it is the notation on the left of \eqref{N4TLequivalence}, and not the one on the right,  which admits a natural generalisation to the quiver theories. 

\paragraph{Link state basis.}
Let us now consider the link states. By analogy with the usual TL picture, we can think of the states as the top half of the Temperley-Lieb generators. However, consistency requires that we also retain the incoming $\bar{Y}$ field in the notation, since we keep the $\bar{Y}$ propagator in the generators, and we want their action on the states to give back the same type of object.  Therefore, the link states are nothing but the $\Ncal=4$ SYM vertices \eqref{N4verticeshol}.

\be
\, .
\ee
Since the two vertices have opposite signs, this leads to the required antisymmetrisation. The state itself is read off at the top, that is, it does not include the auxiliary $Y$ line:\footnote{To make contact with the notation of Section \ref{sec:XXZ}, one identifies $X=\uparrow$, $Z=\downarrow$. }
\be
%
=g(XZ-ZX)\, .
\ee
At $L=3$ we have two states,
\begin{align}
&%
=g\,X(XZ-ZX) \;\;\text{or}\;\; g\,Z(XZ-ZX)
\end{align}
Notice that, just like the standard TL case, the notation does not distinguish between states related by the $\SU(2)$ raising/lowering operator, which in this case exchanges $X \leftrightarrow Z$. Proceeding 
to $L=4$ we find the two states in the doubly-linked sector:

\begin{align}
&%
=g^2X(XZ-ZX)Z-g^2Z(XZ-ZX)X\, .
\end{align}
Acting with the $e_i$ from \eqref{eq:e_i} on the link state, the allowed Wick contractions give four terms, which organise themselves into a link state as well. 
\be
%
\, .
\ee
This reproduces the action 
\eqref{TLaction1} of the Temperley-Lieb generators on the link states, at $q=1$. The other actions can be shown to work in an analogous way. 

Having seen how the Temperley-Lieb algebra at $q=1$ naturally arises from the superspace description of $\Ncal=4$ SYM, and that the 2-site link states correspond to the vertices of the superpotential, we can use the same recipe to
construct the analogous algebra for any supersymmetric theory. In the following we wish to apply this to the $\Ncal=2$ quiver theories, with our main focus being on the $\Zset_2$ orbifold case. 

\section{The $\Zset_2$ quiver theory}\label{sec:orbifold}

In this section we will briefly review some salient aspects of the $\Ncal=2$ quiver theory obtained by $\Zset_2$-orbifolding $\Ncal=4$ SYM and marginally deforming. We refer to the lecture notes \cite{Pomoni:2019oib} for more details. 

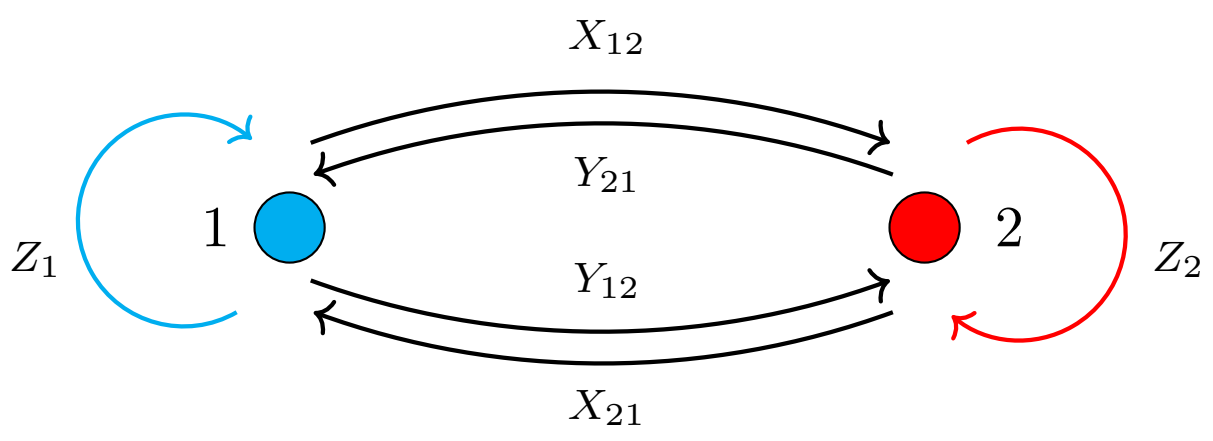
\begin{figure}[h!]
\begin{center}  
\hspace{1cm}\begin{tikzpicture}[scale=0.7]
  \draw[fill=\lcyantwo] (0,0) circle (2ex);
\draw[fill=red] (6,0) circle (2ex);
\draw[->,\lcyantwo,thick] (-0.5,-0.8) arc (-60:-310:1);
\draw[->,red,thick] (6.4,0.8) arc (120:-130:1);
\draw[->,black,thick] (0.2,0.8) arc (110:70:8);
\draw[->,black,thick] (5.7,0.5) arc (70:110:8);
\draw[->,black,thick] (5.7,-0.8) arc (-70:-110:8);
\draw[->,black,thick] (0.2,-0.5) arc (-110:-70:8);
  \node at (-0.7,-0) {$1$};\node at (6.8,-0) {$2$};
  \node at (3,1.8) {$\scriptstyle X_{12}$};\node at (3,-1.7) {$\scriptstyle X_{21}$};
 \node at (3,0.5) {$\scriptstyle Y_{21}$};\node at (3,-0.5) {$\scriptstyle Y_{12}$};
\node at (-2.4,-0.3) {$\scriptstyle Z_{1}$};\node at (8.4,-0.3) {$\scriptstyle Z_{2}$};
\end{tikzpicture}
\end{center}
\caption{The quiver corresponding to the $\Zset_2$ orbifold theory. The nodes correspond to the two $\SU(N)$ gauge groups. The bifundamental fields are represented as arrows between nodes, while the adjoint fields as arrows returning to the same node. For clarity, only the holomorphic content is shown.} \label{Z2quiver}
\end{figure}

To obtain the $\Zset_2$ orbifold, one starts with $\Ncal=4$ SYM with chiral-field action \eqref{eq:N4_SYM_action} and gauge group $\SU(2N)$ and performs the following projection:
\be\label{eq:orbifold}
\gamma\,V\,\gamma^{-1}=V\,,\;\;\gamma\,Z\,\gamma^{-1}=Z\,\;,\;\; \gamma X\gamma^{-1}=-X\,,\;\;\gamma Y\gamma^{-1}=-Y\;,
\ee
where $\gamma=\mathrm{diag}(1,-1)\otimes 1_N$. 
The projected fields take the following form in $N\times N$ blocks:
\be
  V=\twobytwo{V_{1}}{0}{0}{V_{2}}\;,  Z=\twobytwo{Z_{1}}{0}{0}{Z_{2}}\;,\;X=\twobytwo{0}{X_{12}}{X_{21}}{0}\;,\;Y=\twobytwo{0}{Y_{12}}{Y_{21}}{0} \, .
\ee
Therefore one obtains a theory with $\SU(N)_1\times \SU(N)_2$ gauge group, with $V_1,Z_1$ and $V_2,Z_2$ in the adjoint of the respective groups while the $X$ and $Y$ fields transform in the bifundamental representations $\Box\times \bar{\Box}$ and $\bar{\Box}\times\Box$. It is convenient to illustrate the field content as a quiver diagram, as shown in Figure~\ref{Z2quiver}. 

  As the adjoint fields $(V,Z)$ form an $\Ncal=2$ vector multiplet and the bifundamentals $(X_{a,a+1},Y_{a+1,a})$ (and their conjugates) form an $\Ncal=2$ hypermultiplet, the orbifolded theory has $\Ncal=2$ supersymmetry. Furthermore, the theory is finite (all gauge $\beta$ functions and anomalous dimensions of the fundamental fields vanish) and therefore the superconformal symmetry of the classical theory persists in the quantum theory as well \cite{Kachru:1998ys,Lawrence:1998ja}. 

At the orbifold point, the couplings of the two gauge groups are equal. Allowing them to vary independently is a marginal and $\Ncal=2$ supersymmetry-preserving deformation \cite{Kachru:1998ys,Lawrence:1998ja}.  In  $\Ncal=1$ superspace language, the chiral-field part of the action of the marginally deformed theory is obtained by inserting the projected fields into \eqref{eq:N4_SYM_action} and replacing $gV\to\mathrm{diag}(g_1V_1,g_2V_2)$ \cite{Pomoni:2019oib,Gadde:2010zi},
\be\label{eq:Z2_action}
\begin{split}
\mathcal{S}_{\Zset_2}=\int\diff^4x\,\diff^4\theta\,\Big[&\Tr\left(e^{-g_1V_1}\,\overline{Z}_1\,e^{g_1V_1}Z_1\right)+\Tr\left(e^{-g_2V_2}\,\overline{Z}_2\,e^{g_2V_2}Z_2\right)\\
&+\Tr\left(e^{-g_2V_2}\,\overline{X}_{21}\,e^{g_1V_1}X_{12}\right)+\Tr\left(e^{-g_1V_1}\,\overline{X}_{12}\,e^{g_2V_2}X_{21}\right)\\
&+\Tr\left(e^{-g_2V_2}\,\overline{Y}_{21}\,e^{g_1V_1}Y_{12}\right)+\Tr\left(e^{-g_1V_1}\,\overline{Y}_{12}\,e^{g_2V_2}Y_{21}\right)\Big]\\
&+\left(\int\diff^4x\,\diff^2\theta\,\Wcal_{\Zset_2}+\text{h.c.}\right),
\end{split}
\ee
where $\overline{X}_{21}\equiv X_{12}^\dagger$ etc., and as before we suppress the vector kinetic and ghost terms. The superpotential is 
\be \label{WZ2}
   \Wcal_{\Zset_2}=i g_1~ \Tr(Y_{21}Z_{1}X_{12}-X_{21}Z_{1}Y_{12})- i g_2~ \Tr(X_{12}Z_{2}Y_{21} -  Y_{12}Z_{2} X_{21}) \, ,
 \ee
 with the traces taken in the respective gauge groups. 
 
At the orbifold point $g_1=g_2=g$ this is simply the projection of the $\Ncal=4$ superpotential. Away from it, $\Ncal=2$ supersymmetry requires the coupling multiplying the terms with $Z_a$ to be the gauge coupling $g_a$ of the node $a$, since $(V_a,Z_a)$ form an $\Ncal=2$ vector multiplet.

Given a theory with superpotential $\Wcal$, the procedure of integrating out the $F$ auxiliary fields described in the previous section leads to the so-called $F$-term contribution to the scalar potential, which takes the form
\be
\Vcal_F=\sum_i \frac{\partial \bar{\Wcal}}{\partial \bar{\Phi}^i}\frac{\partial{\Wcal}}{\partial \Phi^i}
\ee
Another contribution to the scalar potential, known as the $D$-terms, comes from integrating out another auxiliary field inside the vector multiplet. However, for the holomorphic sectors we will consider, these terms do not contribute. The  $F$- and $D$- terms for the $\Zset_2$ quiver played an important role in \cite{Bertle:2024djm}, where it was shown that even though the $\SU(4)$ R-symmetry of $\Ncal=4$ SYM is broken, the Lagrangian of the quiver theory is still invariant under the action of a quantum-deformed version of $\SU(4)$ (more specifically, a twisted groupoid naturally defined in terms of the path algebra of the quiver in Figure~\ref{Z2quiver}). 
   
The spin-chain Hamiltonian modelling the one-loop dilatation operator of the
planar limit of this quiver theory was constructed in \cite{Gadde:2009dj,Gadde:2010zi} for the full sector originating from the $\SU(4)$ sector of the mother $\Ncal=4$ SYM theory. Here we will just focus on two holomorphic $\SU(2)$-like sectors, the XY and XZ sectors. They were discussed in great detail in \cite{Pomoni:2021pbj} from the perspective of the coordinate Bethe ansatz, with further results on the XZ sector appearing in \cite{Bozkurt:2024tpz,Bozkurt:2025exl}. Referring to those works for more details, below we briefly summarise some essential aspects required to follow the construction of the dynamical Temperley-Lieb algebras of these sectors in the sections to follow.

\subsection{The dense XY sector} \label{ReviewXY}

This sector consists of the two bifundamental fields $X$ and $Y$. It preserves the $\SU(2)$ symmetry of the mother $\Ncal=4$ SYM theory. In the notation of \cite{Pomoni:2021pbj}, where $\kappa=g_2/g_1$, the two-site Hamiltonian is 
\be
\label{eq:HXY}
\Hcal_1
=
\left(\begin{array}{cccc}
 0 & 0 & 0 & 0 \\
 0 & \kappa^{-1} & -\kappa^{-1} & 0 \\
 0 & -\kappa^{-1} & \kappa^{-1} & 0 \\
 0 & 0 & 0 & 0 \\
\end{array}
\right)
\;\;,
\;\;
\Hcal_2
=
\left(\begin{array}{cccc}
 0 & 0 & 0 & 0 \\
 0 & \kappa & -\kappa & 0 \\
 0 & -\kappa & \kappa & 0 \\
 0 & 0 & 0 & 0 \\
\end{array}
\right)
\;, \; \text{in the basis} \left(\begin{array}{c}XX\\XY\\YX\\YY\\ \end{array}\right)_{a=1,2} \, .
\ee
We are using a dynamical notation, where the subscripts indicate the first node index, e.g. $(XY)_1=X_{12}Y_{21}$ while $(XY)_2=X_{21}Y_{12}$. $\Hcal_{1,2}$ also denote the action on states with node index 1 or 2. 

The fact that both fields are bifundamental means that the node index strictly alternates as one moves along the chain: if the node index is 1 for odd sites, it will be 2 for even sites.  In that case, we have two copies of the XXX model Hamiltonian, where the ferromagnetic coupling alternates from $\kappa^{-1}$ to $\kappa$ for odd-even and even-odd sites, respectively.  

Despite the alternation, this sector has global $\SU(2)$ symmetry for generic $\kappa$, and therefore we do not expect to see any energy splittings between descendants in different magnon sectors, i.e.\ between members of the same $\SU(2)$ multiplet, which are obtained from its highest-weight state by repeatedly acting with the lowering operator and therefore have different numbers of magnons. In other words, the standard $\SU(2)$ Schur-Weyl duality holds and the states can be organised into representations of the symmetric group. We will discuss the dynamical Temperley-Lieb algebra related to this sector in Section~\ref{sec:XYsector}. 

\begin{figure}[t!]
  \begin{center}
    \begin{tikzpicture}
      \node at(0,0){\includegraphics[width=8cm]{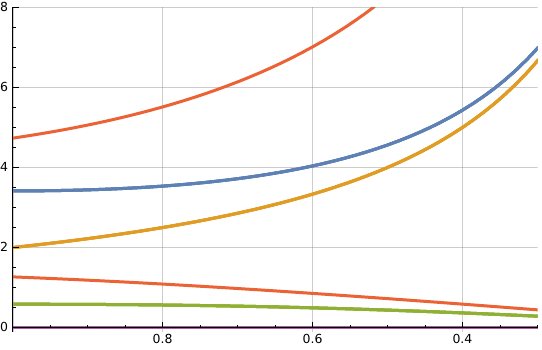}};
\node at (4.3,-2.2){$\kappa$};\node at (-4.3,2.4){$E$};\end{tikzpicture}
  \caption{The spectrum of the open $\Zset_2$ XY-sector Hamiltonian for $L=4$. The two singlet states are denoted in red, the spin-1 states in green, orange and blue, while the spin-2 (BPS) states in purple. Note that the energies at $\kappa=1$ are the same as those of the XXZ model at $q=1$ (see Figure~\ref{GraphL4XXZ}). As one deforms, all degeneracies are preserved, however the evolution of the eigenvalues is different to the XXZ case.} \label{GraphL4XY}
  \end{center}
\end{figure}

Figure~\ref{GraphL4XY} shows the spectrum of the open chain in the XY sector, for $L=4$, as one deforms away from the XXX point $\kappa=1$. 

In order to obtain the form \eqref{eq:HXY} in \cite{Pomoni:2021pbj}, the Hamiltonian one obtains from the gauge theory was rescaled by a factor of $g_1g_2$. For our current purposes, it is useful to reinstate this factor and write
\begin{equation}
\label{eq:HXYg1g2}
\Hcal_1
=
\begin{pmatrix}
0  & 0  & 0 & 0 \\
0  &  g_1^2 &  -g_1^2&  0 \\
 0 & -g_1^2  &  g_1^2  & 0  \\
0  & 0  & 0  & 0
\end{pmatrix}
\, , \quad
\Hcal_2
=
\begin{pmatrix}
0  & 0  & 0 & 0 \\
0  &  g_2^2 &  -g_2^2  &  0 \\
 0 & -g_2^2    &  g_2^2  & 0  \\
0  & 0  & 0  & 0
\end{pmatrix}
\;, \quad \text{in the basis} \left(\begin{array}{c}XX\\XY\\YX\\YY\\ \end{array}\right)_{a=1,2} \, .
\end{equation}
An advantage of keeping the factors of $g_1,g_2$ is that one can smoothly take the limit $\kappa\ra0$, which corresponds to Superconformal QCD, where one of the two gauge groups becomes global \cite{Gadde:2009dj,Gadde:2010zi}. But our main current motivation is that it is the Hamiltonian \eqref{eq:HXYg1g2} which generalises to the case of $\Zset_K$ orbifolds. 

To conclude our summary of the XY sector, let us consider the energy spectrum for $L=6$, separately for each magnon sector. We denote by $M$ the number of magnons, i.e.\ the number of $Y$ fields in a vacuum of $X$ fields (and, in the XZ sector below, the number of $Z$ fields). The XY sector, just like the XXZ chain, has spin-flip symmetry, so the sectors with $M$ and $L-M$ magnons have the same spectrum. For the example of $L=6$ we can therefore restrict our plots to $M=1,2,3$. %
To avoid overcrowded plots, Figures \ref{fig:m2XY} and \ref{fig:m3XY} only show the eigenvalues of the highest-weight vectors in the two- and three-magnon sectors. 

\begin{center}
  \captionsetup{type=figure}
  \captionsetup[subfigure]{skip=2pt}

  \setcounter{subfigure}{0}
  \begin{subfigure}{0.48\linewidth}
    \includegraphics[width=\linewidth]{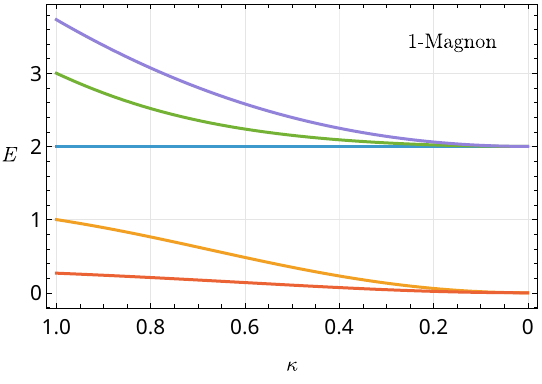}
    \caption{}\label{fig:m1XY}
  \end{subfigure}\hfill
  \begin{subfigure}{0.48\linewidth}
    \includegraphics[width=\linewidth]{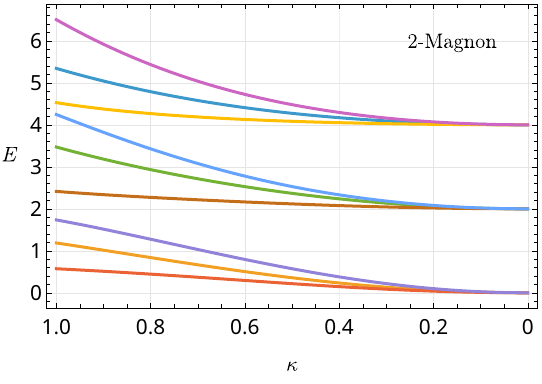}
    \caption{}\label{fig:m2XY}
  \end{subfigure}

  \vspace{1em}

  \begin{subfigure}{0.48\linewidth}
    \includegraphics[width=\linewidth]{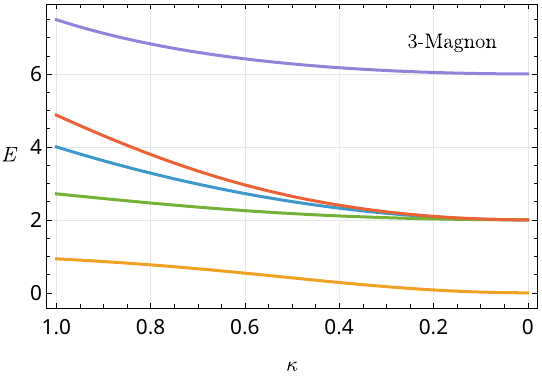}
    \caption{}\label{fig:m3XY}
  \end{subfigure}

  \nopagebreak
  \caption{The spectrum of the XY-sector Hamiltonian \eqref{eq:HXYg1g2} for $L=6$. Here we have fixed $g_1=1$ and plot for $\kappa=g_2$.  }
  \label{fig:allmagnonXYL6}
\end{center}

In Section~\ref{sec:XYsector}, we will construct the dynamical Temperley-Lieb algebra corresponding to the XY sector and (among other applications) reproduce the above spectrum in a diagrammatic way. 

\subsection{The dilute XZ sector}\label{sec:diluteXZ}

This sector contains one bifundamental and one adjoint field, taken to be $X$ and $Z$. These would form an $\SU(2)$ doublet in the mother $\Ncal=4$ theory, which naively\footnote{As shown in \cite{Bertle:2024djm} a groupoid version of $\SU(2)$ does persist.} is not present after orbifolding, as the fields are in different representations of the two gauge groups. In the notation of \cite{Pomoni:2021pbj}, the two-site Hamiltonian is
\begin{equation}
\label{eq:HXZ}
\Hcal_1
=
\begin{pmatrix}
0  & 0  & 0 & 0 \\
0  &  \kappa &  -1&  0 \\
 0 & -1  &  \kappa^{-1}  & 0  \\
0  & 0  & 0  & 0
\end{pmatrix}
\, , \quad
\Hcal_2
=
\begin{pmatrix}
0  & 0  & 0 & 0 \\
0  &  \kappa^{-1} &  -1 &  0 \\
 0 & -1   &  \kappa  & 0  \\
0  & 0  & 0  & 0
\end{pmatrix}
\;, \quad \text{in the basis} \left(\begin{array}{c}XX\\XZ\\ZX\\ZZ\\ \end{array}\right)_{a=1,2} \, ,
\end{equation}
with $\kappa=g_2/g_1$ as before, and where the subscripts indicate the first node index, so now $(XZ)_1=X_{12}Z_2$ while $(XZ)_2=X_{21}Z_1$. 

The resemblance to the Temperley-Lieb Hamiltonian \eqref{eq:TL_ham} is clear, and indeed for a single $X$ magnon propagating in a vacuum made up of $Z$ fields one obtains precisely the same dispersion relation as one would in the XXZ model. However, the $S$-matrix for two-magnon scattering (computed in \cite{Gadde:2010zi}) is different to XXZ, and in particular does not satisfy the standard YBE. Furthermore, one can also consider $Z$ magnons in a vacuum composed of $X$ fields, and the resulting dispersion relation is naturally parametrised by elliptic functions, as discussed in \cite{Pomoni:2021pbj}. 

In order to express the Hamiltonian in the above notation with $\kappa$, one needs to divide the gauge theory dilatation operator by a  factor of $g_1g_2$. To make the link to the superspace description more immediate, and also for the extension to $\Zset_K$ orbifolds, where the $\kappa$ notation is not as useful, it is convenient to re-introduce these factors. So in the following we will be using the XZ sector Hamiltonian in the form
\begin{equation}\label{eq:HXZg1g2}
\Hcal_1
=\begin{pmatrix}
0  & 0  & 0 & 0 \\
0  &  g_2^2 &  -g_1g_2&  0 \\
 0 & -g_1g_2  &  g_1^2  & 0  \\
0  & 0  & 0  & 0
\end{pmatrix}
\, , \quad
\Hcal_2
=
\begin{pmatrix}
0  & 0  & 0 & 0 \\
0  &  g_1^2 &  -g_1g_2 &  0 \\
 0 & -g_1g_2   & g_2^2  & 0  \\
0  & 0  & 0  & 0
\end{pmatrix}
\;, \; \text{in the basis} \left(\begin{array}{c}XX\\XZ\\ZX\\ZZ\\ \end{array}\right)_{a=1,2} \, .
\end{equation}
An immediate consequence of the
-- at least naively -- broken $\SU(2)$ symmetry is that different components of $\SU(2)$ multiplets acquire different energies as one deforms away from the orbifold point. As an example, for $L=3$ one finds the following energies for the two components of one spin-$\half$ doublet (labelled by its Young tableau, cf.\ Appendix~\ref{app:young}), both of which reduce to $E=1$ at the orbifold point $\kappa=1$. 
\be
\begin{tikzpicture}[scale=0.7,baseline=0.2cm]
  \draw[-] (0,0.5)--(0.5,0.5)--(0.5,1)--(0,1)--(0,0.5);
  \draw[-] (0.5,0.5)--(1,0.5)--(1,1)--(0.5,1)--(0.5,0.5);
   \draw[-] (0,0)--(0.5,0)--(0.5,0.5)--(0,0.5)--(0,0);
  \node[scale=0.8,violet] at (0.25,0.75) {1}; \node[scale=0.8,violet] at (0.75,0.75) {2};
    \node[scale=0.8,violet] at (0.25,0.25) {3};
\node at (-0.3,0.8){``};\node at (1.3,0.8){''};
\end{tikzpicture}=\left\{\begin{array}{ll}Z_1Z_1X_{12}+(1-\kappa) Z_1X_{12}Z_2 -\kappa X_{12}Z_2Z_2\;& \;\;E=\frac1\kappa-1+\kappa\;\\ \\
X_{12}X_{21}Z_1-Z_1 X_{12}X_{21} \; &\;\;E=\frac{1}{\kappa}
\end{array}\right.
\ee
The quotes are intended to alert the reader that we don't have an independent definition of this Young tableau for $\kappa\neq1$, but that it should reduce to the $\SU(2)$ case as $\kappa\ra1$. In the spin-chain context, the above splitting means that descendants of states in different magnon sectors will have different energies. This is unlike the case of XXZ where, as we saw, the quantum version of Schur-Weyl duality, or in other words the commuting action of the Hamiltonian with the $U_q(\mathfrak{sl}(2))$-coproduct, guarantees that all descendants have the same energy. 

Generically, the only exception to the energy splittings within multiplets is for the BPS states, which have $E=0$.  As supersymmetry prevents these states from acquiring an anomalous dimension, the whole BPS multiplet remains at $E=0$ as one deforms away from the orbifold point. At the orbifold point these states are totally symmetrised, and they retain this structure in the deformed theory, with additional factors of $\kappa$ appearing \cite{Gadde:2010zi}, similarly to the $q$-deformed case \eqref{eq:link_state_l4}. 

\begin{figure}[h!]
  \begin{center}
    \begin{tikzpicture}
      \node at(0,0){\includegraphics[width=8cm]{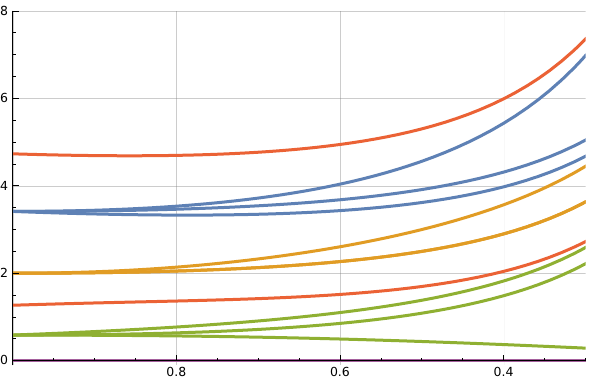}};
\node at (4.3,-2.2){$\kappa$};\node at (-4.3,2.4){$E$};\end{tikzpicture}
  \caption{The spectrum of the $\Zset_2$ XZ-sector Hamiltonian for $L=4$. The two singlet states are denoted in red, the spin-1 states in green, orange and blue, while the spin-2 (BPS) states in purple. The energies at $\kappa=1$ are the same as those of the XXZ model at $q=1$ (see Figure~\ref{GraphL4XXZ}). As one deforms, the BPS states stay BPS, and two of the orange spin-1 states have equal energies $\kappa+\kappa^{-1}$, but all other degeneracies are lifted.  Note that three energies (the lower blue, orange and upper green curves) are the same as those of the XXZ case: These are the states with a single-$X$ magnon in the $Z$ vacuum, whose dispersion relation is the same as XXZ.}  \label{GraphL4XZ}
  \end{center}
\end{figure}

Figure~\ref{GraphL4XZ} shows the spectrum in the XZ sector as one deforms away from the XXX point $\kappa=1$, for the $L=4$ open chain. Note the splitting of all spin-1 energy levels.

For the XXZ chain, the eigenvalues of $M$ and $L-M$ magnons coincide due to spin-flip symmetry. This is no longer the case in the XZ sector. In contrast to XXZ, where both states are on an equal footing, here $X$ and $Z$ are genuinely different fields: the former transforms in the bifundamental and the latter in the adjoint representation. The plot below illustrates this effect for $L=6$, where we show side by side the spectra for $M=1$ and $M=5$, and for $M=2$ and $M=4$. The differences are clearly visible, and they persist in the limit $\kappa\to 0$, which corresponds to $\mathcal{N}=2$ superconformal QCD.

For even $M$ (and even $L$) the number of $X$ fields is even, so a state starts and ends with the same colour. The chain then has a parity symmetry $P$, which reflects it, i.e.\ reverses the order of the fields and reads the gauge indices in the opposite direction, for instance $P\big(X_{12}Z_2X_{21}Z_1\big)=Z_1X_{12}Z_2X_{21}$. Parity maps each such sector to itself and commutes with the Hamiltonian for all $g_1,g_2$, so we show the $M=2$ and $M=4$ spectra separately for parity-even ($P=+1$) and parity-odd ($P=-1$) states. For odd $M$ the first and last colours differ, and reflection maps a state to the $\Zset_2$-conjugate sector. Combining it with $\Zset_2$ conjugation also exchanges $g_1\leftrightarrow g_2$, so at fixed couplings there is no parity symmetry, and the odd-$M$ spectra are not split.

\begin{center}
  \captionsetup{type=figure}
  \captionsetup[subfigure]{skip=2pt}

  \setcounter{subfigure}{0}
  \begin{subfigure}{0.48\linewidth}
    \includegraphics[width=\linewidth]{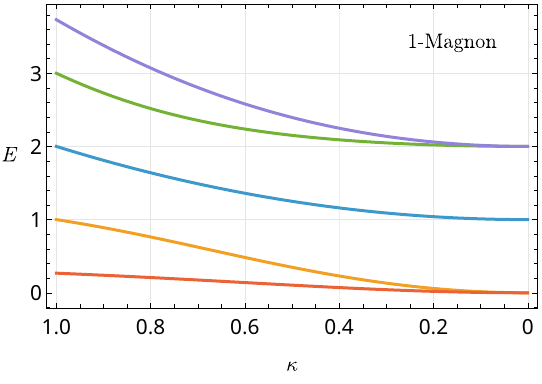}
    \caption{}\label{fig:m1XZ}
  \end{subfigure}\hfill
  \begin{subfigure}{0.48\linewidth}
    \includegraphics[width=\linewidth]{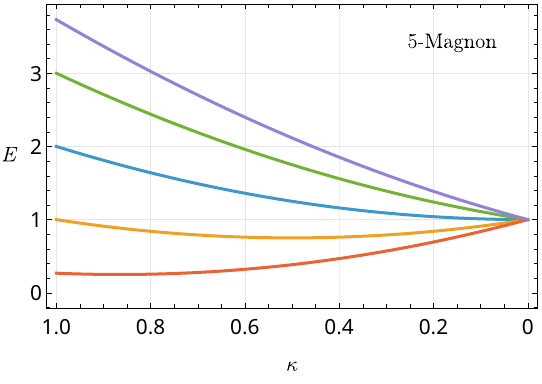}
    \caption{}\label{fig:m5XZ}
  \end{subfigure}

  \vspace{1em}

  \begin{subfigure}{0.48\linewidth}
    \includegraphics[width=\linewidth]{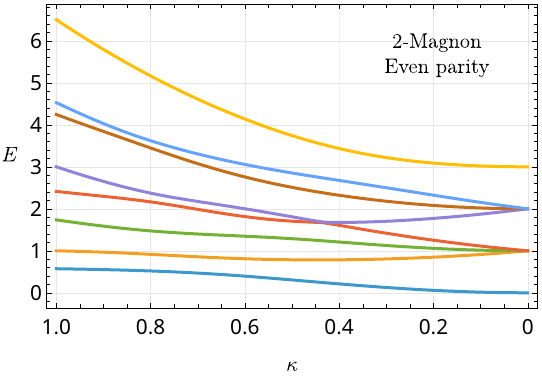}
    \caption{}\label{fig:m2XZeven}
  \end{subfigure}\hfill
  \begin{subfigure}{0.48\linewidth}
    \includegraphics[width=\linewidth]{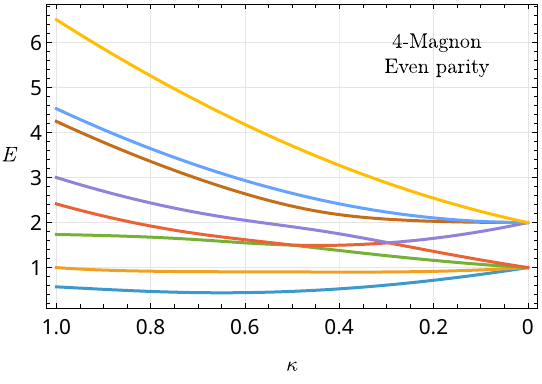}
    \caption{}\label{fig:m4XZeven}
  \end{subfigure}

  \vspace{1em}

  \begin{subfigure}{0.48\linewidth}
    \includegraphics[width=\linewidth]{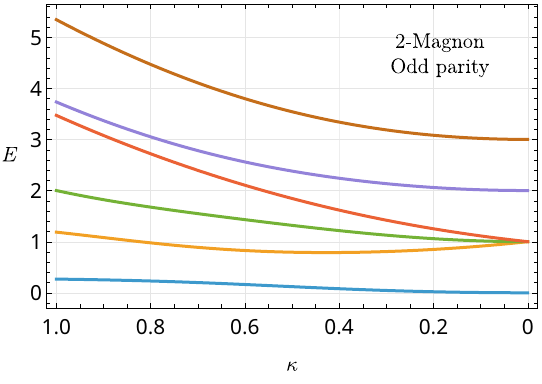}
    \caption{}\label{fig:m2XZodd}
  \end{subfigure}\hfill
  \begin{subfigure}{0.48\linewidth}
    \includegraphics[width=\linewidth]{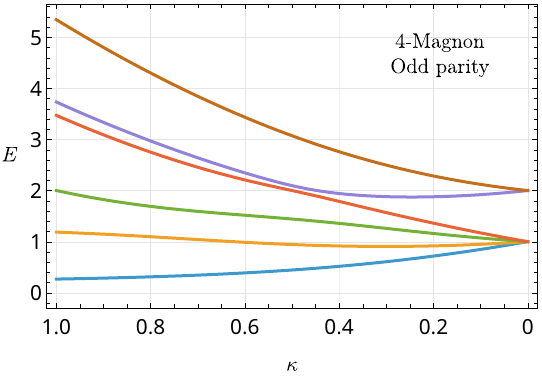}
    \caption{}\label{fig:m4XZodd}
  \end{subfigure}

  \vspace{1em}

  \begin{subfigure}{0.48\linewidth}
    \includegraphics[width=\linewidth]{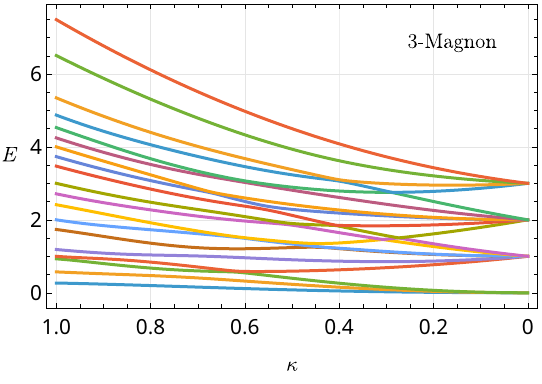}
    \caption{}\label{fig:m3XZ}
  \end{subfigure}

  \nopagebreak
  \caption{The spectrum of the XZ-sector Hamiltonian for $L=6$, with each plot illustrating a different magnon sector. The even-$M$ sectors are split by parity, with $P=+1$ in (c), (d) and $P=-1$ in (e), (f). For these plots we fix $g_1=1$ and plot $\kappa=g_2$.}
  \label{fig:allmagnonXZL6}
\end{center}

\begin{figure}[h!]
  \centering \includegraphics[width=0.7\linewidth]{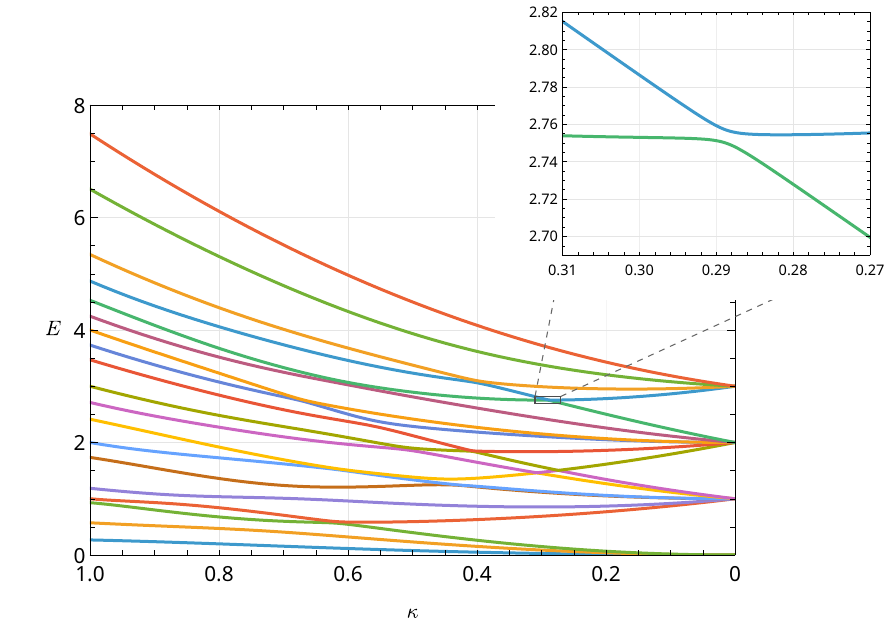}
  \caption{Spectrum of the 3-magnon XZ-sector Hamiltonian for $L=6$. The inset shows that although the lines approach each other, they do not actually cross. This is consistent with the fact that the energies in this sector are roots of a 19-degree polynomial, which does not factorise.}
  \label{fig:threeMagnonZoom}
\end{figure}

It is notable that for $M=2,3$ there are several points where eigenvalues seem to cross. However, by zooming in the plots, it is clear that all crossings are avoided.\footnote{With two exceptions: in Figure~\ref{fig:m4XZeven} there are two coinciding eigenvalues at $E=3/2$ for 
$\kappa=1/2$ and at $E=1.55102$ for $\kappa=0.29544$}  In  Figure~\ref{fig:threeMagnonZoom} we show one such avoided crossing  for the $M=3$ case. Of course, separating the $M=2,4$ plots according to parity sector was crucial to see this, as nothing prevents eigenvalues from different parity sectors from crossing.  

In Section \ref{sec:XZsector} we will study the Temperley-Lieb algebra which arises for the XZ sector of the $\Zset_2$ quiver theory, which will allow us to find the energies in a simple diagrammatic way. More importantly, through the TL construction we will be able to clearly pinpoint the origin of the above splittings in terms of mixing between sectors with different symmetry structures. 

\section{The dynamical Temperley-Lieb algebra in the $\mathbb{Z}_2$ XY sector }\label{sec:XYsector}

In Section~\ref{sec:N=4SYM} we saw how the diagrammatic Temperley-Lieb algebra generators are essentially the same as the superspace diagrams of $\Ncal=4$ SYM. In this and the next section we will follow the same steps to define the diagrammatic Temperley-Lieb algebra corresponding to the $\Zset_2$ quiver SCFT. 
We start with the XY sector, which is the simpler of the two sectors discussed in Section \ref{sec:orbifold}, as it preserves the $\SU(2)$ symmetry of the mother theory.

As reviewed in Section \ref{ReviewXY}, the XY sector is composed  of the bifundamentals $X_{12},Y_{12}$ and their $\Zset_2$ conjugates $X_{21},Y_{21}$. It can be described by a spin chain where the node index alternates between even and odd sites, as does the prefactor of the Hamiltonian. However, recall from the general discussion of the link state basis in Section \ref{sec:XXZ} that the number of links (pairs of antisymmetrised indices) is related to the $\SU(2)$ spin of a given state. 
Since the XY sector has an unbroken $\SU(2)$ symmetry, we therefore expect that the Hamiltonian preserves the number of links, and we will explicitly verify this below. However, as we will also see, the alternation of node indices will lead to a generalisation of the Temperley-Lieb algebra, where the elements (and therefore also the fugacity) will acquire dependence on the node index. We will express this dependence in a very intuitive way by introducing a colouring of the regions in the Temperley-Lieb diagrams. 

\subsection{Diagrams from superspace}\label{subsec:fromsuperspace-YXsector}

From the superpotential \eqref{WZ2} we see that the vertices relevant for the XY sector are those with an incoming adjoint field $\bar{Z}$ and two outgoing bifundamental fields $X,Y$:
\be
=-g_2\,.\label{eq:A2[0]forXY}
\ee
As in Section~\ref{sec:N=4SYM}, if the order, read anticlockwise, is $XYZ$ then the sign is positive, while if it is $XZY$ the sign is negative. The coupling constant $g_1$ or $g_2$ depends on whether there is a $Z_1$ or $Z_2$ in the vertex. We also have the conjugate vertices, again with the $X$ and $Y$ fields taken as incoming and the $\bar{Z}$ outgoing:
\be
%
=-g_2\, . \label{eq:vertices_upsidedown}
\ee
The computation of the Hamiltonian and Temperley-Lieb elements follows closely that for $\Ncal=4$ SYM in Section~\ref{sec:N=4SYM}. 
The main difference is that the fields now carry the indices of two different gauge groups. In the spirit of the dynamical Yang-Baxter equation \cite{Felder:1994be}, where the dynamical parameter is associated to regions on the plane, we colour the regions of a diagram according to the nodes: blue for node 1 and pink for node 2 (corresponding to the blue and pink index lines in the double-line notation used above). Crossing a solid line (a bifundamental field) swaps the colour, while crossing a dashed line (an adjoint field) preserves the colour. In the XY sector the line joining the two vertices of a generator is the adjoint field $Z$, which we draw as a dashed line, so all the colours are fixed once the colour of the outer region is chosen. We therefore have \emph{two} TL generators, one for each colour of the outer region:
\begin{subequations}
\label{eq:XY_TL_gens}
\begin{align}
e^{(1)} = 

 \end{align}
\end{subequations}
Summing over the allowed vertices, we can see that the sum of closed loops gives $-2g_a^2$ times the $\bar{Z}$-identity strand. For instance, for $a=1$, we have 
\be
 %
\,~\,.
\ee
This yields the following diagrammatic identities
\be
%
\,~\,,
 \ee
 which imply that the generators \eqref{eq:XY_TL_gens} satisfy
 \begin{equation}
 \label{eq:XYquadratic}
%
\, . 
\end{equation}
We conclude that the quadratic Temperley-Lieb relations are modified:
 \be \label{XYquadratic}
 e_1^{(1)}e_1^{(1)}=-2g_1^2 e_1^{(1)} \quad \text{and}\quad  e_1^{(2)}e_1^{(2)}=-2g_2^2 e_1^{(2)}.
\ee
We see that the coefficient in the quadratic Temperley-Lieb relation depends on the configuration, which in our notation translates to the colours of the regions. 

\paragraph{Dynamical notation.}In this section, we keep track of the colour to the left of the first site where the Temperley-Lieb elements act by explicitly indicating the node, see \eqref{eq:XY_TL_gens}. This notation is very intuitive and useful for concrete computations, but becomes unwieldy for the study of more formal aspects of the algebra. A powerful alternative is to use ``dynamical" notation, where the  algebra elements are taken to depend on a dynamical parameter $\lambda$, whose value at each site is linked to the colour of the region to its left (in the convention we use). As one progresses along the chain, its value is shifted according to the weights of the fields that are crossed, ensuring that the correct elements appear in expressions without having to explicitly keep track of the colours. This notation is motivated by the dynamical Yang-Baxter equation \cite{Felder:1994be}, which provides a quantum-group interpretation of elliptic models such as those related to the XYZ chain \cite{BaxterBook} and the Andrews-Baxter-Forrester SOS model \cite{Andrews:1984af}. In the $\Zset_2$ case there are two nodes and correspondingly the dynamical parameter is identified mod 2. We can therefore take the two distinct values of $\lambda$ to be 1 and 2. Since crossing either an $X$ or $Y$ field flips the colour, in the $\Zset_2$ XY sector the behaviour of the colours (and therefore of the dynamical parameter) is strictly alternating between even and odd sites. However the dynamical formalism extends to all the other cases we will consider. 

As our focus in this and the following sections is on establishing the properties of our algebras diagrammatically, we will keep our notation light, and not explicitly use the dynamical notation. Readers who are more interested in this aspect (and more formal applications) are referred to Section~\ref{sec:dynamical}. However, in view of this underlying structure, we will refer to relations such as \eqref{XYquadratic} as dynamical. 

Let us now move on to the cubic relations, which are obtained by considering diagrams like 
\be \label{XYcubicsuperspace}
\, .
\ee 
Here the incoming fields at sites 1 and 2 (not shown) can be in either of the two legal combinations, and similarly for the outgoing fields at sites 1 and 2. Fixing the incoming field at site 3 to be $X_{12}$, the only nonzero combination of vertices is the one indicated in the dashed box. Therefore the diagram on the left is equivalent to the value  of these vertices ($g_1^2g_2^2$) times a single TL generator, times the $X$ identity strand. Fixing the incoming field at site 3 to be $Y_{12}$ leads to the same result, but with the $Y$ identity strand.

The same cubic relation as in \eqref{XYcubicsuperspace} can be expressed in the dynamical Temperley-Lieb picture as 
\be
\, \, ,
\ee
where the field at site 3 can be either a $X_{12}$ or $Y_{12}$. 
We can similarly consider the analogue of \eqref{XYcubicsuperspace} that corresponds to the $e_2e_1e_2$ cubic relation. Translated to Temperley-Lieb language, it gives
\be%
\, \, .
\ee
The field at site 1 can be either an $X_{12}$ or $Y_{12}$. The coefficient is the same as for the $e_1e_2e_1$ relation. 

\be \label{XYcubic}
e_1^{(1)}e_2^{(2)}e_1^{(1)}=g_1^2g_2^2 ~e_1^{(1)}\quad \text{and}\quad
e_2^{(2)}e_1^{(1)}e_2^{(2)}=g_1^2g_2^2 ~e_2^{(2)} \, ,
\ee
and their $\Zset_2$ conjugates. Crucially, the coefficient in the cubic relation is $\Zset_2$ invariant and can be absorbed by the global rescaling $e^{(a)}\ra g_1g_2 \tilde{e}^{(a)}$. Doing so, we can summarise the relations satisfied by the rescaled TL generators in the XY sector as 
\be  \label{eq:XYTLrescaled}
\tilde{e}_i^{(a)}\tilde{e}_i^{(a)}=-2\kappa^{-s(a)} \tilde{e}_i^{(a)}\, ,  \quad \tilde{e}_i^{(a)}\tilde{e}_{i\pm 1}^{(a+1)}\tilde{e}_i^{(a)}=~\tilde{e}_i^{(a)}\, ,  \quad \tilde{e}_i^{(a)}\tilde{e}_{j}^{(b)} = \tilde{e}_{j}^{(b)} \tilde{e}_i^{(a)}\, ,\; |i-j|>1\;, 
\ee
where $\kappa=g_2/g_1$, $s(1)=1$ and $s(2)=-1$, and $a$ is identified mod 2. 
 
In words, the quadratic relation has become dynamical, while the cubic relation has not. These relations align with the dynamical TL algebras recently discussed in \cite{Ren:2023mtn}, as will become more apparent when written in terms of a dynamical parameter in \eqref{eq:XY_dTL_relations_K2}. In Section~\ref{subsec:ZK-XY} we will see that for $K>2$ both the quadratic and cubic relations will become dynamical, which to the best of our knowledge define a new type of dynamical TL algebra. In the following, we will continue to use the unrescaled generators, as those appear directly in the Hamiltonian. 

\subsection{Link state basis and diagonalisation}
\label{sec:basisXY}

As in Section~\ref{sec:N=4SYM}, the link states are the top halves of the TL generators, i.e.\ they are given by the vertices  \eqref{eq:A2[0]forXY}, where the vertical dashed line indicates the incoming $\bar{Z}$ field. Summing over the two orderings of the outgoing fields, we can then define the links as
\begin{align}
A_1&=A_1^{[0]}=\begin{tikzpicture}[baseline=-0.55cm,scale=1]
      \filldraw[\lcyan] (-0.25,0)--(-0.25,-1)--(0.5,-1)--(0.5,0)--(-0.5,0);
      \filldraw[\lcyan] (0.5,0)--(0.5,-1)--(1.25,-1)--(1.25,0)--(0.5,0);
      \filldraw[pink] (0,0) arc (-180:0:0.5);
         \draw[-,thick,gray] (-0.25,0)--(1.25,0);
        \draw[-,thick,black] (0,0) arc (-180:0:0.5);
        \draw[-,gray,thick,dashed] (0.5,-0.5)--(0.5,-1);
        \filldraw[red,thick] (0,0) circle (1pt); \filldraw[red,thick] (1,0) circle (1pt);
\end{tikzpicture}=g_1\left(Y_{12}X_{21}- X_{12}Y_{21}\right),\\
A_2&=A_2^{[0]}=\begin{tikzpicture}[baseline=-0.55cm,scale=1]
      \filldraw[pink] (-0.25,0)--(-0.25,-1)--(0.5,-1)--(0.5,0)--(-0.5,0);
      \filldraw[pink] (0.5,0)--(0.5,-1)--(1.25,-1)--(1.25,0)--(0.5,0);
      \filldraw[\lcyan] (0,0) arc (-180:0:0.5);
         \draw[-,thick,gray] (-0.25,0)--(1.25,0);
        \draw[-,thick,black] (0,0) arc (-180:0:0.5);
        \draw[-,gray,thick,dashed] (0.5,-0.5)--(0.5,-1);
        \filldraw[red,thick] (0,0) circle (1pt); \filldraw[red,thick] (1,0) circle (1pt);
\end{tikzpicture}=g_2\left(Y_{21}X_{12}- X_{21}Y_{12}\right).
\end{align}
The lower index indicates the first (and last) colour, 1 for blue and 2 for pink. The upper index indicates the level of nestedness, in this case zero. 

Using these links, we can straightforwardly construct a basis for the action of the Hamiltonian at each length. 
The procedure is very similar to what was described for XXZ in Section \ref{sec:XXZ}, but now with the addition of colour. The idea is to distribute links and vertical lines in all possible ways, without the links crossing each other or crossing lines. One should first write all possible states with no links, then all possible states with one link, then with two links, etc, and fill the rest with lines. The assigned colours should flip every time a line is crossed. To clarify the construction, we will look at the cases $L=4$ and $L=6$ in some detail.\footnote{We note that for the open chain, there is nothing special for either even or odd $L$. However, when going from $L=4$ to $L=5$, not much changes. The case $L=6$ is just a more interesting example than $L=5$.}

Clearly, for open chains a state starting with colour 1 (blue) can never mix with a state starting with colour 2 (pink). Therefore, in the examples below, we will write the basis and the Hamiltonians only for states starting with colour 1. The corresponding quantities for the states starting with colour 2 are obtained by their $\Zset_2$ conjugates, in other words by swapping $1\leftrightarrow 2$ (blue $\leftrightarrow$ pink) everywhere.

\subsubsection{Example: $L=4$}

\paragraph{No links.} For the case with no links, the state is composed of vertical lines 
\be 
\scalebox{0.7}{\begin{tikzpicture}[baseline=-0.6cm,scale=1]
     \filldraw[\lcyan] (-0.25,-1) rectangle (0,0);
     \filldraw[pink] (0,-1) rectangle (1,0);
     \filldraw[\lcyan] (1,-1) rectangle (2,0);
     \filldraw[pink] (2,-1) rectangle (3,0);
     \filldraw[\lcyan] (3,-1) rectangle (3.25,0);
     \draw[-,thick,gray] (-0.25,0)--(3.25,0);
     \foreach \x in {0,...,3} \draw[thick] (\x,0)--(\x,-1);
     \foreach \x in {0,...,3} \filldraw[red,thick] (\x,0) circle (1pt);
\end{tikzpicture}}\,.
\ee
This one diagram can represent five different states: 
\begin{itemize}
    \item the pseudo-vacuum: $X_{12}X_{21}X_{12}X_{21}$,
    \item the fully symmetrised state with one $Y$ and three $X$ (1-Magnon BPS state),
    \item the fully symmetrised state with two $Y$ and two $X$ (2-Magnon BPS state),
    \item the fully symmetrised state with three $Y$ and one $X$ (3-Magnon BPS state),
    \item the 4-magnon state: $Y_{12}Y_{21}Y_{12}Y_{21}$.
\end{itemize}

Since the Hamiltonian is a sum of antisymmetrisers (the generators in \eqref{eq:XY_TL_gens}), it gives zero when acting on any fully symmetrised state. Therefore, $E=0$ has degeneracy equal to 5. Additionally, we have to consider the $\mathbb{Z}_2$ conjugate states. They start with colour 2 (pink) instead, and are also annihilated by the TL generators.

\paragraph{One link.} We now have to distribute one link in four sites. There are three ways (in general $L-1$ ways) to do that:
\begin{align}
 & v_{1}^{(1)}\equiv\scalebox{0.7}{
}\;.\label{eq:v3^1-L=4XY}
\end{align}
Similarly to the previous case, a pair of vertical lines can represent any symmetrised state. Therefore, the states above are a good basis to describe a 1-magnon sector, a 2-magnon sector or a 3-magnon sector, as follows
\begin{align}
v_{1}^{(1)}&\equiv\scalebox{0.7}{}=\;\left\{\begin{array}{c}
 A_1X_{12}X_{21}\\
 A_1(Y_{12}X_{21}+X_{12}Y_{21})\\
 A_1Y_{12}Y_{21}
 \end{array}%
\right.\;.\label{eq:v3^1-L=4XYa}
\end{align}

The Hamiltonian is given by
\be
H^{(1-\text{link})}=%
\, ,\label{eq:HforXY-1-link}
\ee
whose eigenvalues are 
\be
E_1=2g_1^2\,,\qquad
E_{\pm}=g_1^2+g_2^2\pm\sqrt{g_1^4+g_2^4}\,.\label{eq:eigenvalues-1-link}
\ee
Above we see one antisymmetriser  and two vertical lines in each state. These vertical lines can describe $XX$, $YX+XY$ and $YY$, as seen in equations \eqref{eq:v1^1-L=4XYa}-\eqref{eq:v3^1-L=4XYa}. So, each of the eigenvalues in \eqref{eq:eigenvalues-1-link} has degeneracy equal to 3.
Furthermore, we have to write the $\mathbb{Z}_2$ conjugate basis. That is made by swapping $1\leftrightarrow 2$ (blue $\leftrightarrow$ pink) everywhere. We see that the last two eigenvalues in \eqref{eq:eigenvalues-1-link} are invariant under $g_1 \leftrightarrow g_2$, while the first is not. Therefore, the following is true
\begin{center}
\renewcommand{\arraystretch}{1.4}
\begin{tabular}{c|c@{\qquad}c@{\qquad}c|c}
Eigenvalue & Degeneracy & & Eigenvalue & Degeneracy\\
\cline{1-2}\cline{4-5}
$2g_1^2$ & 3 & & $2g_2^2$ & 3\\
$g_1^2+g_2^2+\sqrt{g_1^4+g_2^4}$ & 3 & $\xleftrightarrow[]{\quad\mathbb{Z}_2\quad}$ & $g_1^2+g_2^2+\sqrt{g_1^4+g_2^4}$ & 3\\
$g_1^2+g_2^2-\sqrt{g_1^4+g_2^4}$ & 3 & & $g_1^2+g_2^2-\sqrt{g_1^4+g_2^4}$ & 3
\end{tabular}
\end{center}

Now let us discuss how to graphically compute the reduced Hamiltonian for one link \eqref{eq:HforXY-1-link}. As an example, let us obtain the first line. First, we act with $e_1^{(1)}$ on the state $v_1$. Then we use the fact that the pink bubble surrounded by blue gives $-2 g_1^2$ times the identity strand
\be
e_1\left(\scalebox{0.78}{%
}\,.\label{eq:e1(v1)--M=1L=4XY}
\ee
Notice that we chose to act with $e_1^{(1)}$ instead of $e_1^{(2)}$ because the state starts with blue and then changes to pink and that is exactly what $e_1^{(1)}$ has (see \eqref{eq:XY_TL_gens}). This is a useful feature of the construction: the colours automatically take care of the computation, since only operators with the correct colour act non-trivially on a given state.

Now, let us act with $e_2^{(2)}$ on the state $v_1$. The two lower vertices are valued $g_1$ and $g_2$. Therefore, we can straighten the first line and obtain that $v_1$ became $v_2$ but with a coefficient $g_1g_2$ 
\be
e_2\left(\scalebox{0.8}{
}\,.\label{eq:e2(v1)--M=1L=4XY}
\ee
Finally, $e_3$ acts non-trivially only on sites 3 and 4, which specifically for $v_1$ means it gives zero. This is because it is an antisymmetriser acting on a fully symmetrised state. 

Therefore, recalling that $\mathcal{H}_{i,i+1}=-e_i$, the computation above gives exactly the same as the first line in \eqref{eq:HforXY-1-link}.

\paragraph{Two links.} Moving on to two links, which for $L=4$ corresponds to the half-filling sector (two X fields and two Y fields), we have two states that mix under the action of the Hamiltonian:
\begin{align}
v_1^{(2)}&=
\scalebox{0.8}{\begin{tikzpicture}[baseline=-0.6cm]
  \filldraw[\lcyan] (-0.25,-1) rectangle (3.25,0);
  \filldraw[pink] (0,0) arc (-180:0:0.5);
  \filldraw[pink] (2,0) arc (-180:0:0.5);
 \draw[-,thick] (-0.25,0)--(3.25,0);
 \draw[-,thick] (0,0) arc (-180:0:0.5);
 \draw[-,thick] (2,0) arc (-180:0:0.5);
  \draw[-,gray,thick,dashed] (0.5,-0.5)--(0.5,-1); \draw[-,gray,thick,dashed] (2.5,-0.5)--(2.5,-1);
   \filldraw[red,thick] (0,0) circle (1pt); \filldraw[red,thick] (1,0) circle (1pt); \filldraw[red,thick] (2,0) circle (1pt);
   \filldraw[red,thick] (3,0) circle (1pt);
\end{tikzpicture}}=A_1A_1=g_1^2(Y_{12}X_{21}-X_{12}Y_{21})(Y_{12}X_{21}-X_{12}Y_{21})\, ,\\
v_2^{(2)}&=A^{[1]}_1=\scalebox{0.8}{\begin{tikzpicture}[baseline=-0.7cm]
  \filldraw[\lcyan] (-0.25,0)--(3.25,0)--(3.25,-1.25)--(-0.25,-1.25)--(-0.25,0);
  \filldraw[\lcyan](0.5,-0.5)--(2.5,-0.5)--(2.5,-1.25)--(0.5,-1.25)--(0.5,-0.5);
  \begin{scope}\clip (0,0)--(1.5,0)--(1.5,-1)--(0,-1)--(0,0);  \filldraw[pink] (0,0) arc (-180:0:1.5cm and 1cm);\end{scope};
  \begin{scope}\clip (1.5,0)--(3,0)--(3,-1)--(1.5,-1)--(1.5,0);  \filldraw[pink] (0,0) arc (-180:0:1.5cm and 1cm);\end{scope};
    \filldraw[\lcyan] (1,0) arc (-180:0:0.5);
  \draw[-,thick] (-0.25,0)--(3.25,0);
  \draw[-,thick] (0,0) arc (-180:0:1.5cm and 1cm);
  \draw[-,thick] (1,0) arc (-180:0:0.5);
  \draw[-,gray,thick,dashed](1.5,-0.5)--(1.5,-1);
     \draw[-,gray,thick,dashed](0.5,-0.75)--(0.5,-1.25);  \draw[-,gray,thick,dashed](2.5,-0.75)--(2.5,-1.25); 
  \filldraw[red,thick] (0,0) circle (1pt); \filldraw[red,thick] (1,0) circle (1pt); \filldraw[red,thick] (2,0) circle (1pt);
   \filldraw[red,thick] (3,0) circle (1pt);
\end{tikzpicture}}=g_1\left(Y_{12}A_2X_{21}-X_{12}A_2Y_{21}\right).
\end{align}
Here the $[1]$ indicates a nested link. As expected, there is no need for symmetrised pairs for the action of the Hamiltonian to close. Notice that, if we had just followed the gauge theory prescription, we would have an overall factor equal to $g_1^2g_2$ instead of $g_1$. Here and in what follows in this section, we normalise the states (by dividing by $g_1g_2$ to an appropriate power) such that the Hamiltonian is quadratic.

Acting with the TL generators we easily find for the half-filling sector:
\be
H^{(2-\text{links})}=\begin{pmatrix}
4g_1^2 & -g_1g_2 \\
-2g_1g_2 & 2g_2^2
\end{pmatrix}\, .
\ee

This  Hamiltonian can be graphically obtained in the same way as the one in \eqref{eq:HforXY-1-link}. In particular, the factor of 4 in the first line arises because both the action of $e_1$ and $e_3$ in this case give $-2g_1^2$ times the state. At the same time, the $-g_1g_2$ comes from acting with $e_2$ on $v_1^{(2)}$, which leads to $v_2^{(2)}$ and a coefficient.

The energy spectrum of course agrees with explicit diagonalisation of the Hamiltonian for $L=4$. Notice that $H^{(2-\text{links})}$ is not symmetric, even though the gauge theory has real anomalous dimensions. This is an artefact of the normalisation of the link states: in terms of $\tilde{v}_2^{(2)}=v_2^{(2)}/\sqrt{2}$ the  Hamiltonian becomes
\be
\tilde{H}^{(2-\text{links})}=
\, .
\ee
We will meet the same phenomenon in the XZ sector, see Section~\ref{subsec:nonHermiticity}. As we prefer to avoid the square roots in the Hamiltonian, we stick with the original normalisation of the link states. 

The spectrum of the XY sector at $L=4$ was plotted in Figure~\ref{GraphL4XY}, and of course agrees with the diagonalisation of the above reduced Hamiltonians. 

\subsubsection{Example: $L=6$}

As a final, less trivial example, let us move on to the half-filling sector at length 6. The relevant states are
\begin{align}
v_1^{(3)}&=\scalebox{0.8}{%
}=g_1\left(Y_{12}A_2^{[1]}X_{21}-X_{12}A_2^{[1]}Y_{21}\right)\, ,
\end{align}
where in the last two states, the square brackets indicate the two possibilities for double nesting. The action of the Hamiltonian in the basis $v_i^{(3)}$ is 
\be
H^{(3-\text{links})}=%
\, ,
\ee
which again agrees with diagonalisation in the monomial basis (and the plot of the spectrum in Figure~\ref{fig:allmagnonXYL6}). 

Proceeding to more sites, one would encounter multi-nested states $A_a^{[n_1,n_2,n_3,\ldots]}$, with $n_j\geq0$, which can be defined in a natural way by iterating the states above. 
We conclude that the dynamical TL algebra and associated link basis provides a useful reformulation of the spectral problem in the XY sector, in that it allows us to focus on sectors with a specific symmetry structure (number of antisymmetrised vs. symmetrised pairs). The action of the dynamical TL algebra in this sector is very similar to that of the standard TL algebra, just with additional powers of $g_1,g_2$ appearing according to the modified quadratic relations \eqref{XYquadratic}. 

Eventually, in Section \ref{sec:twist}, we will see that this algebra fits within the dynamical Tem\-per\-ley-Lieb framework recently introduced in \cite{Ren:2023mtn}.

\section{The dynamical Temperley-Lieb algebra in the $\mathbb{Z}_2$ XZ sector }\label{sec:XZsector}

We now turn to the XZ sector, composed of the bifundamentals $X_{12},X_{21}$ and the adjoints $Z_1,Z_2$, and follow the same steps as for the XY sector in Section~\ref{sec:XYsector}. The XZ sector is richer: the line joining the two vertices of a TL generator is now a bifundamental field, so not all colours are fixed, and, since the $\SU(2)$ symmetry is broken, besides the $g_a$-antisymmetrised links (which we call A-links) it turns out that we will also need $g_a$-symmetrised ones (which we call S-links). As we will see, in this case it is the cubic, rather than the quadratic\footnote{For $\mathbb{Z}_K$ with $K>2$ both the cubic and the quadratic relation will become dynamical. See Sections~\ref{sec:Zk-XZ-sector} and \ref{sec:twist} for more details.}, relation that becomes dynamical. 

Furthermore, we show that rewriting the model in terms of this new dynamical Tem\-per\-ley-Lieb algebra explains the origin of the energy splitting observed in Section~\ref{sec:diluteXZ}, and makes symmetries such as parity completely manifest.

\subsection{Diagrams from superspace}\label{subsec:fromsuperspace-XZsector}

From the superpotential \eqref{WZ2} we see that there are now four holomorphic vertices, which we again arrange with an incoming $\bar{Y}$ field and outgoing $X,Z$ fields:
\be
=-g_2\, . \label{eq:vertices}
\ee
Similarly, for their conjugates we take the $X,Z$ fields as incoming and the $\bar{Y}$ outgoing:
\be
%
=-g_2\, . \label{eq:vertices_upsidedown_XZ}
\ee
As before, if the order, read anticlockwise, is $XYZ$ then the sign is positive, while if it is $XZY$ the sign is negative. The coupling constant $g_1$ or $g_2$ depends on whether there is a $Z_1$ or $Z_2$ in the vertex. 

The crucial difference to the XY sector is that the line joining the two vertices is now the bifundamental field $\bar{Y}$, so crossing it changes the node index from $1$ to $2$ and vice versa. So, unlike the $\Ncal=4$ SYM case where indicating the $\bar{Y}$ line was optional, and it could be suppressed to recover the standard TL notation, here the $\bar{Y}$ line needs to be kept. We will draw it as a solid line to indicate that it separates domains of different colour. Colouring the regions according to the nodes as in Section~\ref{subsec:fromsuperspace-YXsector}, we again have \emph{two} TL generators. They are defined as
\begin{subequations}
\label{eq:XZ_TL_gen1} 
\begin{align}
e^{(1)}=
\;.
 \end{align}
\end{subequations}
Contrary to the XY sector, here not all colours are fixed. This is indicated by the white regions, in which there are multiple options for the gauge node index. Each allowed choice of filling in the colours can be identified with one of the individual Feynman diagrams, for instance\be
%
\;.
 \ee
 As the indices of the fields have to match when acting consecutively, the colours of the regions also have to agree in order to get a nonzero result. This often selects a unique colouring of the white regions, but if more possibilities are allowed then they are summed over. 

 An important example is the case where we act twice with the generators \eqref{eq:XZ_TL_gen1} and \eqref{eq:XZ_TL_gen2}. Isolating the closed loop, we find that it receives contributions from two possible internal colours. Summing over these two allowed contributions, we find that the sum of closed loops gives $-(g_1^2+g_2^2)$ times the $\bar{Y}$-identity strand. 
\be

\ee
Clearly, the $\Zset_2$ conjugate diagrams produce the same coefficient. So, in the graphical dTL language, we have the identities
\begin{align}
&%
\; . \label{eq:bubbleXZb}
\end{align}

We see that the loop fugacity in this case is $\beta=-g_1^2-g_2^2$, which is  $\Zset_2$-invariant, and defines a non-dynamical quadratic relation. %

We now proceed to consider the cubic relations, which are obtained by considering diagrams such as
 \be \label{XZcubicFeynmanX}
%
\, .
\ee
 By fixing the incoming field at site 3 to be $X_{21}$, the only nonzero combination of vertices is the one indicated in the dashed box. Therefore the diagram on the left is equivalent to the value of the four vertices in the box, times a single TL generator times the $X$ identity strand. We can represent this in our dynamical TL diagrams as
\be
%
\label{eq:cubicXZrelation1-Z2}
\ee
The colour of the region in the middle can only be pink, as the last colour is blue and the colour must flip when crossing any $Y$ line.

Now let us look at the same cubic relation, but in the case where the field at site 3 is a $Z$ instead of an $X$. Then we write
\be  \label{XZcubicFeynmanZ}
%
\ee
Again, there is only one possible combination of vertices within the dashed box. So the diagram on the left is equivalent to the value of the fixed vertices times a single TL generator times the $Z$ identity strand. However, this time there are two vertices with $Z_1$'s and two with $Z_2$'s, giving a coefficient of $g_1^2g_2^2$. The other choices of incoming/outgoing fields at sites 1,2  give the same coefficients as above. Therefore, in dynamical TL notation we have
\be%
\ee
Given that the field at site 3 is a $Z_2$, the colour of the region in the middle can only be blue, as the last colour is pink and the colour must flip when crossing any $Y$ line. 

The difference between \eqref{XZcubicFeynmanX} and \eqref{XZcubicFeynmanZ} comes down to how many $X$ fields (and therefore alternations of colour) there are in the three sites. The same pattern appears when looking at the $e_2e_1e_2$ cubic relation. 

We conclude that the XZ-sector dilute dynamical Temperley-Lieb algebra satisfies the relations:
\be \label{XZTLrelations}
e_i^2=-(g_1^2+g_2^2) e_i\;\;,\;\; e_ie_{i\pm 1}e_i=\left\{\begin{array}{l}g_{a+1}^4 e_i\;\;\; \text{for two} \;X\; \text{fields on the three sites}\\
g_1^2g_2^2 e_i\;\;\;\text{for one} \;X\; \text{field on the three sites}\end{array}\right.\;\;,
\ee
as well as $e_ie_j=e_je_i \;,\;|i-j|>1$. Here $a$ is the node index to the left of the three sites $(i,i+1,i+2)$, and depends on the field content on the first $i-1$ sites. 
We can also define the globally rescaled versions $\tilde e_i=e_i/(g_1g_2)$, using which the relations can be naturally expressed in terms of $\kappa$:
\be
\label{eq:XZTL_relations_rescaled}
\tilde e_i^2=-(\kappa+\kappa^{-1})\tilde e_i\;,\;\tilde{e}_i\tilde{e}_{i\pm 1}\tilde{e}_i=\left\{\begin{array}{ll}\kappa^{2 s(a)}  \tilde{e}_i\;\;\; &\text{for two} \;X\; \text{fields on the three sites}\\
\tilde{e}_i\;\;\;&\text{for one} \;X\; \text{field on the three sites}\end{array}\right.\;,
\ee
with $s(a)=(-1)^{a+1}$ with $a$ the node index to the left, in other words $s(1)=1$, $s(2)=-1$.

As discussed after \eqref{XYquadratic}, instead of explicitly keeping track of colours it is also convenient to use a dynamical notation for the algebra generators, in which they depend on a dynamical parameter which tracks the colour/node index as one proceeds along the chain. The main difference to the XY sector is that here one sometimes crosses bifundamental $X$ fields, which flip the colour, and sometimes adjoint $Z$ fields, which do not. Therefore the colour (and dynamical parameter) at a given site is determined by the first gauge index at the left end of the chain, together with the number of $X$ fields one has crossed to reach that site. As our goal in this section is to study the spectrum of the XZ sector diagrammatically, we will defer the full development of the dynamical notation to Section~\ref{sec:dynamical}.  

Borrowing a term from the related loop model description of the above Temperley-Lieb algebra (for more details of which we refer to Section~\ref{sec:loop}), we will call the above model \emph{dilute}. In the present context this simply refers to the colour staying the same when $Z$ fields are crossed, which in the dynamical language means that the $Z$ fields have weight zero and therefore do not shift the dynamical parameter. 

There exist other definitions of dilute Temperley-Lieb algebras (such as the one described in \cite{Belletete:2013zla}), however the algebra defined here appears to be new. In particular the dynamical nature of the cubic terms is a very novel feature which (among other aspects) will clearly have implications for the braiding relations that typically follow from the TL cubic relation, something which we will not study in this work. We will however discuss the XZ-sector TL relations in a more general setting and slightly more formal language in Section~\ref{subsec:quiver-to-Ren}, see \eqref{eq:gen-dTL}. 

\subsection{Link states in the XZ sector}\label{sec:guideXZ}

From the gauge theory description we learned that we can define the following states
\begin{align}
A^{[0]}_{12}\equiv \scalebox{0.8}{
}&=g_2 X_{21}Z_1+g_1 Z_2X_{21}\,,\label{eq:AandS_21}
\end{align}
where $A$ denotes $g_a$-antisymmetrisation and $S$ denotes $g_a$-symmetrisation. The lower indices indicate the colours at the two ends of the link, so $A_{12}^{[0]}$ and $S_{12}^{[0]}$ start with blue and end with pink, while $A_{21}^{[0]}$ and $S_{21}^{[0]}$ do the opposite. When a diagram starts and ends with the same colour we write a single index, as for $A_1$ in the XY sector. The upper index again indicates the level of nestedness. For this reason, from now on we will refer to half-circle links as A-links and half-square links as S-links.

In this section we explain how to read these states and construct their multi-site generalisations. The  $g_a$ coefficients in the A-links above are derived from the gauge theory. Reading a vertex anticlockwise, leads to a ``$+\,g_a $" for XYZ and a $``-\,g_a "$ for XZY (see equations \eqref{eq:vertices} and \eqref{eq:vertices_upsidedown_XZ}).  The S-links are chosen such that they are annihilated by the TL generator acting on their two sites (see Appendix~\ref{app:kappa-sym}) and are a new feature of this dynamical TL. The full list of links needed to describe a $\mathbb{Z}_2$-XZ sector is
\begin{align}
\scalebox{0.8}{
}&=\, g_2\, .\label{eq:guide-for-links}
\end{align}

Recall that dashed lines separate regions of the \textit{same colour} (corresponding to adjoint fields in the gauge theory), while solid lines separate regions of \textit{different colours} (corresponding to bifundamentals).  For example, $\raisebox{-0.2ex}{\scalebox{0.3}{%
}}$  is given by
\be
   \scalebox{0.8}{%
}\, .\label{eq:detailed_diagram}
\ee
Next, recalling that blue represents index~1 and pink represents index~2, we read off the  colours along the top of each diagram. At each dot, if the colour is the same on both sides, the dot contributes a $Z$; if the colour changes, it contributes an $X$. For example, the first diagram is proportional to $Z_1 X_{12}$: at the first dot the colour stays blue, giving $Z_1$, and at the second dot it changes from blue to pink, giving $X_{12}$. Combining this with the vertex value derived above, we obtain
\be
\scalebox{0.8}{%
}=-g_1Z_1X_{12}\,.
\ee
Similarly, the second one in \eqref{eq:detailed_diagram} is given by $g_2X_{12}Z_2$\, .

The diagrams above correspond to the level 0 of nesting. When more sites are included, higher levels of nesting appear. Fortunately, all expressions and coefficients can be determined from the map in \eqref{eq:guide-for-links} and the rule in \eqref{eq:detailed_diagram}. For example, the level-1 A-link is obtained as follows
\begin{align}
  A_1^{[1]}\equiv \scalebox{0.8}{%
}\right)\label{eq:A11b}\\
&= \frac{1}{g_2^2}\left(g_1g_2^2\,X_{12}A_{21}^{[0]}Z_1-g_1g_2^2\,Z_1A_{12}^{[0]}X_{21}\right)\label{eq:A11c}\\
&= g_1\left(X_{12}A_{21}^{[0]}Z_1-Z_1A_{12}^{[0]}X_{21}\right)\,.\label{eq:A11d}
\end{align}
Notice that reading the first diagram  in \eqref{eq:A11b} from left to right, first blue becomes pink when passing the node, so that is $X_{12}$. Next, we see  exactly the definition of the A-link $A_{21}^{[0]}$ ($\raisebox{-0.2ex}{\scalebox{0.3}{%
}}$). Finally, in the last node, we have blue on both sides, so that is a $Z_{1}$. The coefficient $g_1g_2^2$ comes from the vertices as summarised in \eqref{eq:guide-for-links}. The extra coefficient $g_2^{-2}$ is introduced to make sure that later, all basis vectors will have the same power in $g$, which ensures that every reduced Hamiltonian is quadratic in $g$.\footnote{By \emph{reduced} (in contrast to \emph{full}) we refer to a Hamiltonian of a specific magnon sector.}

Analogously, for the S-link we have
\begin{align}
  S_1^{[1]}\equiv \scalebox{0.8}{%
}\right)\label{eq:S11b}\\
&= \frac{1}{g_1^2}\left(g_1^2g_2\,X_{12}A_{21}^{[0]}Z_1+g_1^2g_2\,Z_1A_{12}^{[0]}X_{21}\right)\label{eq:S11c}\\
&= g_2\left(\,X_{12}A_{21}^{[0]}Z_1+\,Z_1A_{12}^{[0]}X_{21}\right)\,,\label{eq:S11d}
\end{align}
where this time we used also the second column in \eqref{eq:guide-for-links}.
Notice that S-links  have no counterpart in the standard TL algebra. They are a genuinely new feature of this novel dynamical TL algebra.

In this section we showed how to read and use the link diagrams with white (unresolved colour) regions, and how to build some simple states using both the A- and S-links and a representation of the bifundamental and adjoint fields in terms of colours. In the same way that we used \eqref{eq:AandS_12}-\eqref{eq:AandS_21} to construct \eqref{eq:A11d} and \eqref{eq:S11d}, we can now use those, together with \eqref{eq:A11d} and \eqref{eq:S11d}, to construct the next level of nesting.
In particular, in Section~\ref{sec:basis}, we will use the information above together with the two-site BPS states
\be
\scalebox{0.7}{\begin{tikzpicture}[baseline=-0.6cm,scale=1]
     \fill[\lcyan] (-0.25,-1) rectangle (0,0);
     \fill[pink] (0,-1) rectangle (1,0);
     \fill[\lcyan] (1,-1) rectangle (1.25,0);
     \draw[-,thick,gray] (-0.25,0)--(1.25,0);
     \foreach \x in {0,1} \draw[thick] (\x,0)--(\x,-1);
     \foreach \x in {0,1} \filldraw[red,thick] (\x,0) circle (1pt);
\end{tikzpicture}}= X_{12}X_{21}\,,\quad
\scalebox{0.7}{\begin{tikzpicture}[baseline=-0.6cm,scale=1]
     \fill[pink] (-0.25,-1) rectangle (0,0);
     \fill[\lcyan] (0,-1) rectangle (1,0);
     \fill[pink] (1,-1) rectangle (1.25,0);
     \draw[-,thick,gray] (-0.25,0)--(1.25,0);
     \foreach \x in {0,1} \draw[thick] (\x,0)--(\x,-1);
     \foreach \x in {0,1} \filldraw[red,thick] (\x,0) circle (1pt);
\end{tikzpicture}}= X_{21}X_{12}\,,\quad
\scalebox{0.7}{\begin{tikzpicture}[baseline=-0.6cm,scale=1]
     \fill[\lcyan] (-0.25,-1) rectangle (1.25,0);
     \draw[-,thick,gray] (-0.25,0)--(1.25,0);
     \foreach \x in {0,1} \draw[thick,dashed] (\x,0)--(\x,-1);
     \foreach \x in {0,1} \filldraw[red,thick] (\x,0) circle (1pt);
\end{tikzpicture}}= Z_1Z_1\,,\quad
\scalebox{0.7}{\begin{tikzpicture}[baseline=-0.6cm,scale=1]
     \fill[pink] (-0.25,-1) rectangle (1.25,0);
     \draw[-,thick,gray] (-0.25,0)--(1.25,0);
     \foreach \x in {0,1} \draw[thick,dashed] (\x,0)--(\x,-1);
     \foreach \x in {0,1} \filldraw[red,thick] (\x,0) circle (1pt);
\end{tikzpicture}}= Z_2Z_2\,,\label{eq:XX-and-ZZ-states}
\ee
to construct a link basis for both $L=4$ and $L=6$ Hamiltonians. For $L=6$, in particular, we will see diagrams for higher level of nesting, including $A^{[2]}$ and $A^{[1,1]}$ in \eqref{eq:basisXZ-L6-M3-true-magnons} and  $S^{[2]}$ and $S^{[1,1]}$ in \eqref{eq:basisXZ-L6-M3-descendants}.

\subsection{Link-state basis}\label{sec:basis}

We now turn to the construction of the basis, giving explicit details for $L=4$ and $L=6$ sites. Unlike in the XY sector, the $SU(2)$ symmetry of the mother theory is broken in the XZ sector. As a consequence, the spectrum of the Hamiltonian with $M$ magnons no longer coincides with that of $L-M$ magnons. This is illustrated in Figure~\ref{fig:allmagnonXZL6} for $L=6$, where the spectra for $M=1$ and $M=5$ are plotted side by side, as are those for $M=2$ and $M=4$.

The basis we construct below makes this behaviour manifest. We denote its states by $v_j^{(M)}$, where $M$ is the number of magnons and $j$ labels the position of the state within the basis. As in the XY sector, for open chains, states starting with colour 1 (blue) never mix with states starting with colour 2 (pink). We therefore only give the basis, states and spectrum for states starting with colour 1. The remaining ones follow by exchanging $1 \leftrightarrow 2$ (blue $\leftrightarrow$ pink) everywhere.

Apart from the vacuum states, here we will only focus on states that include at least one A-link, which (as we will see) results in them acquiring a nonzero eigenvalue. For each magnon number, there also exists a BPS state, with $E=0$. These states can be thought of as descendants of the vacuum states, and do not mix with any states involving links. We discuss them separately in Appendix \ref{app:kappa-sym}.  

Let us start with the first non-trivial example: an open chain with four sites ($L=4$). 

\subsubsection{Example: $L=4$}

We organise this section according to increasing magnon number $M$.

\paragraph{$\mathbf{M=0}$.}

We start with the X-vacuum. Since $X$ transforms in the bifundamental, we have an index 1 and 2 and the  colours alternate. Consequently, it can be represented as
\be
v_{1}^{(0)}\equiv\scalebox{0.7}{
}= X_{12}X_{21}X_{12}X_{21}\,.
\ee

\paragraph{$\mathbf{M=1}$.} We start increasing the magnon number by adding $Z_a$ fields to our chain. For one magnon ($M=1$), the basis is constructed by putting one single A-link, in all possible positions, and keeping $X_{a,a+1}$ everywhere else. The basis is $(L-1)$-dimensional and given by
\begin{align}
 v_{1}^{(1)}\equiv\scalebox{0.7}{%
}&= X_{12}X_{21}A_{12}^{[0]}\;.\label{eq:v3^1-L=4}
\end{align}
Notice that contrary to the XY sector, the two solid vertical lines can only represent $X_{12}$ and $X_{21}$. We will see the effects of this soon. 

\paragraph{$\mathbf{M=2}$.} Now, for $M=2$, we start by writing what we will call \textbf{true magnon states}. Those are the ones that can be written with $M$ A-links and $L-2M$ lines, such that the links never cross. For $L=4$, $M=2$ is the half-filling case. So, there are no extra lines. The true magnon states are then written as
\begin{align}
v_{1}^{(2)}\equiv\scalebox{0.8}{
}&=g_1\left(X_{12}A_{21}^{[0]}Z_1-Z_1A_{12}^{[0]}X_{21}\right)\equiv A_1^{[1]}\;,\label{eq:v2^2-L=4}
\end{align}
while the three descendants are given by
\begin{align}
v_{3}^{(2)}\equiv\scalebox{0.8}{%
}&=S_{12}^{[0]}A_{21}^{[0]}\;.\label{eq:v5^2-L=4}
\end{align}
The states $v_3^{(2)}$, $v_4^{(2)}$ and $v_5^{(2)}$ are the descendants of $v_1^{(1)}$, $v_2^{(1)}$ and $v_3^{(1)}$, and graphically are obtained by connecting the two vertical lines by a horizontal line, producing a half-square. 

The normalisation was chosen to make sure that all states in a given basis have the same total power of $g$. We refer to equations \eqref{eq:A11a}-\eqref{eq:A11d} and explanation thereafter for more details. 

The states that include S-links will play a fundamental role in our construction. The fact that they behave so differently from the ones in \eqref{eq:v1^1-L=4}-\eqref{eq:v3^1-L=4}, will later explain the energy splitting observed in Figure~\ref{fig:allmagnonXZL6}. See Section~\ref{sec:acting-on-states} for more details on how this happens.

\paragraph{$\mathbf{M=3}$.}  Up to the half-filling, the number of links was \textit{increasing} by one, every time we increased $M$ by one. After the half-filling, the number of links starts \textit{reducing} by one instead. For $L=4$, $M=3$, we have to distribute the A-links in all possible ways, and put $Z$'s on the remaining sites as follows  
\be\begin{split}
 v_{1}^{(3)}\equiv\scalebox{0.7}{
}&= Z_{1}Z_{1}A_{12}^{[0]}.
\end{split}\ee
Notice that since the $Z$ transforms in the adjoint, the colour before and after such a $Z$ line remains the same.

\paragraph{$\mathbf{M=4}$.} For $M=4$, the only option for $L=4$ is having a $Z$ at every site

\be
v_{1}^{(4)}\equiv\scalebox{0.8}{%
}= Z_1Z_1Z_1Z_1.
\ee

\subsubsection{Example: $L=6$}

We now consider one more example, $L=6$, to clarify some of the details. For $M=0$ and $M=1$ the construction generalises straightforwardly from the $L=4$ case, so we simply list the states. We then focus on $M=2$ and $M=3$.

\paragraph{$\mathbf{M=0}$.} For the pseudo-vacuum, it is again an alternating state represented as 
 
\be
v_{1}^{(0)}\equiv\scalebox{0.7}{%
}= X_{12}X_{21}X_{12}X_{21}X_{12}X_{21}.
\ee

\paragraph{$\mathbf{M=1}$.} For $M=1$, the ($L-1$)-dimensional basis can be written by putting one A-link in all possible positions as follows
\begin{align}
v_{1}^{(1)}\equiv\scalebox{0.6}{%
}&=X_{12}X_{21}X_{12}X_{21}A_{12}^{[0]}\;.\label{eq:L=6M=1}
\end{align}

\paragraph{$\mathbf{M=2}$.} The non-BPS states for $M=2$ are written in a 14-dimensional basis. That contains the following nine true magnon states (six simple ones and three nested states) 
\begin{align}
v_{1}^{(2)}\equiv\scalebox{0.6}{%
}&=X_{12}X_{21}A_{12}^{[0]}A_{21}^{[0]},\label{eq:XZ-L6-M2-true-magnon}
\end{align}
and five descendants
\begin{align}
v_{10}^{(2)}\equiv\scalebox{0.57}{%
}&= X_{12}S_{2}^{[1]}X_{21}.\label{eq:XZ-L6-M2-descendants}
\end{align}

The descendants can also be obtained graphically. The idea is to put an A-link in every possible nearest-neighbour configuration like in \eqref{eq:L=6M=1} and then connect the vertical lines on either side of the A-link by an S-link. For the cases where the links are not at the boundary, it is straightforward and we obtain states $v_{11}^{(2)}$-$v_{13}^{(2)}$. However, it is clear that for an open chain it is not possible to do that for $v_1^{(1)}$ and $v_5^{(1)}$. Instead, we connect the two lines on either side of them, which produces states $v_{10}^{(2)}$ and $v_{14}^{(2)}$, respectively. 

\paragraph{$\mathbf{M=3}$.} 
Let us start by the true magnon states, that have three ($=M$) A-links and no S-links
\begin{align}
v_{1}^{(3)}\equiv\scalebox{0.6}{
}&=g_2 X_{12}A_2^{[1]}Z_2-g_1 Z_1A_{1}^{[1]}X_{12}\equiv A_{12}^{[2]}\;.\label{eq:basisXZ-L6-M3-true-magnons}
\end{align} 
Notice that the coefficients are determined by the gauge theory, and in particular, what $v_4^{(3)}$ really represents is
\begin{align}
   \scalebox{0.7}{%
}\right),
\end{align}
which when simplified gives exactly equation \eqref{eq:basisXZ-L6-M3-true-magnons}.
The analogous computation leads to $A_{12}^{[2]}$. Notice that, as already mentioned, we always choose the normalisation such that all states in a basis have the same total power in $g$.

Notice that the true magnon states are usually the only ones that appear in a standard Temperley-Lieb algebra, but here there are many other states we need to construct.

In this particular example, the next states in the basis are the ones with two ($=M-1$) A-links. They are nine states and are given by
\begin{align}
v_{6}^{(3)}\equiv\scalebox{0.6}{
}&= S_{12}^{[0]}A_{2}^{[1]}.\label{eq:basisXZ-L6-M3-descendants}
\end{align}
They are obtained as descendants of the $M=2$ states in \eqref{eq:XZ-L6-M2-true-magnon}. In practice, graphically, those come from closing each pair of vertical lines in \eqref{eq:XZ-L6-M2-true-magnon}, transforming them into S-links. 

Finally, we consider the case with only one $(=M-2)$ A-link(s), and two S-links. These can be obtained as descendants from \eqref{eq:XZ-L6-M2-descendants} and are given by
\begin{align}
v_{15}^{(3)}\equiv\scalebox{0.6}{%
}&=  S_{12}^{[0]}A_{21}^{[0]}S_{12}^{[0]}.
\end{align}

\noindent \textbf{Comment about the basis:} At this moment, one could wonder why our basis includes the state $v_{17}^{(3)}$ as given by
\be
\scalebox{0.7}{%
}\; ,
\ee
instead of the one below with nested S-links
\be
\scalebox{0.7}{%
}\; .
\ee
The answer is that these two states are not linearly independent in the basis and can be expressed in terms of each other. More specifically,  we have  the following identity
\begin{align}
\scalebox{0.6}{%
}\;.\label{eq:doubly-symmetric-state}
\end{align}

In other words, the state with nested S-links can be written in terms of  the non-nested state \inl{%
} and other states that are already in the basis. We therefore choose to use the simpler state. We expect this property to generalise, so that the basis can always be chosen such that S-links only appear in the outermost layer. 

\paragraph{$\mathbf{M=4}, \, \mathbf{M=5}$ and $\mathbf{M=6}$.} The basis for these magnon sectors can be obtained by taking states $M=2$, $M=1$ and $M=0$, respectively, and replacing each solid vertical line ($X$) in them by a dashed vertical line ($Z$), while keeping all links as they are. However, this does not mean that the same spectrum is obtained for those sectors. 

In the next section we will see how to obtain such a spectrum using the graphical structure of the link basis.

\subsection{Action of the Hamiltonian}\label{sec:acting-on-states}

Let us now consider the action of the dynamical TL operators on the states, as we did for the XY sector in Section~\ref{sec:basisXY}. The new feature is that the colours of the generators are not fixed, but are determined by the state they act on. To start, let us  consider $M=1$, $L=4$ (see \eqref{eq:v1^1-L=4}-\eqref{eq:v3^1-L=4}) and act with  $e_1$ on $v_1^{(1)}$
\be
e_1\left(\scalebox{0.74}{
}\,,\label{eq:e1(v1)--M=1L=4}
\ee
where we used that the bubble is equal to $-(g_1^2+g_2^2)$ (see eq.~\eqref{eq:bubbleXZa}).
\noindent Acting with $e_2$ on the same state on the other hand gives 
\be
e_2\left(\scalebox{0.8}{%
}\,.\label{eq:e2(v1)--M=1L=4}
\ee
Furthermore, acting with $e_3$ results in zero, since it is an antisymmetriser acting on a symmetrised part of the state. Together, these give 
\be
H^{(1)}(v_1^{(1)})=(g_1^2+g_2^2)v_1^{(1)}-g_2^2v_2^{(1)}\,. 
\ee
Doing the same with states $v_2^{(1)}$ and $v_3^{(1)}$ leads to the following Hamiltonian
 \be
H^{(1)}=%
\,.
 \ee
whose eigenvalues are  $\{g_1^2+g_2^2,g_1^2+g_2^2\pm \sqrt{g_1^4+g_2^4}\}$.

We just computed the spectrum for the one-magnon case. But as already mentioned, contrary to XXZ model and the $\mathbb{Z}_2$-XY sector, the spectrum of the XZ sector is not the same for $M$ and $L-M$ magnons. So, let us see what we get for $M=3$. Let us start by acting on $v_1^{(3)}$ with $e_1$ 
\be
e_1\left(\scalebox{0.74}{%
}\,.\label{eq:e1(v1)--M=3L=4}
\ee
As expected, this acts diagonally on the state, with a coefficient $-(g_1^2+g_2^2)$ generated by the bubble.
Now, acting with $e_2$ on the same state gives
\be
e_2\left(\scalebox{0.8}{%
}\, .\label{eq:e2(v1)--M=3L=4}
\ee
As in $M=1$, this moves the link to the right by one, but with a different coefficient, $g_1g_2$. Finally, $e_3$ acts as zero on that state. 

Acting on $v_2^{(3)}$ and $v_3^{(3)}$ in the same way allows us to write the Hamiltonian for the three magnon sector as
\begin{equation}
    H^{(3)}=%
,
\end{equation}
whose eigenvalues are given by $\{g_1^2+g_2^2,g_1^2\pm \sqrt{2}g_1g_2+g_2^2\}$.  As we can see, one of the eigenvalues overlaps with $M=1$, but the other two do not.

Let us now do a few examples for $L=4$, $M=2$.  As we know, acting with $e_1$ or $e_3$ in $v_1^{(2)}$ gives automatically the state back with an eigenvalue $-(g_1^2+g_2^2)$ (see the example in eq. \eqref{eq:e1(v1)--M=1L=4}). Alternatively, acting with $e_2$ on such a state gives  the following\begin{align}
e_2(v_1^{(2)})&=e_2\left(\,\scalebox{0.78}{%
} = g_2^2 v_2^{(2)}.
\end{align}
In the first step we act with a dynamical TL generator $e_2$, by stacking it at the top of $v_1^{(2)}$. Such a generator is initially left uncoloured, because the  colours it will assume depend on the vector it is acting on. That is what we do next step: write it as a sum of the two possible colour combinations   (see eq. \eqref{eq:e2(v1M2)twoterms}). Next, we just made them smooth by removing the middle horizontal lines and gluing them together. Finally, we compute the diagrams using the prescription in Section~\ref{sec:guideXZ}, which leads to equation \eqref{eq:e2(v1M2)}. Finally, we recognise the state as proportional to $v_2^{(2)}$.

Therefore, acting with the Hamiltonian on $v_1^{(2)}$ gives
\be
H^{(2)}(v_1^{(2)})=2(g_1^2+g_2^2)v_1^{(2)}-g_2^2v_2^{(2)}, \label{eq:H2-acton-v1^2-XZ-Z2}
\ee
which correspond to a diagonal term, plus a shift of the link to the right. 

Now let us act with $e_1$ on $v_2^{(2)}$
\begin{align}
    e_1(v_2^{(2)})&=e_1\left(\scalebox{0.8}{
}\,.\label{eq:e1(v2M2)g}
\end{align}
In the first step, we act with the $e_1$ operator by stacking it on the top of $v_2^{(2)}$. We then write it as a sum of the allowed colours. However, as soon as we put  colours on the state, we have to remember that the definition of $v_2^{(2)}$ contains a normalisation $1/(g_2^2)$ that was added by hand before. From eq. \eqref{eq:e1(v2M2)b} to \eqref{eq:e1(v2M2)c}, we just remove the horizontal lines and glue the operator and the state together. Next, we write down the mathematical expression corresponding to the two diagrams, and simplify it. At this step, we notice that what remains on sites 3 and 4, isn't either only an A or an S, but a linear combination of both. Finally, we determine the coefficients and write the final expression.

The action of $e_3$ on $v_2^{(2)}$ is obtained by a very similar computation. The difference is that the linear combination will now involve sites 1 and 2, instead of 3 and 4. As a consequence, instead of state $v_3^{(2)}$ it is $v_5^{(2)}$  that appears. Finally, the action of $e_2$ on $v_2^{(2)}$ is given by $e_2(v_2^{(2)})=-(g_1^2+g_2^2)v_2^{(2)}$, naturally obtained by using that the bubble has the value $-(g_1^2+g_2^2)$.

Altogether, we find
\be
H^{(2)}v_2^{(2)}=-\frac{4g_1^2g_2^2}{g_1^2+g_2^2}\,v_1^{(2)} + (g_1^2+g_2^2)\,v_2^{(2)} +\frac{g_1g_2(g_1^2-g_2^2)}{g_1^2+g_2^2}\,v_3^{(2)} -\frac{g_1g_2(g_1^2-g_2^2)}{g_1^2+g_2^2}\,v_5^{(2)}.\label{eq:H2-acton-v2^2-XZ-Z2}
\ee

Now let us focus on $v_3^{(2)}$. If we act with $e_1$ on it, we obtain again a bubble and therefore $e_1(v_3^{(2)})=-(g_1^2+g_2^2)v_3^{(2)}$. If we act with $e_3$ instead, it is an antisymmetriser acting on a symmetrised state, which gives zero. The only non-trivial term therefore comes from $e_2$ acting on $v_3^{(2)}$
\begin{align}
    e_2(v_3^{(2)})&=e_2\left(\, \scalebox{0.78}{
}\,.\label{eq:e2(v3M2)L=4e}
\end{align}
Already in eq. \eqref{eq:e2(v3M2)L=4b} and then in eq. \eqref{eq:e2(v3M2)L=4c}, it is clear that the answer will be some combination between states $v_2^{(2)}$ and $v_4^{(2)}$. By comparing eqs \eqref{eq:e2(v3M2)L=4c} and \eqref{eq:e2(v3M2)L=4d}, we obtain the values of $\alpha$ and $\beta$ then written in the last line.

This result, together with  \eqref{eq:H2-acton-v1^2-XZ-Z2} and \eqref{eq:H2-acton-v2^2-XZ-Z2}, and the action on the remaining two states, altogether lead to the following two magnon Hamiltonian 
\be
H^{(2)}=%
\,,\label{eq:HXZ-L=4-M=2}
\ee
whose eigenvalues are $\{g_1^2+g_2^2\pm g_2\sqrt{g_1^2+g_2^2},E_3,E_4,E_5\}$  with
\begin{align}
E_k={}&\frac{4}{3}(g_1^2+g_2^2)+\frac{2}{3}\sqrt{(g_1^2+g_2^2)(g_1^2+4g_2^2)}\nonumber\\
&\times\cos\!\left[\frac{1}{3}\arccos\!\left(\frac{2(g_1^2+g_2^2)^3+9g_2^2(g_1^2+g_2^2)^2-27g_2^2(g_1^2-g_2^2)^2}{2\big[(g_1^2+g_2^2)(g_1^2+4g_2^2)\big]^{3/2}}\right)-\frac{2\pi k}{3}\right],\nonumber\\
&\qquad k=3,4,5\,.
\end{align}

As we have seen, this Hamiltonian was obtained directly from the TL links. But it has an even more remarkable property: the true magnon states (\eqref{eq:v1^2-L=4}-\eqref{eq:v2^2-L=4}) and the three descendants (\eqref{eq:v3^2-L=4}-\eqref{eq:v5^2-L=4}) mix. This explains the energy splitting shown in the plots in Figure~\ref{fig:allmagnonXZL6}.  In the XXZ model this is impossible because of the $U_q(\mathfrak{sl}(2))$ symmetry. But the  $\SU(2)$ was broken in the XZ sector therefore  allowing such behaviour. 

Contrary to the monomial basis, the TL basis, makes this property manifest, since the effect is clearly connected with the need to introduce the S-links and treat them differently from the $XX$ and $ZZ$ states.

Additionally, notice that the mixing disappears at the orbifold point ($g_2=g_1$).

\subsection{Transforming to a manifestly hermitian Hamiltonian}\label{subsec:nonHermiticity}

As in the XY sector, $H^{(2)}$ is non-Hermitian, even though the gauge theory has only real anomalous dimensions. Again, the non-Hermiticity is just an artificial effect caused by the choice of basis. For this case, for example, rescaling the states in the following way,
\begin{align}
    & \tilde{v}_1^{(2)}=v_{1}^{(2)}~, && \tilde{v}_4^{(2)}=\frac{\sqrt{g_1^2+g_2^2}}{2g_2}v_{4}^{(2)}~,\nonumber\\
    & \tilde{v}_2^{(2)}=\frac{\sqrt{g_1^2+g_2^2}}{2g_1}v_{2}^{(2)}~, && \tilde{v}_5^{(2)}=\frac{1}{\sqrt{2}}v_{5}^{(2)}~,\nonumber\\
    & \tilde{v}_3^{(2)}=\frac{1}{\sqrt{2}}v_{3}^{(2)}~,
\end{align}
leads to the following Hermitian reduced Hamiltonian
\be
\tilde{H}^{(2)}=\begin{pmatrix}
2(g_1^2+g_2^2) & -\frac{2g_1g_2^2}{\sqrt{g_1^2+g_2^2}} & 0 & 0 & 0 \\
-\frac{2g_1g_2^2}{\sqrt{g_1^2+g_2^2}} & g_1^2+g_2^2 & \frac{g_2(g_1^2-g_2^2)}{\sqrt{2(g_1^2+g_2^2)}} & 0 & -\frac{g_2(g_1^2-g_2^2)}{\sqrt{2(g_1^2+g_2^2)}} \\
0 & \frac{g_2(g_1^2-g_2^2)}{\sqrt{2(g_1^2+g_2^2)}} & g_1^2+g_2^2 & -g_2\sqrt{\frac{g_1^2+g_2^2}{2}} & 0 \\
0 & 0 & -g_2\sqrt{\frac{g_1^2+g_2^2}{2}} & g_1^2+g_2^2 & -g_2\sqrt{\frac{g_1^2+g_2^2}{2}} \\
0 & -\frac{g_2(g_1^2-g_2^2)}{\sqrt{2(g_1^2+g_2^2)}} & 0 & -g_2\sqrt{\frac{g_1^2+g_2^2}{2}} & g_1^2+g_2^2
\end{pmatrix}.
\ee

Since they are related by similarity transformation $\tilde{H}^{(2)}=SH^{(2)}S^{-1}$, with 
\be
S=\mathrm{diag}\left(1,\frac{\sqrt{g_1^2+g_2^2}}{2 g_1},\frac{1}{\sqrt{2}},\frac{\sqrt{g_1^2+g_2^2}}{2 g_2},\frac{1}{\sqrt{2}}\right).
\ee

\subsection{Parity-manifest basis}
\label{subsec:parity}

It is clear from the spectrum, that the characteristic polynomial of certain Hamiltonians factorises (remember that all eigenvalues are real), which indicates the existence of a symmetry in the Hamiltonian. Notice that states $v_3^{(2)}$ and $v_5^{(2)}$ in $M=2$, $L=4$ are related by reflection, i.e.\ by the parity $P$ defined in Section~\ref{sec:diluteXZ}. In this subsection, we show how to write a basis where the block structure of the Hamiltonian is manifest.

\paragraph{$\mathbf{L=4}$ and $\mathbf{M=2}$.}

The characteristic polynomial for this case factorises into one of order three, times one of order two:
\begin{align}
&P=\left(-E^3+4\left(g_1^2+g_2^2\right)E^2-\left(5g_1^4+9g_1^2g_2^2+4g_2^4\right)E+2g_1^2\left(g_1^4+2g_1^2g_2^2+3g_2^4\right)\right)\nonumber\\
&\times\left(E^2-2\left(g_1^2+g_2^2\right)E+g_1^2\left(g_1^2+g_2^2\right)\right)\;.
\end{align}
This becomes manifest in the Hamiltonian once we work in a parity-manifest basis, in which the states split into a 3-dimensional subspace that is even under parity
\begin{align}
u_{1}^{(2,+)}&\equiv v_{1}^{(2)} =\scalebox{0.8}{
}\;,
\end{align}
and a two-dimensional one that is odd under parity
\begin{align}
u_1^{(2,-)}&\equiv v_{4}^{(2)}=\scalebox{0.8}{%
}\;.
\end{align}

The reduced Hamiltonian in this basis is block diagonal and  given by 

\be
H^{(2)}=%
\, .
\ee

The presence of parity was already discussed in \cite{Bozkurt:2024tpz,Bozkurt:2025exl}. But contrary to the monomial basis, the TL basis makes it manifest. 

As discussed in Section~\ref{sec:diluteXZ}, parity requires the states to start and end with the same colour, so for even $L$ and odd $M$ there is no parity symmetry at fixed couplings (see also Appendix~\ref{app:spectrum}).

\paragraph{$\mathbf{L=6}$ and $\mathbf{M=2}$.}

Another example where this parity symmetry manifests is the $L=6$, $M=2$ case. It is clear that eight states are positive under parity
\begin{align}
u_{1}^{(2,+)}&\equiv v_{1}^{(2)}+v_{6}^{(2)}=\,\scalebox{0.65}{
}\;,
\end{align}
and the remaining six states are negative under parity
\begin{align}
u_{1}^{(2,-)}&\equiv v_{1}^{(2)}-v_{6}^{(2)}=\,\scalebox{0.65}{%
}\;.
\end{align}
The Hamiltonian for this case is written in Appendix \ref{app:spectrum}. 

\section{Generalisation to $\Zset_K$ orbifolds}\label{sec:Zk}

The $\Zset_2$ quiver theory of the previous sections is the simplest member of a family of $\Ncal=2$ superconformal quiver theories, the abelian $\Zset_K$ orbifolds of $\Ncal=4$ SYM and their marginal deformations. In this section we show that the Temperley-Lieb structure of the XY and XZ sectors extends to this whole family, with one colour for each node of the quiver.

\subsection{$\Zset_K$ orbifolds}\label{subsec:ZK-orbifolds}

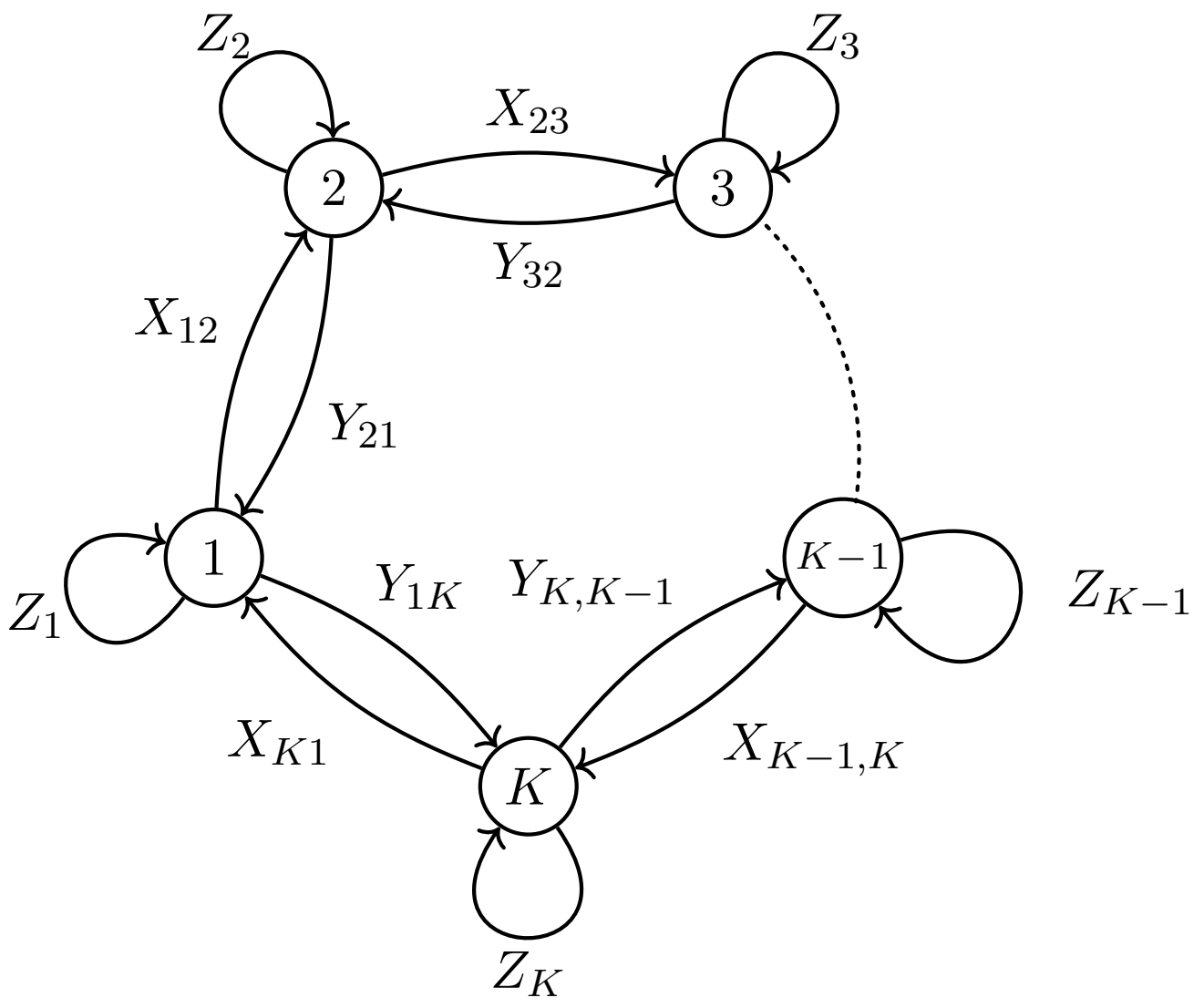
\begin{figure}[t]
 \begin{center}
   \begin{tikzpicture}[scale=0.8,
     v/.style={circle,draw,thick,minimum size=7mm,inner sep=0pt},
     e/.style={->,thick}]
  \def\R{3}
  \node[v] (n1)  at (198:\R) {$1$};
  \node[v] (n2)  at (126:\R) {$2$};
  \node[v] (n3)  at (54:\R)  {$3$};
  \node[v,minimum size=8.5mm] (nk1) at (-18:\R) {\scriptsize $K\!-\!1$};
  \node[v] (nk)  at (-90:\R) {$K$};
  \draw[e] (n1) to[bend left=15] node[auto] {$X_{12}$} (n2);
  \draw[e] (n2) to[bend left=15] node[auto] {$Y_{21}$} (n1);
  \draw[e] (n2) to[bend left=15] node[auto] {$X_{23}$} (n3);
  \draw[e] (n3) to[bend left=15] node[auto] {$Y_{32}$} (n2);
  \draw[e] (nk1) to[bend left=15] node[auto] {$X_{K-1,K}$} (nk);
  \draw[e] (nk) to[bend left=15] node[auto,pos=0.6,xshift=3pt] {$Y_{K,K-1}$} (nk1);
  \draw[e] (nk) to[bend left=15] node[auto] {$X_{K1}$} (n1);
  \draw[e] (n1) to[bend left=15] node[auto,pos=0.4,xshift=-3pt] {$Y_{1K}$} (nk);
  \draw[thick,dotted,line cap=round] (44:\R) arc (44:-8:\R);
  \draw[e] (n1)  to[out=233,in=163,looseness=8] (n1);
  \draw[e] (n2)  to[out=161,in=91,looseness=8]  (n2);
  \draw[e] (n3)  to[out=89,in=19,looseness=8]   (n3);
  \draw[e] (nk1) to[out=17,in=-53,looseness=8]  (nk1);
  \draw[e] (nk)  to[out=-55,in=-125,looseness=8] (nk);
  \node at (198:4.7) {$Z_1$};
  \node at (126:4.7) {$Z_2$};
  \node at (54:4.7)  {$Z_3$};
  \node at (-13:5.6) {$Z_{K-1}$};
  \node at (-90:4.7) {$Z_K$};
\end{tikzpicture}
 \end{center}
 \caption{The quiver corresponding to the $\Zset_K$ orbifold theory. Node $a$ corresponds to colour $a$ (indices mod $K$); $X_{a,a+1}$ and $Y_{a+1,a}$ connect neighbouring nodes and $Z_a$ is the adjoint at node $a$.} \label{Fig:Zkquiver}
\end{figure}

The construction parallels the one of Section~\ref{sec:orbifold}. One starts from $\Ncal=4$ SYM with gauge group $\SU(KN)$ and projects onto the fields invariant under
\be
\gamma\,V\,\gamma^{-1}=V\,,\quad \gamma\,Z\,\gamma^{-1}=Z\, , \quad \gamma\,X\,\gamma^{-1}=\omega^{-1}X\,,\quad \gamma\,Y\,\gamma^{-1}=\omega\,Y\,,
\ee
where $\gamma=\mathrm{diag}(1,\omega,\dots,\omega^{K-1})\otimes 1_N$ and $\omega=e^{2\pi i/K}$, which for $K=2$ reduces to \eqref{eq:orbifold}. This leaves the gauge group $\SU(N)^K$, with $V_a$ and $Z_a$ in the adjoint of the $a$-th factor, while the surviving components $X_{a,a+1}$ and $Y_{a+1,a}$ of $X$ and $Y$ are bifundamentals. Here and in the following, colour indices are understood mod $K$. After marginally deforming, each node comes with its own coupling $g_a$, and the superpotential reads
\be\label{WZK}
\Wcal_{\Zset_K}=i\sum_{a=1}^K g_a\,\Tr\big(Z_aX_{a,a+1}Y_{a+1,a}-Z_aY_{a,a-1}X_{a-1,a}\big)\,,
\ee
which for $K=2$ reproduces \eqref{WZ2}. The orbifold point, where all the $g_a$ are equal, corresponds to the $\Zset_K$ orbifold of $\Ncal=4$ SYM. The quiver of the general $\Zset_K$ orbifold is shown in Figure~\ref{Fig:Zkquiver}.

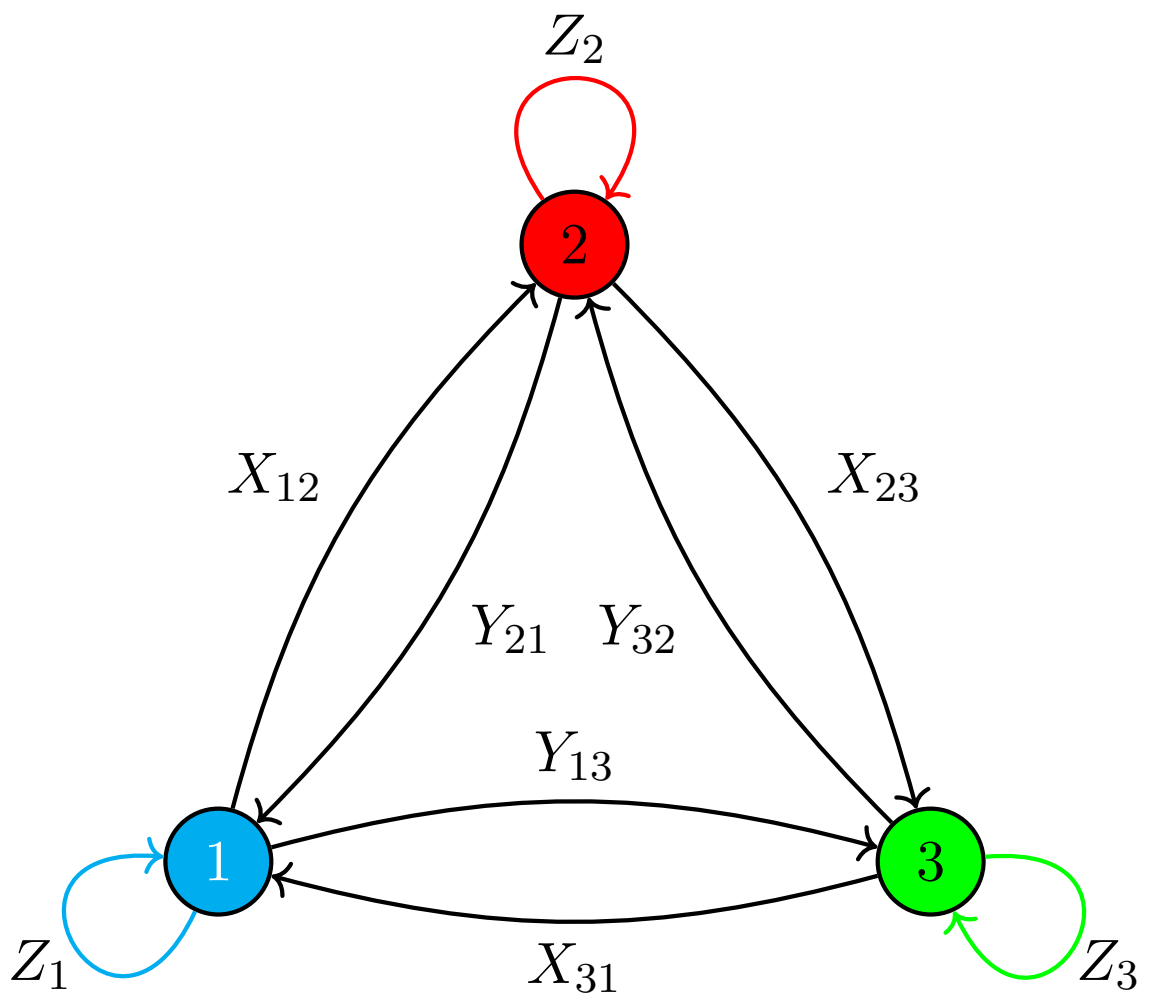
\begin{figure}[h]
 \begin{center}
   \begin{tikzpicture}[scale=0.8,
     v/.style={circle,draw,thick,minimum size=7mm,inner sep=0pt},
     e/.style={->,thick}]
  \def\R{3.4}
  \node[v,fill=\lcyantwo,text=white]  (n1) at (210:\R) {$1$};
  \node[v,fill=red]   (n2) at (90:\R)  {$2$};
  \node[v,fill=green] (n3) at (-30:\R) {$3$};
  \draw[e] (n1) to[bend left=15] node[auto] {$X_{12}$} (n2);
  \draw[e] (n2) to[bend left=15] node[auto] {$Y_{21}$} (n1);
  \draw[e] (n2) to[bend left=15] node[auto] {$X_{23}$} (n3);
  \draw[e] (n3) to[bend left=15] node[auto] {$Y_{32}$} (n2);
  \draw[e] (n3) to[bend left=15] node[auto] {$X_{31}$} (n1);
  \draw[e] (n1) to[bend left=15] node[auto] {$Y_{13}$} (n3);
  \draw[e,\lcyantwo] (n1) to[out=245,in=175,looseness=8] (n1);
  \draw[e,red] (n2) to[out=125,in=55,looseness=8]  (n2);
  \draw[e,green] (n3) to[out=5,in=-65,looseness=8]   (n3);
  \node at (210:5.1) {$Z_1$};
  \node at (90:5.1)  {$Z_2$};
  \node at (-30:5.1) {$Z_3$};
\end{tikzpicture}
 \end{center}
 \caption{The quiver corresponding to the $\Zset_3$ orbifold theory.} \label{Fig:Z3quiver}
\end{figure}

As an example, consider the $\Zset_3$ theory. The holomorphic fields are
\be
Z=\threebythree{Z_1}{0}{0}{0}{Z_2}{0}{0}{0}{Z_3}\;,\;X=\threebythree{0}{X_{12}}{0}{0}{0}{X_{23}}{X_{31}}{0}{0}\;,\;Y=\threebythree{0}{0}{Y_{13}}{Y_{21}}{0}{0}{0}{Y_{32}}{0}\, ,
\ee
and they can be represented in the quiver diagram of Figure~\ref{Fig:Z3quiver}. Since we now have three nodes, we will obtain a three-colour version of the dynamical TL algebra.

For $K\geq3$, $X$ and $Y$ connect neighbouring nodes in opposite directions, so the $\SU(2)$ symmetry of the XY sector is broken; as we show in Section~\ref{subsec:ZK-XY}, a generalisation of the Temperley-Lieb structure persists.

\subsection{The XY sector}\label{subsec:ZK-XY}

The XY sector is built from the bifundamentals $X_{a,a+1}$ and $Y_{a+1,a}$.\footnote{Note that contrary to $K=2$ there are no $X_{a+1,a}$ and $Y_{a,a+1}$ fields, which is special for $K=2$ due to the topology of the $\mathbb{Z}_2$-quiver.} Reading a state from left to right, every $X$ takes us one node forward on the quiver of Figure~\ref{Fig:Zkquiver} and every $Y$ one node back, so a state is a walk on the quiver. For $\Zset_3$, for example,
\be
X_{12}\,X_{23}\,Y_{32}\,X_{23}\,X_{31}
\label{eq:stateexampleZ3}
\ee
is the walk $1\to2\to3\to2\to3\to1$. The colour of the region to the left of site $i$ is
\be\label{eq:ZK-XY-colour}
a_i=a_1+n_X-n_Y \mod K\,,
\ee
where $a_1$ is the colour at the far left and $n_X$ and $n_Y$ count the $X$ and $Y$ fields to the left of site $i$. 

Notice that for $\mathbb{Z}_2$, because of the $\SU(2)$ symmetry, it is impossible to distinguish an $X$ from a $Y$ just by looking at the diagram of the state,  because they are represented in the same way.  However, as mentioned above, that $\SU(2)$ is broken for $K>2$. As a consequence, $X$ only moves us forward by one node, while $Y$ does the same in the opposite direction. So if, for instance for $\mathbb{Z}_3$ we see a  state of the form
\be
\scalebox{0.7}{\begin{tikzpicture}[baseline=-0.6cm,scale=1]
     \filldraw[\lcyan] (-0.25,-1) rectangle (0,0);
     \filldraw[pink] (0,-1) rectangle (1,0);
     \filldraw[\lgreen] (1,-1) rectangle (2,0);
     \filldraw[pink] (2,-1) rectangle (3,0);
     \filldraw[\lgreen] (3,-1) rectangle (4,0);
     \filldraw[\lcyan] (4,-1) rectangle (4.25,0);
     \draw[-,thick,gray] (-0.25,0)--(4.25,0);
     \foreach \x in {0,...,4} \draw[thick] (\x,0)--(\x,-1);
     \foreach \x in {0,...,4} \filldraw[red,thick] (\x,0) circle (1pt);
\end{tikzpicture}}\, ,
\ee
we immediately know it corresponds to \eqref{eq:stateexampleZ3}. Furthermore, if we wanted for instance to change the second $X$ by a $Y$ and keep all other fields in their positions, we would have instead
\be
X_{12}Y_{21}Y_{13}X_{31}X_{12}
\ee
which is then represented as 
\be
\scalebox{0.7}{\begin{tikzpicture}[baseline=-0.6cm,scale=1]
     \filldraw[\lcyan] (-0.25,-1) rectangle (0,0);
     \filldraw[pink] (0,-1) rectangle (1,0);
     \filldraw[\lcyan] (1,-1) rectangle (2,0);
     \filldraw[\lgreen] (2,-1) rectangle (3,0);
     \filldraw[\lcyan] (3,-1) rectangle (4,0);
     \filldraw[pink] (4,-1) rectangle (4.25,0);
     \draw[-,thick,gray] (-0.25,0)--(4.25,0);
     \foreach \x in {0,...,4} \draw[thick] (\x,0)--(\x,-1);
     \foreach \x in {0,...,4} \filldraw[red,thick] (\x,0) circle (1pt);
\end{tikzpicture}}\,,
\ee
which is substantially different from the first.

As for $\Zset_2$, the Hamiltonian comes from the F-terms. The F-term of $Z_a$ only sees pairs of fields that leave node $a$ and come back to it, and writing
\be
H=-\sum_{i=1}^{L-1}e_i
\ee
we find
\begin{align}\label{eq:eXY}
& e_i^{(a)}\,X_{a,a+1}X_{a+1,a+2}=0\,,\qquad e_i^{(a)}\,Y_{a,a-1}Y_{a-1,a-2}=0\,,\nonumber\\
& e_i^{(a)}\,X_{a,a+1}Y_{a+1,a}=-g_a^2\,X_{a,a+1}Y_{a+1,a}+g_a^2\,Y_{a,a-1}X_{a-1,a}\,,\nonumber\\
& e_i^{(a)}\,Y_{a,a-1}X_{a-1,a}=g_a^2\,X_{a,a+1}Y_{a+1,a}-g_a^2\,Y_{a,a-1}X_{a-1,a}\,,
\end{align}
where $a$ is the colour to the left of site $i$. For $K=2$ this is \eqref{eq:HXYg1g2}. What is new for $K\geq3$ is that $e_i$ changes the colour of the region between the two sites, from $a+1$ to $a-1$ and back. For $K=2$ these are the same colour.

The generators satisfy
\begin{align}\label{eq:ZK-XY-rels}
e_i^{(a)}e_i^{(a)}&=-2g_a^2\,e_i^{(a)}\,,\nonumber\\
e_i^{(a)}e_{i\pm1}^{(a')}e_i^{(a)}&=g_a^2\,g_{a'}^2\,e_i^{(a)}\,,\\
e_ie_j&=e_je_i\,,\qquad |i-j|>1\,,\nonumber
\end{align}
where $a'$ is the colour seen by the generator $e_{i\pm1}$. On the states where the cubic product is non-zero, $a$ and $a'$ are the colours at the two ends of the three sites, so $a'$ is either $a+1$ or $a-1$. For $K=2$, $a$ and $a'$ are always $1$ and $2$, and we get back \eqref{XYquadratic} and \eqref{XYcubic}. For $K\geq3$ the cubic coefficient depends on the colours, so the cubic relation becomes dynamical, as will also happen in the XZ sector below (and already happened for $\mathbb{Z}_2$-XZ in Section~\ref{sec:XZsector}).

As in the $\mathbb{Z}_2$ case, all the generators mentioned above can be derived directly from the Feynman diagrams. Let us focus on  $K=3$ as example. 

From the superpotential \eqref{WZK} we see that the vertices relevant for the XY sector are those with an incoming adjoint field $\bar{Z}$ and two outgoing bifundamental fields $X$, $Y$:
\begin{align}
\scalebox{0.8}{
}&=-g_3\,.\label{eq:vertexZ3-XY_2}
\end{align}

The main difference with respect to $\mathbb{Z}_2$ is that the fields now carry the indices of three different gauge groups. We colour the regions of a diagram according to the nodes: blue for node 1, pink for node 2 and green for node 3 (corresponding to the blue, pink and green index lines in the double-line notation used above). Crossing a dashed line (an adjoint field) preserves the colour, while crossing a solid line (a bifundamental field) changes the colour. When crossing $X$ fields, the colour changes cyclically as \{blue $\rightarrow$ pink, pink $\rightarrow$ green, green $\rightarrow$ blue\}, while crossing $Y$ fields results in the opposite order. In the XY sector the line joining the two vertices of a generator is the adjoint field $Z$, which we draw as a dashed line, therefore, the colour outside  the links is the same on both sides. However, contrary to the $K=2$ case, the colours inside the links are \emph{not} fixed once the colour of the outer region is chosen. We therefore have \emph{three} TL generators, one for each colour of the outer region:
\begin{subequations}
\label{eq:XY_TL_gens_Z3}
\begin{align}
e^{(1)} = 
\,.
\end{align}
\end{subequations}
Summing over the allowed vertices, we can see that the sum of closed loops gives $-2g_a^2$ times the $Z$-identity strand. For instance, for $a=1$, we have 
\be
 %
\,~\,.
\ee
This yields the following diagrammatic identities
\begin{gather}
%
\,~\,.
\end{gather}

The two-site A-links and S-links are therefore described as
\begin{align}
A^{[0]}_{1}\equiv \scalebox{0.8}{%
}&=g_3 Y_{32}X_{23}+g_3 X_{31}Y_{13}\,,\label{eq:AandS_c_Z3XY}
\end{align}
while two-site BPS states are represented as
\begin{align}
\scalebox{0.7}{%
}&=Y_{32}Y_{21}\,.
\end{align}

In order to study more than two sites we need to introduce nested states. To do that, it is convenient to summarise all links in one place first
\begin{align}
\scalebox{0.8}{%
}&=\, g_3\, . \label{eq:all-links-Z3-XY}
\end{align}

But first let us use these links to graphically show the cubic relations in equation \eqref{eq:ZK-XY-rels}. If we compute  $e_1^{(1)}e_2^{(2)}e_1^{(1)}$, we obtain
\be%
\, \, .
\ee

 In a certain sense, the nested states in the $\mathbb{Z}_K$-XY sector with $K>2$ share more properties with the XZ sector than with $\mathbb{Z}_2$-XY one. In particular, the states now also contain the white colour and require the introduction of S-links. This difference can be traced back to the underlying symmetry: while the $\mathbb{Z}_2$-XY sector possesses $\mathrm{SU}(2)$ symmetry, this symmetry is broken in both the $\mathbb{Z}_K$-XY model for $K>2$ and the XZ sector for all $K\geq 2$.

The antisymmetrised nested link in the XY sector is defined as
\begin{align}
  A_1^{[1]}\equiv \scalebox{0.8}{
}\right)\label{eq:A11b_Z3XY}\\
&= \frac{1}{g_1^2}\left(g_1^2g_3\,Y_{13}A_{3}^{[0]}X_{31}-g_1^2g_2\,X_{12}A_{2}^{[0]}Y_{21}\right)\label{eq:A11c_Z3XY}\\
&= g_3 \, Y_{13}A_{3}^{[0]}X_{31}-g_2\,X_{12}A_{2}^{[0]}Y_{21}\,,\label{eq:A11d_Z3XY}
\end{align}
where the $1/g_1^{2}$ was introduced to make the total power of $g$ the same as in the state of the form $A_1^{[0]}A_1^{[0]}$ (\raisebox{-0.2ex}{\scalebox{0.3}{%
}}) . Making all states in a given basis have the same power in $g$,  guarantees that the  Hamiltonian of that magnon-sector will be quadratic in $g$. In the computation, we used the values of the links from \eqref{eq:all-links-Z3-XY} and the form of the antisymmetrised states from \eqref{eq:AandS_b_Z3XY} and \eqref{eq:AandS_c_Z3XY}.  We also read the diagrams on the top, for instance, in the first one, blue becomes green ($1\rightarrow 3$) which corresponds to a $Y_{13}$, then green becomes white and green again, which is exactly $A_{3}^{[0]}$ (see \eqref{eq:AandS_c_Z3XY}) and finally, green changes to blue ($3\rightarrow 1$) which corresponds to $X_{31}$. The analogous procedure was applied to the second diagram.

Similarly for the symmetrised nested link (that does not appear for the $\mathbb{Z}_2$-XY-sector)
\begin{align}
  S_1^{[1]}\equiv \scalebox{0.8}{
}\right)\label{eq:S11b_Z3}\\
&= \frac{1}{g_1^2}\left(g_1^2g_3\,Y_{13}A_{3}^{[0]}X_{31}+g_1^2g_2\,X_{12}A_{2}^{[0]}Y_{21}\right)\label{eq:S11c_Z3}\\
&= g_3 \, Y_{13}A_{3}^{[0]}X_{31}+g_2\,X_{12}A_{2}^{[0]}Y_{21}\,.\label{eq:S11d_Z3}
\end{align}

A basis for the two-magnon case in $L=4$, for $a=1$ is then composed of 
\begin{align}
    & v_1^{(2)}=\scalebox{0.8}{%
}=S_1^{[0]}A_{1}^{[0]}\,. \label{eq:XY-Z3-v5}
\end{align}
One can easily obtain the corresponding basis for $a=2$ by shifting all indices ahead by one, namely ${1\rightarrow 2,\, 2\rightarrow 3,\,3\rightarrow 1}$. Doing that once more leads to $a=3$.

Let us now discuss how to go to higher nesting. In order to define $A_1^{[1]}$ in \eqref{eq:A11a_Z3XY}-\eqref{eq:A11d_Z3XY}, we used $A^{[0]}$. So, in practice, we used a recursion relation. We can push that concept further. In the next step we need to nest the following two states 
\be
A_1^{[1]}=\scalebox{0.8}{%
}\label{eq:two-nesting-options-Z3-XY-L4vsL6}\,.
\ee
However, as we will see, we will actually need them with an outer green and pink colour instead of blue. 

Starting with the first diagram in \eqref{eq:two-nesting-options-Z3-XY-L4vsL6} leads to 
\begin{align}
  A_1^{[2]}&\equiv \scalebox{0.8}{%
}\right)\label{eq:A11b_Z3_L6}\\
&= \frac{1}{g_1^4}\left(g_1^3g_3^2\,Y_{13}A_{3}^{[1]}X_{31}-g_1^3g_2^2\,X_{12}A_{2}^{[1]}Y_{21}\right)\label{eq:A11c_Z3_L6}\\
&= \frac{g_3^2}{g_1} \, Y_{13}A_{3}^{[1]}X_{31}-\frac{g_2^2}{g_1}\,X_{12}A_{2}^{[1]}Y_{21}\,,\label{eq:A11d_Z3_L6}
\end{align}
where as before, we notice that for an XY sector, both $X$ and $Y$ are in the bifundamental representation and therefore, have a change of colour at each node. For this reason, if we want the outside colour to be $a$ ($a=1=$ blue in \eqref{eq:A11a_Z3_L6}), the region in between has to  be $a-1 \text{ mod } 3$ ($a-1=3=$ green in the first diagram in \eqref{eq:A11b_Z3_L6}) or $a+1 \text{ mod }3 $  ($a+1=2= $ pink in the second diagram in \eqref{eq:A11b_Z3_L6})). 
We also highlight that when going from $L=4$  (eq. \eqref{eq:A11c_Z3XY}) to $L=6$ (eq. \eqref{eq:A11c_Z3_L6}) the green diagram's coefficient went from $g_1^2g_3$ to $g_1^3g_3^2$. This is because we went from two links of type \raisebox{-0.2ex}{\scalebox{0.3}{\begin{tikzpicture}[baseline=-1cm,scale=1]
      \fill[\lcyan] (-0.25,0) rectangle (1.25,-1);
      \fill[\lgreen] (0,0) arc (-180:0:0.5);
        \draw[thick,black] (0,0) arc (-180:0:0.5);
        \draw[gray,dashed] (0.5,-0.5)--(0.5,-1);
\end{tikzpicture}}} and one of type \raisebox{-0.2ex}{\scalebox{0.3}{\begin{tikzpicture}[baseline=-1cm,scale=1]
  \begin{scope}[rotate around={180:(0.5,-0.5)}]
      \fill[\lgreen] (-0.25,0) rectangle (1.25,-1);
      \fill[\lcyan] (0,0) arc (-180:0:0.5);
        \draw[thick,black] (0,0) arc (-180:0:0.5);
        \draw[gray,dashed] (0.5,-0.5)--(0.5,-1);
  \end{scope}
\end{tikzpicture}}}, to three of type \raisebox{-0.2ex}{\scalebox{0.3}{\begin{tikzpicture}[baseline=-1cm,scale=1]
      \fill[\lcyan] (-0.25,0) rectangle (1.25,-1);
      \fill[\lgreen] (0,0) arc (-180:0:0.5);
        \draw[thick,black] (0,0) arc (-180:0:0.5);
        \draw[gray,dashed] (0.5,-0.5)--(0.5,-1);
\end{tikzpicture}}} and two of type \raisebox{-0.2ex}{\scalebox{0.3}{\begin{tikzpicture}[baseline=-1cm,scale=1]
  \begin{scope}[rotate around={180:(0.5,-0.5)}]
      \fill[\lgreen] (-0.25,0) rectangle (1.25,-1);
      \fill[\lcyan] (0,0) arc (-180:0:0.5);
        \draw[thick,black] (0,0) arc (-180:0:0.5);
        \draw[gray,dashed] (0.5,-0.5)--(0.5,-1);
  \end{scope}
\end{tikzpicture}}}. For the pink diagram the same happened, just replacing 3 by 2 (green by pink). At the same time, to keep the Hamiltonian quadratic, in the first case we normalised by $1/g_1^2$, and in the second by $1/g_1^4$. If we continue to add more and more A-links on the top  and divide by the appropriate normalisation, we can write the following recursion relation for the state starting with $a=1$ (blue) 
\begin{equation}
    A_1^{[r]}=\frac{g_3^r}{g_1^{r-1}}Y_{13}A_3^{[r-1]}X_{31}-\frac{g_2^r}{g_1^{r-1}}X_{12}A_2^{[r-1]}Y_{21}\,, \quad r\ge 1\, . \label{eq:Z3XY-A_1^[r]}
\end{equation}

For the second state in \eqref{eq:two-nesting-options-Z3-XY-L4vsL6}, we write the following diagrams
\begin{align}
  A_1^{[1,1]}&\equiv \scalebox{0.8}{
}\right)\label{eq:A11b_Z3_L6b}\\
&= \frac{1}{g_1^4}\left(g_1^3g_3^2\,Y_{13}A_{3}^{[0]}A_{3}^{[0]}X_{31}-g_1^3g_2^2\,X_{12}A_{2}^{[0]}A_{2}^{[0]}Y_{21}\right)\label{eq:A11c_Z3_L6b}\\
&= \frac{g_3^2}{g_1} \, Y_{13}A_{3}^{[0]}A_{3}^{[0]}X_{31}-\frac{g_2^2}{g_1}\,X_{12}A_{2}^{[0]}A_{2}^{[0]}Y_{21}\,.\label{eq:A11d_Z3_L6b}
\end{align}
However, one can observe that the only difference to construct $A_1^{[1,1]}$ in comparison with $A_1^{[2]}$ is that the known part $A_a^{[1]}$ was replaced by $A_a^{[0]}A_a^{[0]}$, while all coefficients remained the same. Therefore, we can write, for $r_1,r_2\ge 1$
\begin{equation}
    A_1^{[r_1,r_2]}=\frac{g_3^{r_1+r_2}}{g_1^{r_1+r_2-1}}Y_{13}A_3^{[r_1-1]}A_3^{[r_2-1]}X_{31}-\frac{g_2^{r_1+r_2}}{g_1^{r_1+r_2-1}}X_{12}A_2^{[r_1-1]}A_2^{[r_2-1]}Y_{21}\,.
\end{equation}
This formula allows one to write states of the form
\be
\scalebox{0.8}{\begin{tikzpicture}[baseline={([yshift=-0.5ex]current bounding box.center)}]
  \fill[\lcyan] (-0.25,0) rectangle (7.25,-2.02);
  \fill[white] (0,0) arc (-180:0:3.5cm and 1.52cm) -- cycle;
  \draw[-,thick] (-0.25,0)--(7.25,0);
  \draw[-,thick] (0,0) arc (-180:0:3.5cm and 1.52cm);
  \draw[-,thick] (1,0) arc (-180:0:0.5);
  \draw[-,thick] (3,0) arc (-180:0:1.5cm and 1cm);
  \draw[-,thick] (4,0) arc (-180:0:0.5);
  \draw[-,gray,dashed] (1.5,-0.5)--(1.5,-1.25);
  \draw[-,gray,dashed] (4.5,-0.5)--(4.5,-1);
  \draw[-,gray,dashed] (3.5,-0.75)--(3.5,-1.52);
  \draw[-,gray,dashed] (5.5,-0.75)--(5.5,-1.25);
  \draw[-,gray,dashed] (0.5,-0.78)--(0.5,-2.02);
  \draw[-,gray,dashed] (2.5,-1.46)--(2.5,-2.02);
  \draw[-,gray,dashed] (4.5,-1.46)--(4.5,-2.02);
  \draw[-,gray,dashed] (6.5,-0.78)--(6.5,-2.02);
  \foreach \x in {0,...,7} \filldraw[red,thick] (\x,0) circle (1pt);
\end{tikzpicture}}\, ,\label{eq:A1^[1,2]}
\ee
for instance, as
\be
A_1^{[1,2]}=\frac{g_3^3}{g_1^2}Y_{13}A_3^{[0]}A_3^{[1]}X_{31}-\frac{g_2^3}{g_1^2}X_{12}A_2^{[0]}A_2^{[1]}Y_{21}\,.
\ee
In the same spirit we can continue higher and obtain for $r_j\ge 1$, 
\begin{align}
    A_1^{[r_1,\,\cdots\, , r_n]}&=\frac{g_3^{\sum_{j=1}^{n}r_j}}{g_1^{\sum_{j=1}^{n}r_j-1}}Y_{13}A_3^{[r_1-1]}A_3^{[r_2-1]}\cdots A_3^{[r_n-1]}X_{31}\nonumber\\
    &\qquad -\frac{g_2^{\sum_{j=1}^{n}r_j}}{g_1^{\sum_{j=1}^{n}r_j-1}}X_{12}A_2^{[r_1-1]}A_2^{[r_2-1]}\cdots A_2^{[r_n-1]}Y_{21}\, .\label{eq:generalZ3-Alink-recursion}
\end{align}
This formula gives immediately the answer for a diagram like
\be
\scalebox{0.45}{\begin{tikzpicture}[baseline={([yshift=-0.5ex]current bounding box.center)}]
  \fill[\lcyan] (-0.25,0) rectangle (17.25,-2.83);
  \fill[white] (0,0) arc (-180:0:8.5cm and 2.33cm) -- cycle;
  \draw[-,thick] (-0.25,0)--(17.25,0);
  \draw[-,thick] (0,0) arc (-180:0:8.5cm and 2.33cm);
  \draw[-,thick] (1,0) arc (-180:0:0.5);
  \draw[-,thick] (3,0) arc (-180:0:1.5cm and 1cm);
  \draw[-,thick] (4,0) arc (-180:0:0.5);
  \draw[-,thick] (7,0) arc (-180:0:0.5);
  \draw[-,thick] (9,0) arc (-180:0:0.5);
  \draw[-,thick] (11,0) arc (-180:0:2.5cm and 1.36cm);
  \draw[-,thick] (12,0) arc (-180:0:1.5cm and 1cm);
  \draw[-,thick] (13,0) arc (-180:0:0.5);
  \draw[-,gray,dashed] (1.5,-0.5)--(1.5,-1.32);
  \draw[-,gray,dashed] (3.5,-0.75)--(3.5,-1.89);
  \draw[-,gray,dashed] (4.5,-0.5)--(4.5,-1);
  \draw[-,gray,dashed] (5.5,-0.75)--(5.5,-2.18);
  \draw[-,gray,dashed] (7.5,-0.5)--(7.5,-2.31);
  \draw[-,gray,dashed] (9.5,-0.5)--(9.5,-2.31);
  \draw[-,gray,dashed] (11.5,-0.82)--(11.5,-2.18);
  \draw[-,gray,dashed] (12.5,-0.75)--(12.5,-1.25);
  \draw[-,gray,dashed] (13.5,-0.5)--(13.5,-1);
  \draw[-,gray,dashed] (13.5,-1.36)--(13.5,-1.89);
  \draw[-,gray,dashed] (14.5,-0.75)--(14.5,-1.25);
  \draw[-,gray,dashed] (15.5,-0.82)--(15.5,-1.32);
  \draw[-,gray,dashed] (0.5,-0.79)--(0.5,-2.83);
  \draw[-,gray,dashed] (2.5,-1.65)--(2.5,-2.83);
  \draw[-,gray,dashed] (4.5,-2.06)--(4.5,-2.83);
  \draw[-,gray,dashed] (6.5,-2.27)--(6.5,-2.83);
  \draw[-,gray,dashed] (8.5,-2.33)--(8.5,-2.83);
  \draw[-,gray,dashed] (10.5,-2.27)--(10.5,-2.83);
  \draw[-,gray,dashed] (12.5,-2.06)--(12.5,-2.83);
  \draw[-,gray,dashed] (14.5,-1.65)--(14.5,-2.83);
  \draw[-,gray,dashed] (16.5,-0.79)--(16.5,-2.83);
  \foreach \x in {0,...,17} \filldraw[red,thick] (\x,0) circle (1pt);
\end{tikzpicture}}\, .\label{eq:long-diagram-example}
\ee
Notice that our nesting notation with upper labels of type $[r_1,r_2,r_3]$, is no longer unique after a certain ``depth" in the nesting process. This poses a limitation for the generality of the relation \eqref{eq:generalZ3-Alink-recursion}.  It would be interesting to find an analogue of the parenthesis notation introduced in \cite{Read:2007ti}.

Notice that the limitation is only on how to name the states. The actual process used to compute these recursions is completely general, and consists of repeating as often as necessary the same procedure described in the examples \eqref{eq:A11a_Z3_L6}-\eqref{eq:A11d_Z3_L6}. 

Similarly, states that have only the outermost layer symmetric and all other layers antisymmetric (such as  $S_1^{[1]}$), can be systematically written by applying the procedure \eqref{eq:S11a_Z3}-\eqref{eq:S11d_Z3} for more and more levels. Remarkably what one finds for $S_1^{[r_1,\,\cdots\, , r_n]}$ is always the same as its antisymmetric counterpart, but replacing the minus sign by a plus sign. Compare for example, the two states in \eqref{eq:AandS_a_Z3XY}, and state \eqref{eq:S11d_Z3} with its antisymmetric counterpart in \eqref{eq:A11d_Z3XY}. The reason behind this is very simple: in the list of links \eqref{eq:all-links-Z3-XY} the only thing that determines which coupling constant appears in a link is the colour outside the link. The form of the link only changes the sign. This is an important difference with respect to the XZ sector (see Sections~\ref{sec:XZsector} and \ref{sec:Zk-XZ-sector}).

For $K\ge 4$, the procedure simply generalises. We will have more copies of the gauge group, and there is nothing fundamentally different than for $K=3$. One just starts with the links
\begin{align}
\;&=\, g_{a}\, . \label{eq:links-Z3-XY-rows14}
\end{align}
all mod $K$, and the states
\begin{align}
A^{[0]}_{a}\equiv%
\;&=\,g_{a}\, Y_{a,a-1}X_{a-1,a}+g_{a}\, X_{a,a+1}Y_{a+1,a}\,.
\end{align}
and then applies the same steps as for $K=3$.

\subsection{The XZ sector}\label{sec:Zk-XZ-sector}

As in Section~\ref{sec:XZsector}, the generators follow from the cubic vertices of the superpotential \eqref{WZK}. Each vertex contains exactly one adjoint field $Z_a$ and comes with a factor $\pm g_a$, where, as before, the sign is positive if the fields read anticlockwise are in the order $XYZ$ and negative if they are in the order $XZY$. 
Gluing two vertices along a $Y$ line exactly as in Section~\ref{sec:XZsector}, and colouring the regions accordingly, we obtain the generators $e_i^{(a)}$, where as before $a$ is the colour of the region to the left of site $i$. They annihilate two neighbouring $X$ fields as well as two neighbouring $Z$ fields, while on the remaining two states they act as
\begin{align}\label{eq:eZK}
e_i^{(a)}\,X_{a,a+1}Z_{a+1}&=-g_{a+1}^2\,X_{a,a+1}Z_{a+1}+g_ag_{a+1}\,Z_aX_{a,a+1}\,,\nonumber\\
e_i^{(a)}\,Z_aX_{a,a+1}&=g_ag_{a+1}\,X_{a,a+1}Z_{a+1}-g_a^2\,Z_aX_{a,a+1}\,,
\end{align}
which for $K=2$ reduces to the generators of Section~\ref{sec:XZsector}. The open Hamiltonian of the XZ sector is then
\be
H=-\sum_{i=1}^{L-1}e_i\,.
\ee

The generators satisfy
\begin{align}\label{eq:ZK-XZ-rels}
e_i^{(a)}e_i^{(a)}&=-\big(g_a^2+g_{a+1}^2\big)\,e_i^{(a)}\,,\nonumber\\
e_ie_{i\pm1}e_i&=g_{a+1}^2\,g_{a'-1}^2\,e_i\,,\\
e_ie_j&=e_je_i\,,\qquad |i-j|>1\,,\nonumber
\end{align}
where in the cubic relation $a$ and $a'$ are the colours of the regions to the left and to the right of the three sites. With one $X$ field on the three sites $a'=a+1$ and the cubic coefficient is $g_a^2g_{a+1}^2$, while with two $X$ fields $a'=a+2$ and it is $g_{a+1}^4$. For $K=2$ and $a=1$ these are $g_1^2g_2^2$ and $g_2^4$, as in Section~\ref{sec:XZsector}. The loop fugacity depends on the colour of the region in which the loop sits. For $K=2$ it is the same for both colours, while for $K\geq3$ it changes from region to region. The cubic coefficient depends on the colours as well, so the cubic relation is dynamical too. We return to this algebra in \eqref{eq:XZ-ZK-dyn} and \eqref{eq:gen-dTL}. At the orbifold point all the $g_a$ are equal, and the generators reduce to $g^2$ times the $q=1$ Temperley-Lieb generators of Section~\ref{sec:N=4SYM}.

\paragraph{$\mathbf{K=3}$.} Let us focus on the concrete example with $\Zset_3$ XZ sector.
In Section~\ref{subsec:ZK-XY} we saw that for the XY sector, going from $K=2$ to $K>2$ led to fundamental differences due to the breaking of $\mathrm{SU}(2)$. These included the colour of the links no longer being fixed after the outside colour was chosen (which translates into the presence of white in the diagrams) and the need to introduce symmetrised (S-)links. For the XZ, however, these features are already there for  $\Zset_2$, since the $\mathrm{SU}(2)$ was already broken from the beginning. Therefore, the generalisation of XZ from $\Zset_2$ to $\Zset_3$ is very straightforward. The only important difference, already mentioned above, is that now both the quadratic and cubic TL relations are dynamical.

One of course now has three gauge groups  and three corresponding  colours ($1=\,$blue, $2=\,$pink and $3=\,$green). The line joining two vertices is now a solid line, describing $\bar{Y}_{12}$, $\bar{Y}_{23}$ or $\bar{Y}_{31}$ and crossing from one side of the $\bar{Y}$ strand to the other changes the colour.

Colouring the regions according to the nodes for $\Zset_3$ for instance, leads to the following generators
\begin{subequations}\label{eq:TL1-Z3-XZ}
\begin{align}
e^{(1)}=
\,\;,
 \end{align}
 \end{subequations}
 where the field connecting the two A-links is a $\bar{Y}$, which makes the colour increase by one when crossing it from left to right. The coupling constant at each vertex, already in terms of links, is given by
\begin{align}
\scalebox{0.8}{%
}&=\, g_3\, .\label{eq:links-for-Z3-XZ}
\end{align}
As in the $\mathbb{Z}_2$-XZ case (Section~\ref{sec:XZsector}), the A-links just come directly from the superspace diagrams. The S-links are chosen such that they are annihilated  by the Hamiltonian of the sector. In particular, here the S-links always give the coupling constant of the colour that appears only once in the vertex. 

The two-site vectors can then be written as
\begin{align}
A^{[0]}_{12}\equiv \scalebox{0.8}{%
}&=g_3 X_{31}Z_1+g_1 Z_3X_{31}\,,\label{eq:AandS_31_Z3}
\end{align}
The notation with lower indices in $A$ and $S$ indicates the colours appearing on the diagram mod $3$. So, if the diagram starts with blue and ends with green, we write for instance $A_{123}$, but if it starts with blue and ends with blue, we just write $A_1$ even if changed colours many times. Similarly, if it starts with blue but it ends in pink, we write $A_{12}$.

Summing over the allowed vertices, we can see that the sum of closed loops gives a coefficient times the $\bar{Y}$-identity strand. Contrary to the $\Zset_2$ case, this coefficient is now dynamical and given by
\begin{align}
\;.
\end{align}
Acting twice in a row with a dynamical TL generator $e^{(a)}$, makes exactly these bubbles to appear, leading to the first relation in \eqref{eq:ZK-XZ-rels}.

The dynamical cubic TL relations \eqref{eq:ZK-XZ-rels}  again follow directly from the superspace vertices. Recall that looking at the graphical representation \eqref{eq:cubic-Z3-XZ-graphically-1},  the A-link has one $X$ and one $Z$, and the solid line represents an $X$ (in this case $X_{23}$). So, for two $X$'s and one $Z$, starting with colour $a=1$ we find a $g_2^4$ overall coefficient on the rhs
\be
%
\, .\label{eq:cubic-Z3-XZ-graphically-1}
\ee
The reasoning behind this is the following. The presence of the additional $X$ forces the last colour to be green, and since the field in between the links is a $Y$ (which is bifundamental), the middle of the diagram has to be pink. Consequently, all four internal vertices have two pink ($a=2$) regions and therefore each of them contributes with a $g_2$. In the same spirit, for two $Z$'s and one $X$, the fact that $Z$ is in the adjoint forces the last region to remain pink. This makes the middle blue, leading to two vertices contributing a $-g_1^2$ in total and the other two contributing $-g_2^2$, leading to
\be

\ee
The construction of the basis smoothly generalises beyond $K=2$. For four sites, for instance, the true magnon states are given by
\begin{align}
v_{1}^{(2)}\equiv\scalebox{0.8}{%
}&=g_3\,X_{12}A_{23}^{[0]}Z_3-g_1\,Z_1A_{12}^{[0]}X_{23}\equiv A_{123}^{[1]}\;,\label{eq:v2^2-L=4_Z3}
\end{align}
while the three descendants are given by
\begin{align}
v_{3}^{(2)}\equiv\scalebox{0.8}{%
}&=S_{12}^{[0]}A_{23}^{[0]}\;.\label{eq:v5^2-L=4_Z3}
\end{align}
The procedure to construct the states \eqref{eq:v2^2-L=4_Z3} and \eqref{eq:v4^2-L=4_Z3} above is exactly the same as the one used to construct $A_{1}^{[1]}$ in \eqref{eq:A11a}-\eqref{eq:A11d} and $S_{1}^{[1]}$ in \eqref{eq:S11a}-\eqref{eq:S11d}. 
Since the construction for the $\mathbb{Z}_3$-XZ sector is not fundamentally different from the $\mathbb{Z}_2$ case described in Section~\ref{sec:XZsector}, we do not give all details, or higher nesting examples. The reader can straightforwardly obtain the corresponding $\mathbb{Z}_3$ version of the $L=6$ link basis introduced in Section~\ref{sec:basis} by just applying the same steps.  

\paragraph{$\mathbf{K\in \mathbb{Z}_{\ge 2}}$.}

In general, for $\Zset_K$, the vertices involving the nodes $a$ and $a+1$ are given by
\begin{align}
&    \scalebox{0.9}{
}=\, g_a\, ,
\end{align}
mod $ K$. The corresponding two-site states are 
\begin{align}
A^{[0]}_{a,a+1}\equiv \scalebox{0.9}{%
}&=g_a X_{a,a+1}Z_{a+1}+g_{a+1} Z_aX_{a,a+1}\,\quad \text{mod} \, K,\label{eq:S_am1}
\end{align}
As before, these states and links are all we need to construct any other state in the basis for higher number of sites and any number of magnons. 
For four sites, for instance, we can define the new antisymmetrised diagram as 
\begin{align}
A_{a,a+1,a+2}^{[1]}&\equiv \scalebox{0.8}{%
}\nonumber \\
&=g_{a+2}\,X_{a,a+1}A_{a+1,a+2}^{[0]}Z_{a+2}-g_a\,Z_aA_{a,a+1}^{[0]}X_{a+1,a+2}\;.\label{eq:v2^2-L=4_Zk}
\end{align}
The symmetrised nested state for four sites is given by
\begin{align}
S_{a,a+1,a+2}^{[1]}&\equiv\scalebox{0.8}{%
}\nonumber\\
& =g_{a+1}\,X_{a,a+1}A_{a+1,a+2}^{[0]}Z_{a+2}+g_{a+1}\,Z_aA_{a,a+1}^{[0]}X_{a+1,a+2}\;.\label{eq:v4^2-L=4_Zk}
\end{align}
Writing a multi-site basis for $\mathbb{Z}_K$ follows naturally from the already discussed examples.

\section{Dynamical Temperley-Lieb algebra and dynamical twist}\label{sec:twist}
In the quiver theories discussed above, we saw that the Temperley-Lieb generators know about the colours of the regions they act in. Algebras of this kind were recently studied by Ren \cite{Ren:2023mtn}, who called them dynamical Temperley-Lieb algebras. In this section we show that our algebras can be placed into a natural generalisation of this framework. 

\subsection{Colours and dynamical notation} \label{sec:dynamical}

In order to properly describe the definition of a dynamical Temperley-Lieb algebra as in \cite{Ren:2023mtn} we need to introduce some notation. 

\paragraph{Dynamical operators.} As before, $e_i^{(a)}$ is the generator acting on sites $i$ and $i+1$ when the region to the left of site $i$ has colour $a$. Reading the chain from left to right, the colour changes at every bifundamental field ($X$ or $Y$) and stays the same at every adjoint field ($Z$). A state is therefore fixed by the colour $\lambda_0\in\{1,\dots,K\}$ of the leftmost region and its sequence of fields, and we write it as $\ket{\lambda_0,XZ\dots}$.

For the $\Zset_2$ case, for example,
\be
\ket{1,XZXX}=\scalebox{0.8}{\begin{tikzpicture}[baseline=-0.6cm,scale=1]
  \filldraw[\lcyan] (-0.25,-1) rectangle (0,0);
  \filldraw[pink] (0,-1) rectangle (2,0);
  \filldraw[\lcyan] (2,-1) rectangle (3,0);
  \filldraw[pink] (3,-1) rectangle (3.25,0);
  \draw[-,thick,gray] (-0.25,0)--(3.25,0);
  \draw[thick] (0,0)--(0,-1);
  \draw[thick,dashed] (1,0)--(1,-1);
  \draw[thick] (2,0)--(2,-1);
  \draw[thick] (3,0)--(3,-1);
  \foreach \x in {0,...,3} \filldraw[red,thick] (\x,0) circle (1pt);
\end{tikzpicture}}=\ket{X_{12}Z_2X_{21}X_{12}}\,,
\ee
\be
\ket{2,XZXX}=\scalebox{0.8}{\begin{tikzpicture}[baseline=-0.6cm,scale=1]
  \filldraw[pink] (-0.25,-1) rectangle (0,0);
  \filldraw[\lcyan] (0,-1) rectangle (2,0);
  \filldraw[pink] (2,-1) rectangle (3,0);
  \filldraw[\lcyan] (3,-1) rectangle (3.25,0);
  \draw[-,thick,gray] (-0.25,0)--(3.25,0);
  \draw[thick] (0,0)--(0,-1);
  \draw[thick,dashed] (1,0)--(1,-1);
  \draw[thick] (2,0)--(2,-1);
  \draw[thick] (3,0)--(3,-1);
  \foreach \x in {0,...,3} \filldraw[red,thick] (\x,0) circle (1pt);
\end{tikzpicture}}=\ket{X_{21}Z_1X_{12}X_{21}}\,.
\ee
It is also common in the dynamical spin chain literature to introduce an operator $\lambda$ \cite{Niccoli:2012mq,Rubtsov:2009ma} which measures the starting colour, for example 
\begin{equation}
  \lambda\,|\lambda_0,XZ\dots\rangle = \lambda_0\, |\lambda_0,XZ\dots\rangle\,,
\end{equation}
and we will make use of this notation below.

To follow the colour along the chain, we introduce the operator $h$ which acts on local states of the spin chain by 
\begin{equation}
  h(X) = +1,\quad h(Y)=-1,\quad h(Z)=0\,.
\end{equation}
In other words, $h$ measures the colour change as we move past a field on the chain, following the quiver diagram. 

The corresponding operator acting on site $i$ of the chain will be denoted $h_i$. The operator $h^{(i)}$, defined by
\begin{equation}
  h^{(i)}:=h_1+\dots+h_i\,,\qquad h^{(0)}:=0\,,
\end{equation}
then measures the total change in colour (mod $K$) after crossing the first $i$ sites of the chain. As a result, the operator $\lambda+h^{(i)}$ tells us the colour after crossing the first $i$ sites. For example, for $\Zset_2$
\begin{equation}
  (\lambda+h^{(3)})|1,XZZX\rangle = (1+1)|1,XZZX\rangle = 2\,|1,XZZX\rangle\,,
\end{equation}
since after $X_{12}Z_2Z_2$ we are in region 2 (pink):
\be
\ket{1,XZZX}=\scalebox{0.8}{\begin{tikzpicture}[baseline=-0.6cm,scale=1]
  \filldraw[\lcyan] (-0.25,-1) rectangle (0,0);
  \filldraw[pink] (0,-1) rectangle (3,0);
  \filldraw[\lcyan] (3,-1) rectangle (3.25,0);
  \draw[-,thick,gray] (-0.25,0)--(3.25,0);
  \draw[thick] (0,0)--(0,-1);
  \draw[thick,dashed] (1,0)--(1,-1);
  \draw[thick,dashed] (2,0)--(2,-1);
  \draw[thick] (3,0)--(3,-1);
  \foreach \x in {0,...,3} \filldraw[red,thick] (\x,0) circle (1pt);
\end{tikzpicture}}=\ket{X_{12}Z_2Z_2X_{21}}\,.
\ee
All of the colour shifts which had to be tracked by hand in the Temperley-Lieb generators of the previous sections can now be incorporated algebraically. In particular, we define
\be\label{eq:dTL-local}
e_i(\lambda)\equiv e_i^{(\lambda+h^{(i-1)})}\,.
\ee
In this notation the dynamical variable $\lambda$ and the shifts $h^{(i)}$ carry all the colour dependence. Acting on a state, $e_i(\lambda)$ simply picks up the colour of the region to the left of site $i$, i.e.\ the colour after crossing the first $i-1$ sites. For the state above, for example, the region to the left of site $2$ has colour $1+h(X)=2$, so
\begin{equation}
  e_2(\lambda)\,|1,XZZX\rangle = e_2^{(\lambda+h^{(1)})}|1,XZZX\rangle = e_2^{(2)}\,|1,XZZX\rangle\,.
\end{equation}
The alert reader might be worried about ordering ambiguities in the above definitions. The TL generators change the fields they act on (e.g.\ $e_i$ maps $XZ$ to $ZX$), so one may ask whether the colour $\lambda+h^{(i-1)}$ should be measured before or after acting with $e_i$. Since $e_i$ only acts on sites $i$ and $i+1$, it does not change $\lambda$ or the first $i-1$ fields, so it commutes with $\lambda+h^{(i-1)}$ and there is no ambiguity.

\subsection{Dynamical Temperley-Lieb algebras}\label{subsec:dynTL}

As discussed in Section~\ref{sec:XXZ}, in the standard Temperley-Lieb algebra every closed loop gives the same factor $\beta$, $e_ie_i=\beta\,e_i$ \eqref{TLrels}, with $\beta=-(q+q^{-1})$ for XXZ. 

In dynamical algebras on the other hand, the regions of the diagrams are coloured, as in SOS or face models \cite{Pearce:1990ila}, and the factor of a loop may depend on the colour of the regions it sits in. Following this logic, Ren \cite{Ren:2023mtn} defined dynamical Temperley-Lieb algebras by
\begin{align}\label{eq:dTL-relations}
e_i(\lambda)\,e_i(\lambda)&=\beta(\lambda+h^{(i-1)})\,e_i(\lambda)\,,\nonumber\\
e_i(\lambda)\,e_{i\pm1}(\lambda)\,e_i(\lambda)&=e_i(\lambda)\,,\\
e_i(\lambda)\,e_j(\lambda)&=e_j(\lambda)\,e_i(\lambda)\,,\qquad |i-j|>1\,.\nonumber
\end{align}
To spell out the details, a two-site loop bordering a region of colour $a$ gives $\beta(a)$, and the cubic relations are the standard ones \eqref{TLrels}.

\paragraph{Baxterisation.}
Ren's dynamical Temperley-Lieb algebras provide a natural starting point for constructing dynamical integrable systems. Indeed, Ren also showed how to Baxterise these algebras, much as for XXZ at the end of Section~\ref{sec:XXZ}, to produce dynamical R-matrices. One makes the ansatz $\check{R}_{i,i+1}(u,\lambda)=1+x(u)\,e_i(\lambda)$ and imposes the dynamical Yang-Baxter equation in braided form,
\be\label{eq:dYBE}
\check{R}_{i+1,i+2}(u-v,\lambda)\,\check{R}_{i,i+1}(u,\lambda)\,\check{R}_{i+1,i+2}(v,\lambda)=\check{R}_{i,i+1}(v,\lambda)\,\check{R}_{i+1,i+2}(u,\lambda)\,\check{R}_{i,i+1}(u-v,\lambda)\,.
\ee
The shifts of the dynamical variable, which are usually written out in this equation, are hidden inside the generators here: $\check{R}_{i+1,i+2}$ sees the colour $\lambda+h^{(i)}$, which depends on the field at site $i$, and hence on what $\check{R}_{i,i+1}$ has done. 

Ren finds a solution to the dynamical Yang-Baxter equation only if neighbouring regions have the same loop weight \cite[Theorem 2.24]{Ren:2023mtn},
\be\label{eq:Ren-kappa-cond}
\beta(\lambda+h^{(i)})=\beta(\lambda+h^{(i-1)})\,.
\ee Since every node of the quiver can be reached from every other node, the loop weight then has to be the same for all colours, in other words $\beta(\lambda+h^{(i-1)})$ must in fact be a constant, and hence non-dynamical. The Baxterisation therefore only applies to \textit{colour-dependent representations} of the ordinary Temperley-Lieb algebra, as in SOS models. 

On the other hand, truly dynamical Temperley-Lieb algebras with underlying dynamical R-matrices do exist. Ren gives an elliptic example and notes that it cannot be Baxterised in this way \cite[Remark 2.32]{Ren:2023mtn}. In Appendix~\ref{app:Felder} we show that this example appears in the bulk Hamiltonian of the integrable spin chain built from Felder's elliptic dynamical $R$-matrix \cite{Felder:1994be}, together with an extra diagonal term. This term is why Felder's $R$-matrix is not of the form $1+x(u)\,e_i(\lambda)$, so that a truly dynamical TL algebra can underlie a dynamical $R$-matrix without Ren's Baxterisation applying.

\subsection{The quiver sectors as dynamical Temperley-Lieb algebras}\label{subsec:quiver-to-Ren}

We now show how the quiver Temperley-Lieb algebras we have constructed in earlier sections can be put into the dynamical framework of the previous subsection. We start by considering the XY sector with $K=2$.

\paragraph{XY sector, $K=2$.} Here every field is a bifundamental $X$ or $Y$, so $h(X)=+1\equiv-1=h(Y)$ mod 2 and the colours simply alternate along the chain: $\lambda+h^{(i-1)}=\lambda+i-1$ mod 2, independently of the fields. In the notation of Section~\ref{sec:XYsector}, the generators satisfy \eqref{XYquadratic} and \eqref{XYcubic},
\begin{equation}
  e_i^{(a)}e_i^{(a)}=-2g_a^2\,e_i^{(a)}\,,\qquad e_i^{(a)}e_{i\pm1}^{(a\pm1)}e_i^{(a)}=g_1^2g_2^2\,e_i^{(a)}\,,
\end{equation}
so the loop weight depends on the colour, while the cubic coefficient does not. Rescaling all generators by the same constant, $\tilde e_i(\lambda)=e_i(\lambda)/(g_1g_2)$, and using the dynamical generators \eqref{eq:dTL-local}, this immediately gives
\begin{align}
\label{eq:XY_dTL_relations_K2}
\tilde e_i(\lambda)\,\tilde e_i(\lambda)&=\beta\big(\lambda+h^{(i-1)}\big)\,\tilde e_i(\lambda)\,,\nonumber\\
\tilde e_i(\lambda)\,\tilde e_{i\pm1}(\lambda)\,\tilde e_i(\lambda)&=\tilde e_i(\lambda)\,,\\
\tilde e_i(\lambda)\,\tilde e_j(\lambda)&=\tilde e_j(\lambda)\,\tilde e_i(\lambda)\,,\qquad |i-j|>1\,,\nonumber
\end{align}
i.e.\ precisely Ren's dynamical TL relations \eqref{eq:dTL-relations} with $\beta(a)=-2g_a^2/(g_1g_2)$, i.e., using $\kappa=g_2/g_1$,
\be\label{eq:betaXY-K2}
\beta(1)=-\frac{2}{\kappa}\,,\qquad \beta(2)=-2\kappa\,.
\ee
Away from the orbifold point $\kappa \neq 1$ and the two colours have different loop weights, so Ren's Baxterisation does not directly apply. The colour dependence is mild, though. Each generator is just $g_a^2$ times the XXX Temperley-Lieb generator, and, as discussed in Section \ref{ReviewXY}, the Hamiltonian is that of an alternating (or `dimerised') Heisenberg chain. 

\paragraph{XY sector, general $K$.} We now consider general $K>2$. Here $X$ raises the colour and $Y$ lowers it, so the colour seen by a generator depends on the state, and not just on whether $i$ is even or odd. In the dynamical notation \eqref{eq:dTL-local}, the relations \eqref{eq:ZK-XY-rels} read
\begin{align}\label{eq:XY-ZK-dyn}
e_i(\lambda)\,e_i(\lambda)&=-2g^2_{\lambda+h^{(i-1)}}\,e_i(\lambda)\,,\nonumber\\
e_i(\lambda)\,e_{i+1}(\lambda)\,e_i(\lambda)&=g^2_{\lambda+h^{(i-1)}}\,g^2_{\lambda+h^{(i+2)}}\,e_i(\lambda)\,,\nonumber\\
e_{i+1}(\lambda)\,e_i(\lambda)\,e_{i+1}(\lambda)&=g^2_{\lambda+h^{(i-1)}}\,g^2_{\lambda+h^{(i+2)}}\,e_{i+1}(\lambda)\,,\\
e_i(\lambda)\,e_j(\lambda)&=e_j(\lambda)\,e_i(\lambda)\,,\qquad |i-j|>1\,.\nonumber
\end{align}
We see that the cubic relation now also becomes dynamical, so the relations above can be understood as a generalisation of those of Ren \eqref{eq:dTL-relations}. The new colour dependence is set by the colours $\lambda+h^{(i-1)}$ and $\lambda+h^{(i+2)}$ at the far left and the far right of the three sites. For $K=2$ these are always $1$ and $2$, and the cubic coefficient reduces to the constant $g_1^2g_2^2$ of \eqref{XYcubic}.

\paragraph{Dynamical rescaling.} The Temperley-Lieb generators entering the Hamiltonian satisfy the relations \eqref{eq:XY-ZK-dyn}. From a purely algebraic perspective we can actually encode the relations \eqref{eq:XY-ZK-dyn} in terms of simpler objects, related to $e_i(\lambda)$ by what we call a \textit{dynamical rescaling}, in which each generator is divided by the coupling of the colour to its left,
\begin{equation}\label{eq:dyn-rescaling}
  \hat{e}_i(\lambda) = \frac{e_i(\lambda)}{g^2_{\lambda+h^{(i-1)}}}\,.
\end{equation} 
In terms of these objects the relations simplify quite substantially, reducing to
\begin{align}\label{eq:XY-rescaled-rels}
\hat e_i(\lambda)\,\hat e_i(\lambda)&=-2\,\hat e_i(\lambda)\,,\nonumber\\
\hat e_i(\lambda)\,\hat e_{i\pm1}(\lambda)\,\hat e_i(\lambda)&=\hat e_i(\lambda)\,,\\
\hat e_i(\lambda)\,\hat e_j(\lambda)&=\hat e_j(\lambda)\,\hat e_i(\lambda)\,,\qquad |i-j|>1\,.\nonumber
\end{align}
To see this for $\hat e_i\hat e_{i+1}\hat e_i$, note that $e_{i+1}$ is only non-zero on $XY$ and $YX$, so in this product the colour to the left of site $i+1$ equals the colour $\lambda+h^{(i+2)}$ to the right of site $i+2$. The rescaling then removes $g^4_{\lambda+h^{(i-1)}}g^2_{\lambda+h^{(i+2)}}$, leaving $g^{-2}_{\lambda+h^{(i-1)}}e_i(\lambda)=\hat e_i(\lambda)$. The same argument applies to $\hat e_{i+1}\hat e_i\hat e_{i+1}$. The relations \eqref{eq:XY-rescaled-rels}
are just the XXX TL relations, or equivalently XXZ at $q=1$. 

That we obtain the XXX relations is not at all surprising, since the colour dependence of the generators only enters through an overall scale,
\begin{equation}
  e_i^{(a)} = -g_a^2\begin{pmatrix}0&0&0&0\\0&1&-1&0\\0&-1&1&0\\0&0&0&0\end{pmatrix} = g_a^2\, e_i^{\rm XXX}\quad\text{in the basis}\quad\begin{pmatrix}XX\\XY\\YX\\YY\end{pmatrix}\,,
\end{equation}
where $e_i^{\rm XXX}$ is the XXZ generator \eqref{eq:TL_rep} at $q=1$.
The dynamical rescaling simplifies the algebraic relations, but one should be very careful as this rescaling is local on each $e_i$ individually and depends on which fields are sitting on the first $i-1$ sites of the chain. As a result, the effect of this rescaling on the spectrum of the Hamiltonian is extremely non-trivial: indeed $-\sum_{i=1}^{L-1}\hat{e}_i(\lambda)$ is simply the XXX Hamiltonian which has a very different spectrum compared to the XY Hamiltonian. Despite this, the dynamical rescaling is still useful from an organisational point of view, since the algebraic relations are simpler. We will see this further below.

\paragraph{XZ sector.} 
We now turn to the XZ sector and consider general $K$ from the start. As we saw in Section~\ref{sec:Zk}, on sites $i$ and $i+1$ the generator acts as
\be\label{eq:XZ-generator}
e_i^{(a)}=-\begin{pmatrix}0&0&0&0\\0&g_{a+1}^2&-g_ag_{a+1}&0\\0&-g_ag_{a+1}&g_a^2&0\\0&0&0&0\end{pmatrix}\quad\text{in the basis}\quad\begin{pmatrix}XX\\XZ\\ZX\\ZZ\end{pmatrix}\,,
\ee
where the colour indices are fixed by $a$, e.g.\ $XZ$ stands for $X_{a,a+1}Z_{a+1}$ and $ZX$ for $Z_aX_{a,a+1}$. In particular, it annihilates two neighbouring $X$ fields and two neighbouring $Z$ fields.
If we identify $X$ with $\uparrow$ and $Z$ with $\downarrow$, this is $g_ag_{a+1}$ times the XXZ generator \eqref{eq:TL_rep} with
\be\label{eq:qj}
q_a=\frac{g_{a+1}}{g_a}\,.
\ee
So locally every generator is an XXZ generator, but its associated $q$ is set by the colour on its left and changes along the chain:
\be\label{eq:etilde}
e_i^{(a)}=g_ag_{a+1}\,e^{\rm XXZ}_i\vert_{q=q_a}\,.
\ee
For $K=2$ we have $q_1=\kappa$ and $q_2=1/\kappa$, in agreement with \eqref{eq:HXZ}.

In terms of the dynamical generators the relations read
\begin{align}\label{eq:XZ-ZK-dyn}
e_i(\lambda)\,e_i(\lambda)&=-\big(g^2_{\lambda+h^{(i-1)}}+g^2_{\lambda+h^{(i-1)}+1}\big)\,e_i(\lambda)\,,\nonumber\\
e_i(\lambda)\,e_{i+1}(\lambda)\,e_i(\lambda)&=g^2_{\lambda+h^{(i-1)}+1}\,g^2_{\lambda+h^{(i+2)}-1}\,e_i(\lambda)\,,\nonumber\\
e_{i+1}(\lambda)\,e_i(\lambda)\,e_{i+1}(\lambda)&=g^2_{\lambda+h^{(i-1)}+1}\,g^2_{\lambda+h^{(i+2)}-1}\,e_{i+1}(\lambda)\,,\\
e_i(\lambda)\,e_j(\lambda)&=e_j(\lambda)\,e_i(\lambda)\,,\qquad |i-j|>1\,.\nonumber
\end{align}
As in the XY sector \eqref{eq:XY-ZK-dyn}, the cubic coefficient depends on the colours $\lambda+h^{(i-1)}$ and $\lambda+h^{(i+2)}$ to the left and to the right of the three sites. Both are well defined in these products, since neither generator changes the fields to the left of site $i$, and both conserve the number of $X$ fields on the three sites. For $K=2$ we recover \eqref{XZTLrelations}: with one $X$ field on the three sites the two colours differ and the coefficient is $g_1^2g_2^2$, while with two $X$ fields they coincide and the coefficient is $g^4_{\lambda+h^{(i-1)}+1}$.

The cubic relation involves two neighbouring generators, and hence two values of $q$. As in the XY sector, it is convenient to perform a dynamical rescaling
\begin{equation}
  \hat e_i(\lambda)=e_i(\lambda)/g^2_{\lambda+h^{(i-1)}}\,.
\end{equation}
The resulting generators satisfy
\begin{align}\label{eq:gen-dTL}
\hat{e}_i(\lambda)\,\hat{e}_i(\lambda)&=\beta\big(\lambda+h^{(i-1)}\big)\,\hat{e}_i(\lambda)\,,\nonumber\\
\hat{e}_i(\lambda)\,\hat{e}_{i+1}(\lambda)\,\hat{e}_i(\lambda)&=\chi\big(\lambda+h^{(i-1)}\big)\,\hat{e}_i(\lambda)\,,\nonumber\\
\hat{e}_{i+1}(\lambda)\,\hat{e}_i(\lambda)\,\hat{e}_{i+1}(\lambda)&=\chi\big(\lambda+h^{(i-1)}\big)\,\hat{e}_{i+1}(\lambda)\,,\\
\hat{e}_i(\lambda)\,\hat{e}_j(\lambda)&=\hat{e}_j(\lambda)\,\hat{e}_i(\lambda)\,,\qquad |i-j|>1\,,\nonumber
\end{align}
with loop fugacity and cubic coefficient
\be\label{eq:XZ-beta-chi}
\beta(a)=-\big(1+q_a^2\big)\,,\qquad \chi(a)=q_a^2\,.
\ee
All the coefficients are now set by the colour $\lambda+h^{(i-1)}$ at the far left of the sites involved. Now both the loop weight and the cubic coefficient depend on the colour, so the resulting algebra nicely generalises Ren's algebra \eqref{eq:dTL-relations}. Note that the rescaled cubic coefficient depends only on the colour to the left of the three sites, while in \eqref{eq:XZ-ZK-dyn} it also depends on the colour to their right. The rescaling cannot be undone term by term: in $\hat e_i\hat e_{i+1}\hat e_i$ the colour seen by $e_{i+1}$ depends on the field at site $i$, which $e_i$ has turned into a superposition of $X$ and $Z$. The rescaling factor of the middle generator is therefore an operator rather than a number, and it is this factor which hides the dependence on the colour to the right. 

For $K=2$ it is more natural to divide all generators by the same constant $g_1g_2$. Both colours then have the same loop weight, $-(\kappa+1/\kappa)$, while the cubic coefficient is $1$ or $\kappa^{\pm2}$, as found in Section~\ref{sec:XZsector}. So for $K=2$ only the cubic relation is dynamical, which is the opposite of the XY sector. At the orbifold point all the $q_a$ are $1$, and we are back to the standard Temperley-Lieb algebra with $\beta=-2$.

\begin{figure}[t]
\centering
\begin{tikzpicture}[
    every node/.style={align=center, font=\small},
    arr/.style={-{Stealth[length=2.5mm]}, thick, shorten <=3pt, shorten >=3pt},
    lab/.style={font=\footnotesize\itshape, align=center},
  ]
  \node (ren) at (0,0) {Ren's dynamical TL \cite{Ren:2023mtn}\\ loop weight: dynamical\\ cubic coefficient: $1$};
  \node (xy2) at (-3.8,-2.8) {XY sector, $K=2$\\ loop weight: dynamical\\ cubic coefficient: $1$};
  \node (xz2) at (3.8,-2.8) {XZ sector, $K=2$\\ loop weight: constant\\ cubic coefficient: dynamical};
  \node (xyK) at (-3.8,-5.6) {XY sector, general $K$\\ loop weight: dynamical\\ cubic coefficient: dynamical};
  \node (xzK) at (3.8,-5.6) {XZ sector, general $K$\\ loop weight: dynamical\\ cubic coefficient: dynamical};
  \node (xyR) at (-3.8,-8.2) {$\mathrm{TL}$ with $q=1$ (XXX)};
  \node (xzR) at (3.8,-8.2) {$\mathrm{TL}(q_1)\oplus\dots\oplus\mathrm{TL}(q_K)$\\ or $\mathrm{TL}$ with $q=1$};
  \draw[arr] (ren) -- (xy2) node[lab, midway, above left=0pt] {same algebra};
  \draw[arr] (ren) -- (xz2) node[lab, midway, above right=0pt] {roles of loop weight\\ and cubic swapped};
  \draw[arr] (xy2) -- (xyK) node[lab, midway, left=2pt] {$\Zset_K$};
  \draw[arr] (xz2) -- (xzK) node[lab, midway, right=2pt] {$\Zset_K$};
  \draw[arr] (xyK) -- (xyR) node[lab, midway, left=2pt] {dynamical\\ rescaling};
  \draw[arr] (xzK) -- (xzR) node[lab, midway, right=2pt] {dynamical\\ twist};
\end{tikzpicture}
\caption{The algebras of the XY and XZ sectors of the $\Zset_K$ quiver theories, and how they relate to Ren's dynamical Temperley-Lieb algebra \cite{Ren:2023mtn}. Each node lists whether the loop weight and the cubic coefficient depend on the colour. The bottom row shows the ordinary Temperley-Lieb algebras obtained by the dynamical rescaling (Section~\ref{subsec:quiver-to-Ren}) and the dynamical twist. The details of the twist are explained below, in Section~\ref{subsec:dTL-from-XXZ}.}
\label{fig:dTL-summary}
\end{figure}
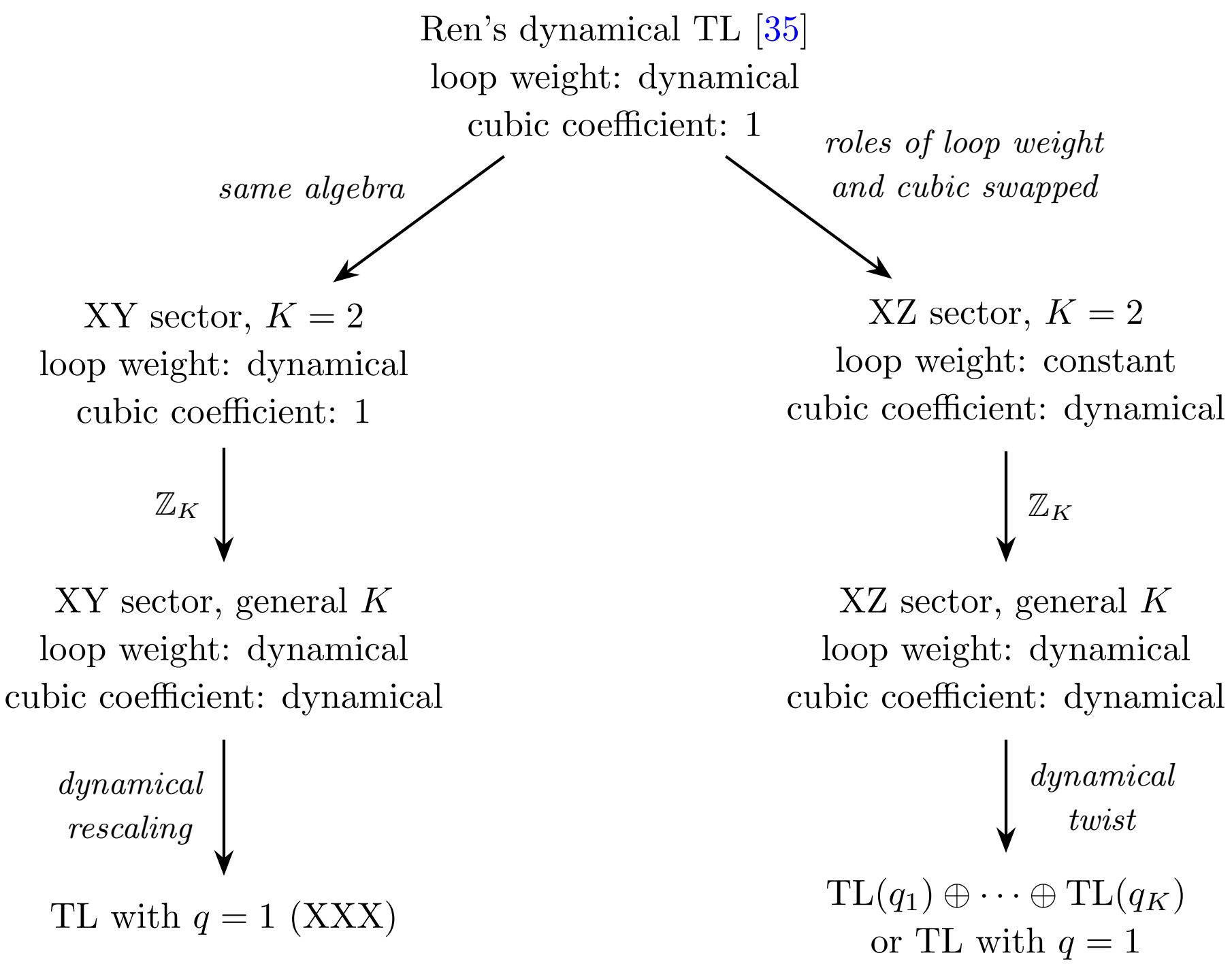

\paragraph{Summary.} The relation between the algebras of the quiver sectors and Ren's dynamical Temperley-Lieb algebra is summarised in Figure~\ref{fig:dTL-summary}. For $K=2$ the XY sector gives exactly Ren's algebra, while the XZ sector has the opposite structure, with a constant loop weight and a dynamical cubic coefficient. For general $K$ both sectors have dynamical loop weights and dynamical cubic coefficients, and so generalise Ren's algebra. In the XY sector the dynamical rescaling reduces the relations to those of the XXX chain, while in the XZ sector we need the dynamical twist of the next subsection.

\subsection{Dynamical Twisting}\label{subsec:dTL-from-XXZ}

By \eqref{eq:etilde}, each XZ TL generator $e_i^{(a)}$ is an XXZ generator with its own $q_a$. We will show that the dynamical TL algebra can then be viewed as a direct sum of different XXZ TL algebras 
\begin{equation}\label{eqn:TLdirect}
\mathrm{TL}(q_1)\oplus \dots \oplus\mathrm{TL}(q_K)\,.
\end{equation}
The XZ Hamiltonian and the dynamical generators $\hat{e}_i(\lambda)$ however are very complicated objects from this point of view as a result of the dynamical dependence in the generators. It turns out though that we can actually make the direct sum structure \eqref{eqn:TLdirect} manifest via a \textit{dynamical twist}.

\paragraph{Twist.} The idea behind the twisting we use is to remove the dynamical dependence of the generators $\hat{e}_i(\lambda)$ such that the resulting generators become block diagonal. Schematically, after twisting we will obtain 
\begin{equation}
  \hat{e}_i(\lambda) \rightarrow {\rm diag}\big(E_i(q_1),\dots,E_i(q_K)\big)\,.
\end{equation}
More precisely, we define new generators $\mathsf{e}_i(\lambda)$ by 
\begin{equation}
  \mathsf{e}_i(\lambda) = \Phi_{i}(\lambda)\mathcal{D}_{i}(\lambda)\hat{e}_i(\lambda)\Phi_{i}(\lambda)^{-1}\,.
\end{equation}
The precise form of the operators $\Phi_i$ and $\mathcal{D}_i$ will be spelled out below. We emphasise that each generator $\hat{e}_i$ is twisted differently, so this is neither a change of basis on the whole Hilbert space nor an algebra homomorphism. As with the dynamical rescaling of the previous subsection, the twist therefore has a highly non-trivial effect on the spectrum of the Hamiltonian. 

\paragraph{Structure of the twist.} 

We now spell out the form of the twist. Locally, each dynamical TL generator $e_i^{(a)}$ can be viewed as an XXZ TL generator with a different $q$-value $q_a$, depending on which fields we crossed on the previous sites. The idea of the twist is to map one such XXZ TL generator to another. First, we argue that such a twist always exists.

We start by comparing two XXZ generators \eqref{eq:TL_rep} with different values of $q$,
\be\label{eq:E-q-qprime}
E(q)=-\begin{pmatrix}0&0&0&0\\0&q&-1&0\\0&-1&q^{-1}&0\\0&0&0&0\end{pmatrix},\qquad
E(q')=-\begin{pmatrix}0&0&0&0\\0&q'&-1&0\\0&-1&q'^{-1}&0\\0&0&0&0\end{pmatrix}.
\ee
For $q+q^{-1}\neq0$ and $q'+q'^{-1}\neq0$, both matrices are diagonalisable and their eigenvalues are
\be
E(q):\ \big\{0,0,0,-(q+q^{-1})\big\}\,,\qquad E(q'):\ \big\{0,0,0,-(q'+q'^{-1})\big\}\,.
\ee
Rescaling $E(q)$ by the diagonal matrix
\be\label{eq:D-qqprime}
\mathcal{D}=\begin{pmatrix}1&0&0&0\\0&1&0&0\\0&0&d&0\\0&0&0&1\end{pmatrix},\qquad d=q\big(q'+q'^{-1}-q\big)\,,
\ee
gives $\mathcal{D}\,E(q)$, with eigenvalues $\{0,0,0,-(q'+q'^{-1})\}$, the same as those of $E(q')$.\footnote{This requires $d\neq0$, i.e.\ $q\neq q'+q'^{-1}$. In that degenerate case one can instead rescale by a multiple of the identity and use a different similarity transformation; we will not need this.} Since two diagonalisable matrices with the same eigenvalues are related by a similarity transformation there is an invertible $\Phi$ with
\be\label{eq:twist-qqprime}
\Phi\,\mathcal{D}\,E(q)\,\Phi^{-1}=E(q')\,.
\ee
So any two-site XXZ generator can be twisted into any other. Explicitly,
\begin{equation}\label{eq:F-qqprime}
  \Phi=\begin{pmatrix}1&0&0&0\\0&1&0&0\\0&q'-q&1&0\\0&0&0&1\end{pmatrix}=1+(q'-q)\,\sigma^-_i\sigma^+_{i+1}=\exp\Big[(q'-q)\,\sigma^-_i\sigma^+_{i+1}\Big]\,,
\end{equation}
where we identify $X=\,\uparrow$ and $Z=\,\downarrow$. It sends $XZ\to XZ+(q'-q)\,ZX$ and leaves the other three states alone. This factorises into a part depending only on $q'$ and a part depending only on $q$,
\begin{equation}
  \Phi=\mathcal{F}(q)^{-1}\,\mathcal{F}(q')\,,\qquad \mathcal{F}(q)=\exp\Big[q\,\sigma^-_i\sigma^+_{i+1}\Big]\,.
\end{equation}

\paragraph{Application to the XZ chain.} Let us now apply this to the dynamical XZ chain. After the dynamical rescaling, each generator is just an XXZ generator, up to normalisation,
\be\label{eq:ehat-XXZ}
\hat e_i(\lambda)=q_{\lambda+h^{(i-1)}}\,E_i\big(q_{\lambda+h^{(i-1)}}\big)\,,
\ee
where $E_i$ is the matrix \eqref{eq:E-q-qprime} acting on sites $i$ and $i+1$. The first generator always sees the leftmost colour, so its value of $q$ is simply $q_\lambda$. For the second generator it depends on what the first field is. If the state starts with a $Z$, the colour does not change and the second generator has the same $q$ as the first. If instead it starts with an $X$, the colour goes up by one and the second generator has a shifted value of $q$. Further along the chain the shift is set by how many $X$ fields we have crossed.

The idea is to undo this shift, at the cost of a twist. For each generator we use \eqref{eq:twist-qqprime}, going from its own $q$ to the leftmost one, $q_\lambda$. This needs
\begin{align}\label{eq:Phi-D-XZ}
\Phi_i(\lambda)&=\exp\Big[\big(q_\lambda-q_{\lambda+h^{(i-1)}}\big)\,\sigma^-_i\sigma^+_{i+1}\Big]=\mathcal{F}_i\big(q_{\lambda+h^{(i-1)}}\big)^{-1}\,\mathcal{F}_i(q_\lambda)\,,\nonumber\\[4pt]
\mathcal{D}_i(\lambda)&=\frac{1}{q_{\lambda+h^{(i-1)}}}\begin{pmatrix}1&0&0&0\\0&1&0&0\\0&0&q_{\lambda+h^{(i-1)}}\big(q_\lambda+q_\lambda^{-1}-q_{\lambda+h^{(i-1)}}\big)&0\\0&0&0&1\end{pmatrix},
\end{align}
where the prefactor in $\mathcal{D}_i$ takes care of the normalisation in \eqref{eq:ehat-XXZ}. Both only act on sites $i$ and $i+1$, so there is no ordering ambiguity. The twisted generators are then
\be\label{eq:twisted-XZ}
\mathsf{e}_i(\lambda)=\Phi_i(\lambda)\,\mathcal{D}_i(\lambda)\,\hat e_i(\lambda)\,\Phi_i(\lambda)^{-1}=E_i(q_\lambda)\,.
\ee
All the dependence on the fields is gone, and only the leftmost colour is left. Whenever the colour to the left of a generator is the leftmost one, for example for the first generator, the twist is trivial. Since none of the generators change the leftmost colour, the Hilbert space splits into $K$ sectors, one for each leftmost colour, and in each of them we simply have the XXZ generators with the corresponding $q$. This makes the direct sum structure manifest,
\be
\mathsf{e}_i={\rm diag}\big(E_i(q_1),\dots,E_i(q_K)\big)\,,
\ee
so the twisted generators satisfy the relations of $\mathrm{TL}(q_1)\oplus\dots\oplus\mathrm{TL}(q_K)$.

\paragraph{Quasi-Hopf structure.} The form of the twist $\Phi_i(\lambda)$ in \eqref{eq:Phi-D-XZ} is reminiscent of the dynamical twists of quantum groups \cite{Babelon:1995rz,Arnaudon:1997hm}. There, a dynamical twist $\mathcal{F}(\lambda)$ of $U_q(\mathfrak{sl}(2))$ satisfies the shifted cocycle condition. In the literature the dynamical variable is usually the rightmost colour, while for us it is the leftmost one, so we relabel the sites to match our conventions. The shifted cocycle condition then reads
\be
\big[(\mathrm{id}\otimes\Delta)\mathcal{F}(\lambda)\big]\,\mathcal{F}_{23}(\lambda+h_1)=\big[(\Delta\otimes\mathrm{id})\mathcal{F}(\lambda)\big]\,\mathcal{F}_{12}(\lambda)\,,
\ee
where $h_1$ is the weight of the first site. It turns $U_q(\mathfrak{sl}(2))$ into a quasi-Hopf algebra \cite{Drinfeld:1989st}, in which associativity only holds up to an operator called the coassociator \cite{Babelon:1995rz},
\be
\Phi_{123}(\lambda)=\mathcal{F}_{23}(\lambda+h_1)^{-1}\,\mathcal{F}_{23}(\lambda)\,.
\ee
Notice that in this coassociator the dependence on the field on site one enters only through the dynamical shift $h_1$.
Compare this with our twist on three sites,
\be
\Phi_2(\lambda)=\mathcal{F}_{2}(q_{\lambda+h_1})^{-1}\,\mathcal{F}_{2}(q_\lambda)\,,
\ee
which also acts on sites $2$ and $3$. The two have exactly the same form. Note, however, that we use the twist in the opposite direction to the usual one. Usually one starts from $U_q(\mathfrak{sl}(2))$ acting on a chain of identical spins, whose commutant is a single XXZ Temperley-Lieb algebra, and twists it into a dynamical object. Here we start from the dynamical chain and twist the dynamics away, ending up with a direct sum of ordinary XXZ Temperley-Lieb algebras, one copy of the standard $U_q(\mathfrak{sl}(2))$ picture for each leftmost colour. As discussed, this has a very non-trivial effect on the spectrum of the Hamiltonian. The deeper implication of this quasi-Hopf structure in our set-up is currently under investigation and we will report on it in the near future. 

\paragraph{Twisting from the XXX spin chain.} We close this section by noting that, in addition to relating the dynamical generators to XXZ ones by a twist, we can also relate them to the XXX spin chain, i.e.\ to the $\mathrm{TL}(q=1)$ algebra of $\Ncal=4$ SYM from Section~\ref{sec:N=4SYM}. Using \eqref{eq:twist-qqprime} with $q'=1$ on each generator, we need
\begin{align}\label{eq:Phi-D-XXX}
\Phi_i(\lambda)&=\exp\Big[\big(1-q_{\lambda+h^{(i-1)}}\big)\,\sigma^-_i\sigma^+_{i+1}\Big]\,,\nonumber\\[4pt]
\mathcal{D}_i(\lambda)&=\frac{1}{q_{\lambda+h^{(i-1)}}}\begin{pmatrix}1&0&0&0\\0&1&0&0\\0&0&q_{\lambda+h^{(i-1)}}\big(2-q_{\lambda+h^{(i-1)}}\big)&0\\0&0&0&1\end{pmatrix},
\end{align}
and the twisted generators are simply the XXX generators,\footnote{As explained below \eqref{eq:D-qqprime}, this form of the twist requires $q_{\lambda+h^{(i-1)}}\neq2$.}
\be\label{eq:twisted-XXX}
\mathsf{e}_i(\lambda)=\Phi_i(\lambda)\,\mathcal{D}_i(\lambda)\,\hat e_i(\lambda)\,\Phi_i(\lambda)^{-1}=E_i(1)\,.
\ee
Now the target does not depend on the leftmost colour either, so all the colour dependence is gone and the twisted generators satisfy the Temperley-Lieb relations at $q=1$. Conversely, every dynamical generator is a twisted XXX generator. As before, each generator is twisted differently, so the map $\hat e_i\mapsto\mathsf{e}_i$ is not an algebra homomorphism.

To summarise, the XZ Hamiltonian is a sum of twisted XXZ (or XXX) Temperley-Lieb generators, each twisted by its own two-site twist. The twists cannot be combined into a single twist of the whole chain, which is why the spectrum of the XZ Hamiltonian differs so much from that of the XXZ and XXX chains.

\section{Loop model representation}\label{sec:loop}
Many versions of the Temperley-Lieb algebra, both dense and dilute, admit an alternative representation in terms of a loop model~\cite{Nienhuis:1982fx,Bloete:1989he}.
These are face-type statistical models on a square lattice, with the Boltzmann weights expressed in terms of tiles with lines connecting the edges. The partition function of the model is computed by considering all configurations subject to matching of the lines on adjacent tiles. In the loop picture, the standard Temperley-Lieb algebra elements can be represented by a single tile: 
\be
U_i=
\begin{tikzpicture}[rotate=-45,baseline=-0.1cm]\fill[\lcyan] (0,1)--(1,1)--(1,0)--(0,0)--(0,1); \draw[-,black!70!green] (0,1)--(1,1)--(1,0)--(0,0)--(0,1);\draw[-,black!70!green] (0,1)--(1,1)--(1,0)--(0,0)--(0,1);\draw[-] (1,0.1) arc (90:180:0.1); \draw[-,line width=0.5mm,black!60!red] (0,0.5) arc (-90:0:0.5); \draw[-,line width=0.5mm,black!60!red] (0.5,0) arc (180:90:0.5);
\end{tikzpicture}\;.
    \ee
This satisfies the usual Temperley-Lieb relations, if one takes closed loops to have fugacity $\beta$. Together with the identity element, the tiles of the loop model take the Baxterised form
\be \label{LoopBaxterised}
\begin{tikzpicture}[rotate=45,baseline=0.6cm]\fill[\lcyan] (0,1)--(1,1)--(1,0)--(0,0)--(0,1);\draw[-,black!70!green] (0,1)--(1,1)--(1,0)--(0,0)--(0,1);\draw[-] (0,0.1) arc (90:0:0.1);\node at (0.5,0.5){$u$}; \end{tikzpicture}=~\begin{tikzpicture}[rotate=45,baseline=0.6cm]\fill[\lcyan] (0,1)--(1,1)--(1,0)--(0,0)--(0,1);\fill[\lcyan] (0,0.5) arc (-90:0:0.5);\fill[\lcyan] (0,0.5)--(0.5,1)--(0,1)--(0,0.5);\fill[\lcyan] (0.5,0) arc (180:90:0.5);\fill[\lcyan](0.5,0)--(1,0)--(1,0.5)--(0.5,0); \draw[-,black!70!green] (0,1)--(1,1)--(1,0)--(0,0)--(0,1);\draw[-,black!70!green] (0,1)--(1,1)--(1,0)--(0,0)--(0,1);\draw[-] (0,0.1) arc (90:0:0.1); \draw[-,line width=0.5mm,black!60!red] (0,0.5) arc (-90:0:0.5); \draw[-,line width=0.5mm,black!60!red] (0.5,0) arc (180:90:0.5);\end{tikzpicture}
+\frac{\sin u}{\sin(u-\eta)}~
\begin{tikzpicture}[rotate=-45,baseline=-0.1cm]\fill[\lcyan] (0,1)--(1,1)--(1,0)--(0,0)--(0,1); \draw[-,black!70!green] (0,1)--(1,1)--(1,0)--(0,0)--(0,1);\draw[-,black!70!green] (0,1)--(1,1)--(1,0)--(0,0)--(0,1);\draw[-] (1,0.1) arc (90:180:0.1); \draw[-,line width=0.5mm,black!60!red] (0,0.5) arc (-90:0:0.5); \draw[-,line width=0.5mm,black!60!red] (0.5,0) arc (180:90:0.5);
\end{tikzpicture}\;,
\ee
where $\beta=-2\cos\eta$,
which satisfies the Yang-Baxter Equation \cite{Owczarek:1987ue}. 
Loop models play an important role in the field of 2d integrable statistical systems, with a major application being the study of the continuum limit and approach to criticality, see e.g. \cite{Fendley:2006ys} for a discussion and the book \cite{guttmanPolygonsPolyominoesPolycubes2009} and review \cite{zinn-justinSixVertexLoopTiling2009} for an overview. Just like Temperley-Lieb algebras, there are many generalisations, for instance to multi-colour  
\cite{Warnaar:1993ka,Grimm:2003am}, as well as dilute models \cite{Warnaar:1992gj,Roche:1992ww,Warnaar:1993zn}. Loop models can also naturally describe generalisations of the TL algebra such as the Birman-Wenzl-Murakami (BWM) algebra and its dilute versions \cite{Grimm:1994kc,Warnaar:1995law}. 

It is straightforward to write a loop model representation of our dynamical Temperley-Lieb algebras. The dynamical nature of the algebra corresponds to different colourings of regions in the tiles (which also need to match for the joining of two tiles to be allowed), while the dilute aspect means that not all edges need to have lines ending on them. 

In this section we will briefly define the loop models for the two sectors (XY and XZ) of the $\Zset_2$ quiver discussed in Sections~\ref{sec:XYsector} and~\ref{sec:XZsector}, and show how one recovers the Temperley-Lieb relations in the loop formalism.

The tiles corresponding to the XY, XZ and YZ $\SU(2)$ sectors can straightforwardly be combined into a dynamical dilute loop model for the  $\SU(3)$-sector. One can also extend the loop model description to the $\Zset_K$ case. Here we just focus on $\Zset_2$.

To construct the $\Zset_2$ loop representation, first of all we need to identify the fields that the tiles act on. Using the same colour scheme as in the previous sections (blue for gauge node 1 and pink for gauge node 2), we have the dictionary:
\be
X_{12}=\begin{tikzpicture}[baseline=-0.1cm]
  \fill[\lcyan](-0.2,-0.7)--(0,-0.7)--(0,0.7)--(-0.2,0.7)--(-0.2,-0.7);
  \fill[pink](0.2,-0.7)--(0,-0.7)--(0,0.7)--(0.2,0.7)--(0.2,-0.7);
  \draw[-,line width=0.5mm,black!60!red] (0,0.7)--(0,-0.7);
  \draw[-,line width=0.5mm,black!60!red] (-0.15,-0.1)--(0,0.05)--(0.15,-0.1);\end{tikzpicture}\;,
Y_{12}=\begin{tikzpicture}[baseline=-0.1cm]
  \fill[\lcyan](-0.2,-0.7)--(0,-0.7)--(0,0.7)--(-0.2,0.7)--(-0.2,-0.7);
  \fill[pink](0.2,-0.7)--(0,-0.7)--(0,0.7)--(0.2,0.7)--(0.2,-0.7);
  \draw[-,line width=0.5mm,black!60!red] (0,0.7)--(0,-0.7);
  \draw[-,line width=0.5mm,black!60!red] (-0.15,0.1)--(0,-0.05)--(0.15,0.1);\end{tikzpicture}\;,
Z_{1}=\begin{tikzpicture}[baseline=-0.1cm]
  \fill[\lcyan](-0.2,-0.7)--(0,-0.7)--(0,0.7)--(-0.2,0.7)--(-0.2,-0.7);
  \fill[\lcyan](0.2,-0.7)--(0,-0.7)--(0,0.7)--(0.2,0.7)--(0.2,-0.7);
  \draw[dashed,line width=0.5mm,black!60!red] (0,0.7)--(0,-0.7);
\end{tikzpicture}\; ,
X_{21}=\begin{tikzpicture}[baseline=-0.1cm]
  \fill[pink](-0.2,-0.7)--(0,-0.7)--(0,0.7)--(-0.2,0.7)--(-0.2,-0.7);
  \fill[\lcyan](0.2,-0.7)--(0,-0.7)--(0,0.7)--(0.2,0.7)--(0.2,-0.7);
  \draw[-,line width=0.5mm,black!60!red] (0,0.7)--(0,-0.7);
  \draw[-,line width=0.5mm,black!60!red] (-0.15,-0.1)--(0,0.05)--(0.15,-0.1);\end{tikzpicture}\;,
Y_{21}=\begin{tikzpicture}[baseline=-0.1cm]
  \fill[pink](-0.2,-0.7)--(0,-0.7)--(0,0.7)--(-0.2,0.7)--(-0.2,-0.7);
  \fill[\lcyan](0.2,-0.7)--(0,-0.7)--(0,0.7)--(0.2,0.7)--(0.2,-0.7);
  \draw[-,line width=0.5mm,black!60!red] (0,0.7)--(0,-0.7);
  \draw[-,line width=0.5mm,black!60!red] (-0.15,0.1)--(0,-0.05)--(0.15,0.1);\end{tikzpicture}\;,
Z_{2}=\begin{tikzpicture}[baseline=-0.1cm]
  \fill[pink](-0.2,-0.7)--(0,-0.7)--(0,0.7)--(-0.2,0.7)--(-0.2,-0.7);
  \fill[pink](0.2,-0.7)--(0,-0.7)--(0,0.7)--(0.2,0.7)--(0.2,-0.7);
  \draw[dashed,line width=0.5mm,black!60!red] (0,0.7)--(0,-0.7);
\end{tikzpicture}\;.
\ee
The tiles will then also carry lines with arrows and coloured regions according to the types of fields involved. 

\subsection{The XY sector}

This is a dense sector, meaning that all edges of each of the six tiles have lines going through them. There will be two $\Zset_2$-conjugate identity operators: 
\be\begin{split}
I^{(1)}&=\left(\TXXXX+\TXYXYid+\TYXYXid+\TYYYY\right)\\
I^{(2)}&=\left(\TpXXXX+\TpXYXYid+\TpYXYXid+\TpYYYY\right)
\end{split}\ee
The two $\Zset_2$-conjugate Temperley-Lieb elements are 
\begin{subequations}\label{loopXYTL}
\begin{align}
U^{(1)}&=\kappa^{-1}\left(-~\TXYXY- ~\TYXYX+~\TXYYX+~\TYXXY\right)\;\\
\intertext{and}
U^{(2)}&=\kappa\left(-~\TpXYXY- ~\TpYXYX+~\TpXYYX+~\TpYXXY\right)
\end{align}\end{subequations}
Similarly to the Temperley-Lieb elements, these tiles are thought of as acting on fields at the bottom and convert them to fields at the top, possibly permuting them in the process. For example, 
\be \label{loopexample}
\;,
\ee
where we also indicated the action of a cubic combination of tiles. 
We take the tiles to have a weight of 1, with their contribution to the overall weight of an expression coming from the coefficients in front. 

To evaluate the quadratic TL relation, we act consecutively by two copies of $U^{(a)}$ by placing them on top of each other. Actions with mismatched arrows or regions will then vanish. An example of two configurations that survive is
\be 
(-\kappa^{-1})^2~%
=
2\kappa^{-2}~\TXYXY
\ee
Here we have focused on the choice of $X_{12}Y_{21}$ as incoming and $X_{12}Y_{21}$ as outgoing fields, as in the first diagram in (\ref{loopexample}). We have indicated the coefficients of these tiles as given in (\ref{loopXYTL}). The tile on the right appears with coefficient $(-\kappa^{-1})$ in $U^{(1)}$ so that factor would need to be factored out to write the right-hand-side in terms of $U^{(1)}$. Doing so, and also checking all other terms,  we find the quadratic relations
\be \label{loopXYquadratic}
(U^{(1)})^2=-2\kappa^{-1} U^{(1)}\quad,\quad(U^{(2)})^2=-2\kappa U^{(2)}\
\ee
We see that  the loop fugacity arises as a combination of the (equal) contributions of the two closed loops.\footnote{In the directed loop description of the XXZ model, the clockwise and anticlockwise loops contribute inverse powers of $q$, leading to $\beta=-(q+q^{-1})$ as we saw in Section~\ref{sec:XXZ}.} 
 
The cubic relations are obtained similarly. To illustrate the procedure, let us just consider a couple of examples. Acting on $X_{12}Y_{21}X_{12}$, we find 
\be
(-\kappa^{-1})(-\kappa)(-\kappa^{-1})
\;,
\ee
where on the left we write the coefficients of the tiles from bottom to top. Exchanging the field on the third site with a $Y_{12}$ we obtain
\be
(\kappa^{-1})(-\kappa)(\kappa^{-1})
%
\ee
We obtain the same answer regardless of the third site, and in all cases we just obtain an identity strand at site 3. Noting that the tile on the right appears with coefficient $(-\kappa^{-1})$ in $U^{(1)}$, the above relations (as well as the ones we do not indicate) can be expressed as  
\be
U^{(1)}_1 U^{(2)}_2 U^{(1)}_1= U^{(1)}_1
\ee
We can also look at the other cubic relation
\be
(-\kappa)(-\kappa^{-1})(-\kappa)%
 \otimes ~\TpXYXY
\ee
which gives 
\be
U_2^{(2)} U_1^{(1)}U_2^{(2)}= U_2^{(2)}
\ee
We see that the region-coloured dense loop representation \eqref{loopXYTL} precisely reproduces the defining relations (\ref{eq:XYTLrescaled}) of the XY-sector Temperley-Lieb algebra. 

Apart from the colours, these tiles are the same as \eqref{LoopBaxterised}. We have just unpacked the different possibilities for incoming and outgoing fields to illustrate how the fugacities arise, and why the quadratic relations are dynamical. We could revert to a version without arrows where $I^{(1,2)}, U^{(1,2)}$ are composed of one tile each. What is different is that we do not claim Baxterisation as in \eqref{LoopBaxterised}. All we can state is that the $U^{(a)}$ would appear at order $u$ in the expansion of a potential dynamical $\check{R}(u,\lambda)$ matrix. However, as discussed in \cite{Pomoni:2021pbj}, the trigonometric formula \eqref{LoopBaxterised} does not lead to the YBE. 
We leave a more detailed study of Baxterisation for this loop model for the future. 

 Given the dense property of the XY sector, it is clear that the generators $U^{(1)}$ and $U^{(2)}$ will alternate at even and odd sites. Therefore, the same will be the case for the quadratic relations \eqref{loopXYquadratic}, i.e. one will alternately obtain $-2/\kappa$ and $-2\kappa$. This property is reminiscent of the two-colour Fuss-Catalan algebra (see e.g. \cite{DiFrancesco:1998hb,Babichenko:2007vz}), although the other relations of that algebra are quite different. 

\subsection{The XZ sector}

Now let us proceed to the XZ sector. As this sector is dilute, some of the tiles will have vacant edges. The identity elements are 
\be\begin{split}    
I^{(1)}=~\TZZZZ+\TXZXZ+\TZXZX+\TXXXX\;,\\
I^{(2)}=~\TpZZZZ+\TpXZXZ+\TpZXZX+\TpXXXX\;.
\end{split}
\ee
The Temperley-Lieb generators for first node being 1 or 2 are: 
\begin{subequations} \label{loopXZTL}
    \begin{align}
U^{(1)}=-\kappa ~\TXZXZ-\frac{1}{\kappa}~\TZXZX+\TXZZX+\TZXXZ\;,\\
\intertext{and}
U^{(2)}=-\frac{1}{\kappa} ~\TpXZXZ-\kappa~\TpZXZX+\TpXZZX+\TpZXXZ\;.
\end{align}
\end{subequations}
The tiles for the YZ sector can simply be found by inverting the arrows. Dilute loop models have a long history and important applications \cite{GrimmPearce93,BehrendPearce97,Vernier:2015ody}. Although there are obvious similarities to known models, our loop model appears to have novel features compared to those in the literature. One difference to many existing models is the absence of tiles of the form \begin{tikzpicture}[rotate=-45,baseline=-0.1cm,scale=0.4]\fill[\lcyan] (0,1)--(1,1)--(1,0)--(0,0)--(0,1); \draw[-,black!70!green] (0,1)--(1,1)--(1,0)--(0,0)--(0,1);\draw[-,black!70!green] (0,1)--(1,1)--(1,0)--(0,0)--(0,1);\draw[-] (1,0.1) arc (90:180:0.1);  \draw[-,line width=0.5mm,black!60!red] (0.5,0) arc (180:90:0.5);
\end{tikzpicture}
and
\begin{tikzpicture}[rotate=-45,baseline=-0.1cm,scale=0.4]\fill[\lcyan] (0,1)--(1,1)--(1,0)--(0,0)--(0,1); \draw[-,black!70!green] (0,1)--(1,1)--(1,0)--(0,0)--(0,1);\draw[-,black!70!green] (0,1)--(1,1)--(1,0)--(0,0)--(0,1);\draw[-] (1,0.1) arc (90:180:0.1); \draw[-,line width=0.5mm,black!60!red] (0,0.5) arc (-90:0:0.5); 
\end{tikzpicture}\;. These tiles would correspond to a $XX\ra ZZ$ or $ZZ\ra XX$ vertex, and these would 
violate the conserved $\Urm(1)$ symmetry of our theory. Extending to an $\SU(3)$ sector, a $XY\ra ZZ$ vertex or its converse would still be disallowed by $\Ncal=2$ supersymmetry, although they could be present in an $\Ncal=1$ setting. 

Perhaps the closest model to the dilute structure of our case is that of \cite{Pearce:2019tvb}, however that model is not dynamical. (And in that work the main interest is in the free-fermionic case $\beta=0$, which cannot be attained in our case where the gauge couplings are real.) 

We can again check the quadratic TL relations. As an example, we have
\be
(-\kappa)^2
=
(1+\kappa^{2})~\TXZXZ
\ee
As the tile on the right-hand-side appears in $U$ with a coefficient of $(-\kappa)$, this needs to be factored out and gives the quadratic relation
\be
(U_1^{(1)})^2=-(\kappa+\kappa^{-1}) U^{(1)}\;,
\ee
in accordance with the XZ-sector quadratic relations \eqref{XZTLrelations}. We can similarly confirm the $\Zset_2$ conjugate relations. 
Let us also give a couple of examples of the cubic relations. Acting on $X_{12}Z_2X_{21}$:
\be
(-\kappa)(-\kappa)(-\kappa)
%
\ee
which is consistent with the relation 
\be
U^{(1)}_1 U^{(2)}_2 U^{(1)}_1=\kappa^2 U^{(1)}\;.
\ee
However, acting on $X_{12}Z_2Z_2$, we find  
\be
(+1)(-\kappa)(-\kappa^{-1})~%
\ee
Since the tile on the right-hand-side appears with a coefficient of 1 in $U^{(1)}$, in this case we get the relation
\be
U_1^{(1)}U_2^{(1)}U_1^{(1)}=U_1^{(1)}
\ee
As an example of the $U_2U_1U_2$ cubic relation, consider acting on $Z_1X_{12}Z_2$:
\be
(-\kappa)(-\kappa^{-1})(-\kappa)
%
 \otimes ~\TXZXZ
\ee
which gives
\be
U^{(1)}_2 U^{(1)}_1 U^{(1)}_2=U^{(1)}_2
\ee
By considering other configurations, we can summarise the cubic relations as
\be
U_i^{(a)} U_{i+1}^{(b)}U_i^{(a)}=\left\{\begin{array}{ll}
\kappa^{2s(a)} U_i^{(a)}\quad&\text{if}\; a(i)=a(i+3)\\
U_i^{(a)} \quad&\text{if}\; a(i)\neq a(i+3)\end{array}\right.
\ee
and 
\be
U_i^{(b)} U_{i-1}^{(a)}U_i^{(b)}=\left\{\begin{array}{ll}
\kappa^{2s(a)} U_i^{(b)}\quad&\text{if}\; a(i-1)=a(i+2)\\
U_i^{(b)} \quad&\text{if}\; a(i-1)\neq a(i+2)\end{array}\right.
\ee
where $s(1)=1, s(2)=-1$, and $a(k)$ indicates the colour of the region to the left of site $k$. These are the same dynamical cubic relations 
as we found from the gauge theory language in Section~\ref{sec:XZsector}, where they were expressed equivalently in terms of the number of $X$ fields.  We therefore confirm that the region-coloured dilute loop model in \eqref{loopXZTL} correctly reproduces the XZ-sector quadratic and cubic relations.

It is important to mention that the non-dynamical cubic relations of the original Temperley-Lieb algebra guarantee that loops of any extent are all weighted by the same coefficient $-(q+q^{-1})$ as the fundamental loop. This scaling invariance plays an important role in obtaining a scale-invariant continuum theory at the critical point. As the cubic relation has become dynamical in the XZ sector, it is important to understand how this affects the existence or not of a critical point. 

\section{Conclusions}\label{sec:conclusions}

\paragraph{Summary.} In this work we showed how the $\Ncal=1$ superspace formulation of the
$\Ncal=2$ quiver theories naturally leads to dynamical generalisations of
the Temperley-Lieb (TL) algebra, in the sense that their defining relations contain `structure constants' which depend in a specific way on the sites on which the operators in these relations act. Just as in $\Ncal=4$ SYM, where they reproduce the standard algebra $\mathrm{TL}(q=1)$, the superspace Feynman diagrams contributing to the one-loop
dilatation operator can be interpreted as Temperley-Lieb diagrams, see
\eqref{eq:XY_TL_gens}, \eqref{eq:XZ_TL_gen1}, \eqref{eq:XY_TL_gens_Z3} and \eqref{eq:TL1-Z3-XZ}. The resulting algebras capture the dynamical nature of the corresponding spin chains through their dependence on the quiver nodes: each face of a diagram is
coloured according to the gauge group whose index line runs through it, and the relations
\eqref{XYquadratic} and \eqref{XZTLrelations} depend on these colours. 
The superspace origin of the algebras also provides a meaning to the lines -- or domain walls -- separating differently-coloured faces, as the bifundamental fields of the quiver theory: crossing a bifundamental line changes the
colour, while crossing an adjoint line does not, see \eqref{XZcubicFeynmanX} and \eqref{XZcubicFeynmanZ}. 

A particularly nice, although not surprising, feature of these algebras is that they can be interpreted as diagrammatic algebras, in which all products can be done by manipulating diagrams. Indeed, depicting their generators in terms of coloured diagrams, such as in \eqref{eq:XY_TL_gens} and \eqref{eq:XZ_TL_gen1}, and following the colour rules automatically takes care of the dynamics, even in cases where multiple colours are possible. 

For the simplest case, the $\Zset_2$ quiver SCFT, we defined the algebras for
the two inequivalent SU(2)-like subsectors: the dense XY sector, built from the two
bifundamental scalars $X$ and $Y$ of the hypermultiplets, and the dilute XZ
sector, built from one bifundamental scalar $X$ and the adjoint scalar $Z$ of
the vector multiplet. 
In the XY sector the loop fugacity depends on the colour,
see \eqref{XYquadratic}, while the cubic relation \eqref{XYcubic} does not, so
that we recover a dynamical Temperley-Lieb algebra of the type introduced
in~\cite{Ren:2023mtn}.
In the XZ
sector it is instead the cubic relation that becomes dynamical, see \eqref{XZcubicFeynmanX}--\eqref{XZTLrelations}, which appears to be a new
feature.

By studying finite-length open chains, we then illustrated how the associated link-state
bases~\cite{Read:2007ti,Ridout:2012gg}, in which the fields are paired up by
non-crossing links (the upper halves of the Temperley-Lieb diagrams), lead to a
better understanding of the spectrum.
In the XY sector, where the SU(2) symmetry that rotates $X$ and $Y$ is unbroken, the Hamiltonian preserves
the number of links, and the couplings enter the energies only through the
colour-dependent loop fugacities. In the XZ sector, the graphical approach
clarifies the energy splittings between states belonging to the same SU(2)
multiplet at the orbifold point ($g_1=g_2$), visible in
Figures~\ref{GraphL4XZ} and~\ref{fig:allmagnonXZL6}. Besides the $g_a$-antisymmetrised A-links,
the basis requires $g_a$-symmetrised S-links, and away from the orbifold point 
the Hamiltonian mixes the magnon states built from A-links only with the
descendants containing S-links. This mixing, which vanishes at $g_1=g_2$, is what
lifts the degeneracy within the multiplets; it is manifest in the link basis but
obscured in the monomial one. The same is true of the parity symmetry of the XZ
chain observed in~\cite{Bozkurt:2024tpz,Bozkurt:2025exl}:
the
parity-even and parity-odd combinations of link states are easily identified,
and in the resulting basis the reduced Hamiltonian is block diagonal.

We derived how our algebras fit into, and extend, the framework of the
dynamical Temperley-Lieb algebras of~\cite{Ren:2023mtn}, in which the colour
dependence is encoded in a dynamical variable and its shifts along the chain,
see \eqref{eq:dTL-local}. For $\Zset_2$ the XY sector gives exactly the algebra
of~\cite{Ren:2023mtn}, while the XZ sector has the opposite structure, with a
constant loop fugacity and a dynamical cubic coefficient. For general $\Zset_K$
both coefficients become dynamical in both sectors, see \eqref{eq:XY-ZK-dyn} and
\eqref{eq:gen-dTL}, so that the quiver algebras are genuine generalisations of
those of~\cite{Ren:2023mtn}. These relations are summarised in
Figure~\ref{fig:dTL-summary}. 

We further showed that the colour dependence can be
removed locally generator by generator. In the XY sector, after a dynamical rescaling
\eqref{eq:dyn-rescaling} the generators satisfy the Temperley-Lieb relations $\mathrm{TL}(q=1)$ of the XXX chain. In the XZ sector each generator is locally an XXZ generator
whose loop fugacity $\beta$ is set by the colour to its left, see \eqref{eq:etilde}. A
dynamical twist \eqref{eq:twisted-XZ} then maps the generators to those of
$\mathrm{TL}(q_1)\oplus\dots\oplus\mathrm{TL}(q_K)$, one copy of the standard
algebra for each colour. Importantly, each generator is twisted differently, so
that this map is neither a change of basis nor an algebra homomorphism, and the
spectrum of the quiver Hamiltonian is not that of a direct sum of $K$ copies of XXZ chains. The twist has precisely the form of the coassociator arising from the dynamical twists of $U_q(\mathfrak{sl}(2))$~\cite{Babelon:1995rz,Arnaudon:1997hm}, which
points to an underlying quasi-Hopf-like structure.

\paragraph{Future directions.} Having established the algebras, an immediate goal is to obtain a better
understanding of the dynamical quantum group-like structure, which we expect to play, in the
quiver theories, a role analogous to that of $U_q(\mathfrak{sl}(2))$ in the XXZ
case. The quasi-Hopf-like structure suggested by the dynamical twist gives a concrete
handle on this question. One expects there to exist a dynamical version of
Schur-Weyl duality. For a recent example of such a construction,  for a different diagrammatic algebra, see \cite{gainutdinov2026cellularalgebraa2webs}.
This would
organise the spectrum of the quiver spin chains in the same way that the
decomposition~\eqref{eq:schur_weyl} does for the XXZ model.

The algebras we constructed can be generalised in several directions. In
Section~\ref{sec:Zk} we showed that the extension from $\Zset_2$ to general
cyclic $\Zset_K$ orbifolds is straightforward, with one colour for each node of
the quiver. The reason is that the Temperley-Lieb generators are fixed by the
superpotential alone, so that the same recipe applies to any quiver gauge theory
with a cubic superpotential, with the details of the theory entering only
through the coefficients in the relations. In particular, one can consider the
non-abelian orbifolds preserving $\Ncal=2$ supersymmetry, whose spin-chain
Hamiltonians were recently studied in~\cite{Bath:2026gcu}. There, after a
rescaling of the generators, the XY-type sector of every ADE quiver is
Pasquier's representation of $\mathrm{TL}(q=1)$ on the affine Dynkin
diagram~\cite{Pasquier:1986jc,Pearce:1990ila}. In addition, a very interesting further
direction is that of $\Ncal=1$ SCFTs, especially the abelian orbifolds studied in~\cite{toappear1,toappear2}. Their quivers
generically contain only bifundamental fields and would therefore more closely
resemble the XY sector of Section~\ref{sec:XYsector}, while marginal
superpotential deformations, such as the Leigh-Strassler deformations \cite{Leigh:1995ep}, introduce
additional factors in the cubic superspace vertices used to define the algebra.
In both cases, the non-abelian $\Ncal=2$ orbifolds and the $\Ncal=1$ theories,
preliminary results show that our construction goes through.

Combining the two inequivalent bosonic $\SU(2)$ sectors yields a closed $\SU(3)$
sector containing all three holomorphic scalars, and repeating our recipe should produce a unifying diagrammatic algebra. 
Going further, by including conjugate fields, fermions, gauge fields and
derivatives, one eventually reaches the full $\mathrm{PSU}(2,2|4)$ sector, which
is non-compact. The starting point for repeating our analysis in these larger
sectors is already available, since the complete one-loop Hamiltonian is known
both for $\Ncal=2$~\cite{Liendo:2011xb} and for $\Ncal=1$
theories~\cite{Liendo:2011wc,toappear1}. We therefore see no obstruction to
extending our construction. Identifying the dynamical extensions of the Temperley-Lieb and related Hecke algebras
that emerge in these larger sectors is a particularly interesting problem for
future work.
Another natural direction is to go beyond one loop, where the dilatation
operator of the XZ sector is known up to three loops~\cite{Pomoni:2011jj}. Since
the dynamical Temperley-Lieb relations are a property of the generators
themselves, it would be very interesting to understand to what extent the
higher-loop Hamiltonians can be expressed in terms of these generators, and what
this implies for the fate of the underlying dynamical symmetry.

Throughout this work we restricted ourselves to the study of open chains, where the action of
the Temperley-Lieb algebra on the multiplets is most transparent. The
gauge-invariant single-trace operators of the quiver theories, however,
correspond to closed, cyclically invariant chains. In the standard case, closing
the chain enlarges the Temperley-Lieb algebra to the affine Temperley-Lieb
algebra~\cite{Martin:1993jka,GrahamLehrer98}, which contains an additional
generator acting on the last and first sites, and in which links can wind around
the chain. For gauge theory applications it is therefore important to construct
the dynamical counterparts of this algebra. In the quiver theories the colouring
of the faces then has to be consistent around the chain, which is the
diagrammatic expression of the gauge invariance of the single-trace operator. 

Diagrammatic algebras have appeared previously in the $\Ncal=4$ SYM context. Specifically, the Brauer algebra, which is in a sense a non-planar extension of the Temperley-Lieb algebra, plays a crucial role in organising the multi-trace BPS sector \cite{Kimura:2007wy}. It would be interesting to make contact with this structure first at the $\Ncal=4$ level, by considering non-planar contributions to the dilatation operator, after which one could contemplate a dynamical extension to the $\Ncal=2$ quiver theories. 

One of the most appealing features of the standard Temperley-Lieb
algebra and its known generalisations is that they can be
Baxterised~\cite{Jones:1990hq}: the algebra itself provides the $R$-matrix, and
with it the integrability of the corresponding spin chain. It is therefore
natural to ask whether the same is true for the dynamical Temperley-Lieb
algebras presented here. The Baxterisation of dynamical Temperley-Lieb algebras
developed in~\cite{Ren:2023mtn} does not directly apply to them, since it
requires relations that are less dynamical than the ones we find away from the
orbifold point. This is in line with the fact that the quiver spin chains are
not integrable by standard criteria~\cite{Gadde:2009dj,Gadde:2010zi}. It does
not close the question, however. As the example of Felder's elliptic
$R$-matrix~\cite{Felder:1994be} in Appendix~\ref{app:Felder} illustrates, an extension of our dynamical Temperley-Lieb algebra is working in the background of the dynamical $R$-matrix. The dynamical twist of Section~\ref{subsec:dTL-from-XXZ}, which
relates the generators of the XZ sector to those of the XXZ model, suggests one
way to look for such a  structure. Whether it exists for the quiver theories, and
what notion of integrability it would imply, is in our view the most exciting
question raised by this work.

We indicated how the different algebras give rise to loop models. The dynamical nature of these algebras introduces a position-dependence that deforms the usual scale invariance of such models, and thereby adds a dynamical component to the physics, which is normally fully determined by topological properties. 
Such dynamical loop models might be useful for applications, e.g. to model certain ring polymers \cite{pearceSolvableCriticalDense2007,guttmanPolygonsPolyominoesPolycubes2009}.
Moreover, given the stringent algebraic constraints it would be interesting to understand what type of field theories one obtains in a suitable continuum limit. 

Finally, it would be interesting to study these dynamical TL algebras as diagrammatic algebras from a more formal perspective, and map out their representation theory. In analogy to results for such algebras this might uncover connections to combinatorics, orthogonal polynomials, knot theory, and cluster algebras. 

\mbox{}

\paragraph{Acknowledgements.} The authors are grateful 
to Raschid Abedin, Jelena Anic, Gleb Arutyunov, Giovanni Felder, Marius de Leeuw, Volker Schomerus and Torben Skrzypek for insightful discussions. We are grateful to Gleb Arutyunov for reading and providing comments on the manuscript.
We would also like to thank
Jarryd Bath, Roland Bittleson, Lewis Cole, Sibylle Driezen, Max Downing, Simon Ekhammar,  Sergey Frolov, Frank G\"ohmann, Adolfo Holguin, Andreas Kl\"umper, Jules Lamers,  Tristan McLoughlin, Adrien Molines, Vera Posch, Sanjaye Ramgoolam and Ryo Suzuki for discussions.

ALR is thankful to the organisers, speakers and participants of the Integrability Workshop in Buggeru, of the InTropea2026 meeting in Tropea, of RAQIS in Annecy and the online VII Young Frontiers Meeting, for numerous discussions. ALR and EP are supported by ERC-2021-CoG - BrokenSymmetries 101044226 and have benefited from
the German Research Foundation DFG under Germany’s Excellence Strategy – EXC 2121
Quantum Universe – 390833306 as well as  SFB 1624 – “Higher structures, moduli spaces and integrability” –
506632645.  ALR is grateful for the hospitality of Perimeter Institute where part of this work was carried out. Research at Perimeter Institute
is supported in part by the Government of Canada through the Department of Innovation, Science and Economic Development and by the Province of Ontario through the Ministry of
Colleges and Universities. This work was supported by a grant from the Simons Foundation (1034867, Dittrich).
PR was funded by the Deutsche Forschungsgemeinschaft (DFG, German Research Foundation) -- 460391856.
KZ would like to thank DESY (Hamburg) and the Quantum Universe network for enabling a research visit which was crucial for the conceptualisation of this project. He is also grateful to NITheCS for travel support to present these results at IGST '26.
R.K.’s research is partially funded
by the Deutsche Forschungsgemeinschaft (DFG, German Research Foundation) – Projektnummer 417533893/GRK2575 “Rethinking Quantum Field Theory”.

We used Claude (Anthropic) only for proof-reading the manuscript and generating some of the diagrams and improving the aesthetic of the plots.

\appendix

\section{Link states and Young tableaux}\label{app:young}

In this appendix we relate the link states of Section~\ref{sec:XXZ} to Young tableaux, for readers more familiar with the latter as the organising principle of $\mathfrak{sl}(2)$ representations (i.e.\ at $q=1$). The link states are simply a different choice of basis, which has the advantage that it can be represented graphically without crossings. We illustrate this with the $L=4$ example discussed there, restricting to the zero-weight sector, i.e.\ the sector with an equal number of up and down spins. First of all, as the representation theory of $U_q(\mathfrak{sl}(2))$ mirrors that of $\mathfrak{sl}(2)$ (for generic $q$) we can still use standard Young tableaux to express its representations. The difference only arises when converting the tableaux to states, where we need to apply $q$-symmetrisation and $q^{-1}$-antisymmetrisation. 

Looking at \eqref{eq:link_state_l4}, we see that the zero-weight vector is equivalent to the totally $q$-symmetrised state. So we have simply:
\be

\ee
As for the weight-zero components of the spin-1 states, we can express weight-zero states encoded in the three standard Young tableaux as\footnote{We use an ``antisymmetrise-last'' prescription to express the tableaux in terms of states.}
\begin{align}
&%
= q~\ket{\uparrow\uparrow\downarrow\downarrow}-\ket{\downarrow\uparrow\downarrow\uparrow}+q^2\ket{\uparrow\downarrow\uparrow\downarrow}-q~\ket{\downarrow\downarrow\uparrow\uparrow}+(q^3-q^{-1})\ket{\uparrow\downarrow\downarrow\uparrow}
\end{align}
The corresponding components of the spin-1 states \eqref{T1} can be expressed in terms of these as 
\be\begin{split}
&%
 \, . \end{split}   
\ee
Similarly, the two standard Young tableaux corresponding to $L=4$ singlets can be written as follows:
\begin{align}
%
&=(\ket{\uparrow\downarrow}-q^{-1}\ket{\downarrow\uparrow})\otimes(\ket{\uparrow\downarrow}-q^{-1}\ket{\downarrow\uparrow})\;.
\end{align}
The two link states \eqref{T2} in the doubly-antisymmetrised sector $T_0$ can be expressed as
\be\scalebox{0.8}{%
\, .
\ee
 So the link states are just a choice of basis which can be expressed graphically without crossings, unlike the Young tableau basis.

\section{BPS states in the XZ sector}\label{app:kappa-sym}

The main focus of this work is on states with nonzero anomalous dimensions, which contain at least one link (antisymmetrised) pair of indices. However, it is important to also consider the BPS sector, i.e. the states which are annihilated by the Hamiltonian. For the XZ sector of the $\Zset_2$ orbifold theory, these states were studied in \cite{Gadde:2010zi}, where a $\kappa$-symmetrisation prescription was proposed for their construction. (As always, $\kappa=g_2/g_1$, the ratio of the two gauge couplings). This was studied from a Hopf-algebra perspective in \cite{Bertle:2024djm}, where it was formalised in terms of a `BPS-coproduct':
\be \label{BPScoproduct}
\Delta(\sigma^\pm)=\sum_{i=1}^L\left(I\otimes \cdots \otimes I\otimes \sigma^\pm \otimes K\otimes \cdots \otimes K\right)\, ,
\ee
where $K$ is an operator expressed via a $\Zset_2$ element $\gamma$ exchanging the node indices as
\be
K=\gamma \kappa^{-s(a)}\;, \; \;\text{where}\;\; s(1)=1 \;\text{and} \;s(2)=-1\, .
\ee
Note that, unlike the $U_q(\mathfrak{sl}(2))$ coproduct \eqref{eq:uqsl2}, this BPS-coproduct does not depend on the diagonal elements $\sigma^z$. On the other hand, it is dynamical as it depends on the node index at each site to the right of the action of $\sigma^\pm$. We emphasise, however, that this is not a full coproduct of some quantum group, as it only applies to the symmetric sector. It is expected that extending to other sectors will require the $\sigma^z$'s.  

Let us call a length $L$ BPS state with $M$ $Z$ fields (i.e. $M$ $Z$ magnons on the $X$-vacuum) $S^{(a)}(L,M)$, where $a$ indicates the first node index. Starting from $M=0$ (the $X$-vacuum), the action of the lowering operator increases the number of $Z$ fields, while starting from $M=L$ (the $Z$-vacuum) the action of the raising operator increases the number of $X$ fields. So for a generic $M$-magnon state we can write:\footnote{For concreteness, we assume $L$ even and hence the last index of the $X$-vacuum state agrees with the first one. The case of $L$ odd proceeds analogously.} 
\be
S^{(1)}(L,M)=\Delta(\sigma^-)^M \ket{X_{12}X_{21}\cdots X_{21}}=\Delta(\sigma^+)^{L-M}\ket{Z_1Z_1\cdots Z_1}\, .
\ee
For $L=2, M=1$, these states are precisely the $S$-link states defined in \eqref{eq:AandS_12},
\eqref{eq:AandS_21}: $S^{(1)}(2,1)=S_{12}^{[0]}$. 

By definition, $\kappa$-symmetrised states are annihilated by the dynamical TL generators, 
\be
e_i(S^{(a)}(L,M))=0\, .
\ee
Although the $\kappa$-symmetrised states do not mix with any of the other XZ-sector basis states, it is still useful to express them as linear combinations of $A$- and $S$-link states. For that, it is necessary to have a diagrammatic notation that is compatible with the one used in Section~\ref{sec:XZsector}, and which therefore generalises the $S$-links to arbitrary length. To identify such a notation, we draw inspiration from the role of the coproduct in the ABA, in which a $Z$ magnon is created by acting with the lowering operator. For simplicity, let us act on the  $X$-vacuum. Then we have:
\be \label{Deltalowering}
\mathcal{B}\ket{X_{12}X_{21}\cdots X_{21}}=
\begin{tikzpicture}[scale=0.6,baseline=0.8cm]
  \draw[->,thick,black!10!green] (1,1)--(1,3);
  \draw[->,thick,black!10!green] (2,1)--(2,3);
 
  \draw[->,thick,black!10!green] (4,1)--(4,3);
  \draw[->,thick,black!10!green] (5,1)--(5,3);
  \draw[->,thick,black!20!red] (0,1.5)--(6,2.5);
  
  \node[scale=0.7] at (1,0.3) {$X_{12}$};
  \node[scale=0.7] at (2,0.3) {$X_{21}$};
  \node[scale=0.7] at (3,0.3) {$\cdots$};
  \node[scale=0.7] at (4,0.3) {$X_{12}$};
  \node[scale=0.7] at (5,0.3) {$X_{21}$};
  \node[scale=0.7] at(-0.6,1.5) {$Z_1$};
\node[scale=0.7] at (6.6,2.5) {$X_{21}$};
\end{tikzpicture}\;.\ee
Similarly, we can create $X$ magnons by acting with the raising operator (which we illustrate on the $Z$-vacuum for simplicity):\footnote{Note that, since the $X$ field shifts the dynamical parameter, to raise the $Z_1$ vacuum state one first needs to act by $\Zset_2$ and apply the raising operator to the $Z_2$-vacuum state.} 
\be \label{Deltaraising}
\mathcal{C}\ket{Z_1Z_1\cdots Z_1}=
\begin{tikzpicture}[scale=0.6,baseline=0.8cm]
  \draw[->,thick,black!10!green] (1,1)--(1,3);
  \draw[->,thick,black!10!green] (2,1)--(2,3);
 
  \draw[->,thick,black!10!green] (4,1)--(4,3);
  \draw[->,thick,black!10!green] (5,1)--(5,3);
  \draw[->,thick,black!20!red] (0,1.5)--(6,2.5);
  
  \node[scale=0.7] at (1,0.3) {$Z_{2}$};
  \node[scale=0.7] at (2,0.3) {$Z_{2}$};
  \node[scale=0.7] at (3,0.3) {$\cdots$};
  \node[scale=0.7] at (4,0.3) {$Z_{2}$};
  \node[scale=0.7] at (5,0.3) {$Z_{2}$};
  \node[scale=0.7] at(-0.6,1.5) {$X_{12}$};
\node[scale=0.7] at (6.6,2.5) {$Z_{2}$};
\end{tikzpicture}\ee
In the above we use standard ABA notation for the magnon creation/annihilation operators. Despite that, as we don't know the $R$-matrix that would define the vertices in \eqref{Deltalowering} and \eqref{Deltaraising}, the above pictures are schematic. However, they are useful in providing intuition to establish a consistent graphical depiction of the action of the coproduct. 

To see this, we now rewrite the above pictures in the same notation we have been using for the Temperley-Lieb basis states. Consider, as above, the lowering operator \eqref{Deltalowering} acting on the $X$-vacuum. To better highlight the quantum numbers of the lowered state (i.e. after the action of $\Delta(\sigma^-))$, it is convenient to convert the outgoing $X$ to an incoming $\bar{X}$ field. Then one can use the fact that the quantum numbers of a $Z$ and $X$ field are the same as a $\bar{Y}$, while an $X$ together with an $\bar{X}$ are neutral so can be removed as far as quantum numbers are concerned. Schematically, for the example of $L=4$,
\be
\begin{split}
&

\end{split}
\ee
We see that in terms of quantum numbers, four $X$ fields have been replaced by one $\bar{Y}$ and two $X$ fields. It is easy to check that the quantum numbers of these fields (at the bottom of the diagram) agree with the new quantum numbers at the top (i.e. one less $X$ and one more $Z$.) One can continue this process of adding a $Z$ and $\bar{X}$. When one runs out of $X$ fields at the bottom, the $Z$ fields cannot combine with anything, while the $\bar{X}$ fields combine with $\bar{Y}$ fields to produce additional $Z$'s. In this way, one obtains states with an increasing number of $Z$ and decreasing number of $\bar{Y}$ fields at the bottom.

By replacing the outgoing $Z$ field in \eqref{Deltaraising} with an incoming $\bar{Z}$, one can similarly consider the raising operator as inserting a $X_{12}$ field on the left and a $\bar{Z}$ field on the right. Acting on the $Z$-vacuum state, the $X$ and one of the $Z$ fields will again lead to a $\bar{Y}$, while the $\bar{Z}$ will annihilate with one of the $Z$'s. So we obtain the following schematic picture (again for $L=4$): 
\be
\begin{split}
  &

\end{split}
\ee
Clearly one can iterate the action of $\Delta(\sigma^+)$, each time replacing a $Z$ by a $\bar{Y}$ while removing one of the other $Z$'s. Once one runs out of $Z$'s, the additional $X$ is kept while the additional $\bar{Z}$ combines with a $\bar{Y}$ to give one more $X$. From then on, repeated action reduces the number of $\bar{Y}$'s at the bottom while increasing the number of $X$'s, until one reaches the $X$-vacuum state. 

We illustrate the result of the above operations on the $L=4$ symmetric multiplet, which has also been considered in \cite{Bertle:2024djm}:\footnote{Note that the normalisation of the $Z$-vacuum used here is not the same as in Section~\ref{sec:XZsector}, but is more natural when applying the $g$-symmetrisation prescription.}
\be
\begin{split}
  &
=\begin{array}{l}g_2^2Z_1Z_1X_{12}X_{21}+g_1g_2Z_1X_{12}Z_2X_{21}+g_2^2Z_1X_{12}X_{21}Z_1\\\;+g_1^2X_{12}Z_2Z_2X_{21}+g_1g_2X_{12}Z_2X_{21}Z_1+g_2^2X_{12}X_{21}Z_1Z_1\end{array}=\begin{array}{l}g_2^3Z_1Z_{1}Z_{1}X_{12}+g_2^2g_1Z_{1}Z_1X_{12}Z_{2}\\\;+g_1^2g_2Z_{1}X_{12}Z_2Z_{2}+g_1^3X_{12}Z_2Z_2Z_2\end{array}%
=g_2^4~Z_1Z_1Z_1Z_1\\
\end{split}
\ee
Note the subtle distinction between the thin gray lines indicating $\bar{Y}$ fields (which are auxiliary as far as the XZ sector goes, and never appear at the top of a state) and solid black lines, indicating $X$ fields. As in the main text, the dashed lines indicate $Z$ fields.

The generalisation of this notation to higher lengths is straightforward. We again emphasise that the above graphical procedure is simply intended to clarify the correct counting of fields at the bottom of the BPS states in each magnon sector, so that we obtain a suitable generalisation of the $S$-link states to $L>2$. The procedure does not by itself compute the coefficients of each monomial, for which we use the BPS-coproduct \eqref{BPScoproduct}, replacing $\kappa=g_2/g_1$. 

As mentioned, the $\kappa$-symmetrised states do not mix with any of the link states under the action of the Hamiltonian, and therefore in principle there is no need to express them in the same graphical notation as the link states. However, an advantage of being able to do so (by following the above steps) is that one can relate them to the ``naive'' descendants of the $X$ or $Z$-vacuum states that one would obtain by connecting sites in a symmetric way via $S$-links. Again for $L=4$, we can for example consider how the half-filling (i.e. two-magnon) $S^{(1)}(4,2)$ state can be related to the two-$S$-link state $S_{12}S_{21}$:
\be\begin{split}
\right)
\end{split}
\ee
The $S^{(1)}(4,2)$ state is highly nonlocal in the sense that all possible positions of $X$ and $Z$ fields are included, while the $S_{12}S_{21}$ state is much more localised as it is made up of two-site objects, the $S$-links. As one increases the length, the $\kappa$-symmetrised states become increasingly complicated, while the naive two-site symmetrised states are simpler to work with, being of the same type as the non-BPS states in the basis. The former (fully $\kappa$-symmetrised) states are of course much more in the spirit of the BPS-coproduct, and ensure that the Hamiltonian gives precisely zero.  The Hamiltonian does not annihilate the $S_{12}S_{21}$ states, and one needs to consider cancellations between all the diagrams on the right-hand-side to see that this is a BPS state. 

One could now combine the $\kappa$-symmetrised states with the link states, to construct states in higher-length sectors, for instance one of the states appearing in the 3-magnon sector at $L=6$ would be:
\be
\begin{tikzpicture}[baseline=-0.5cm,scale=1]
  \filldraw[\lcyan] (-0.25,0)--(0.5,0)--(0.5,-1)--(-0.25,-1)--(-0.25,0);
  \filldraw[pink] (0.5,0)--(2.5,0)--(2.5,-1)--(0.5,-1)--(0.5,0);
  \filldraw[\lcyan] (2.5,0)--(4.5,0)--(4.5,-1)--(2.5,-1)--(2.5,0);
\filldraw[pink] (4.5,0)--(5.25,0)--(5.25,-1)--(4.5,-1)--(4.5,0);
   \filldraw[white] (0,0)--(3,0)--(3,-0.75)--(0,-0.75)--(0,0);
  \draw[-,thick] (-0.25,0)--(5.25,0);
 \draw[-,gray] (0.5,-0.75)--(0.5,-1);   
  \draw[-,gray] (2.5,-0.75)--(2.5,-1);
 \draw[-,thick] (0,0)--(0,-0.75);\draw[-,thick] (1,0)--(1,-0.75);\draw[-,thick] (2,0)--(2,-0.75);\draw[-,thick] (3,0)--(3,-0.75);
 \draw[-,thick] (0,-0.75)--(3,-0.75);
 \filldraw[white] (4,0) arc (-180:0:0.5);
  \draw[-,thick] (4,0) arc (-180:0:0.5);
  \draw[-,gray] (4.5,-0.5)--(4.5,-1);   
\filldraw[red,thick] (0,0) circle (1pt); \filldraw[red,thick] (1,0) circle (1pt); \filldraw[red,thick] (2,0) circle (1pt);
\filldraw[red,thick] (3,0) circle (1pt);\filldraw[red,thick] (4,0) circle (1pt);\filldraw[red,thick] (5,0) circle (1pt);
\end{tikzpicture}
\ee
The advantage of a basis constructed out of such states, compared to that used in Section~\ref{sec:basis}, is that the action of the Hamiltonian trivially gives zero apart from in the neighbourhood of the link state. On the other hand, the action at the interface of the two states (sites 4 and 5 in the above example) would be more involved. So for short chains the construction of Section~\ref{sec:basis} seems more efficient, while the above approach might find use in cases with long chains with few links, for instance as would be the case when considering the coordinate Bethe ansatz in this formalism. 

\section{Spectrum for short chains: $\mathbb{Z}_2$-XZ sector}\label{app:spectrum}

In Section~\ref{sec:acting-on-states} we discussed how to act with the TL generators on the link-state bases of different magnon sectors and generate the corresponding reduced Hamiltonian for that sector. In the process, we understood how the mixing of certain states explains the energy splittings observed in the plots in Figure \ref{fig:allmagnonXZL6}. 
In this appendix, we provide some additional details on the action of the XZ-sector Hamiltonian on the link states.  

First, we list the reduced Hamiltonians for $L=6$ for $M=0,1,2,3$ in the link basis of Section~\ref{sec:basis}. The generalisation to the remaining magnon sectors ($M=4,5,6$) is straightforward.

\paragraph{X-vacuum ($\boldsymbol{M=0}$).}

The $X$-vacuum is a BPS state, and therefore has energy $E=0$.

\paragraph{One-Magnon ($\boldsymbol{M=1}$) sector.}

The reduced Hamiltonian for this $M=1$ sector is given by

\be
H^{(1)}=\begin{pmatrix}
    g_1^2+g_2^2 & -g_2^2 & 0 & 0 & 0\\
    -g_2^2 & g_1^2+g_2^2 & -g_1^2 & 0 & 0\\
    0 & -g_1^2 & g_1^2+g_2^2 & -g_2^2 & 0\\
    0 & 0 & -g_2^2 & g_1^2+g_2^2 & -g_1^2\\
    0 & 0 & 0 & -g_1^2 & g_1^2+g_2^2
\end{pmatrix},
\ee
 whose eigenvalues are $\{g_1^2+g_2^2,g_1^2+g_2^2\pm \sqrt{g_1^4+g_1^2g_2^2+g_2^4},g_1^2+g_2^2\pm \sqrt{g_1^4-g_1^2g_2^2+g_2^4}\}$. The corresponding eigenvectors are 
\begin{align}
    \texttt{v}_1^{(1)}&=\frac{g_1^4}{g_2^4}v_1^{(1)}-\frac{g_1^2}{g_2^2}v_3^{(1)}+v_5^{(1)},\\
    \texttt{v}_{2,3}^{(1)}&=\frac{g_2^2}{g_1^2}v_1^{(1)}\mp \frac{\sqrt{g_1^4+g_1^2g_2^2+g_2^4}}{g_1^2}\,(v_2^{(1)}+v_4^{(1)})+\frac{g_1^2+g_2^2}{g_1^2}v_3^{(1)}+v_5^{(1)},\\
    \texttt{v}_{4,5}^{(1)}&=-\frac{g_2^2}{g_1^2}v_1^{(1)}\pm \frac{\sqrt{g_1^4-g_1^2g_2^2+g_2^4}}{g_1^2}\,(v_2^{(1)}-v_4^{(1)})-\frac{g_1^2-g_2^2}{g_1^2}v_3^{(1)}+v_5^{(1)}.
\end{align}

\paragraph{Two-Magnon ($\boldsymbol{M=2}$) sector.}

The Hamiltonian acts on the states in the following way:
\begin{align}
H^{(2)}\left(v_{1}^{(2)}\right) &= 2\left(g_1^2+g_2^2\right)v_1^{(2)} - g_1^2\,v_2^{(2)} - g_2^2\,v_7^{(2)}, \nonumber\\
H^{(2)}\left(v_{2}^{(2)}\right) &= -g_1^2\,v_1^{(2)} + 2\left(g_1^2+g_2^2\right)v_2^{(2)} - g_2^2\left(v_3^{(2)}+v_4^{(2)}\right), \nonumber\\
H^{(2)}\left(v_{3}^{(2)}\right) &= -g_2^2\,v_2^{(2)} + 2\left(g_1^2+g_2^2\right)v_3^{(2)} - g_2^2\,v_5^{(2)}, \nonumber\\
H^{(2)}\left(v_{4}^{(2)}\right) &= -g_2^2\,v_2^{(2)} + 2\left(g_1^2+g_2^2\right)v_4^{(2)} - g_2^2\,v_5^{(2)} - g_1^2\,v_8^{(2)}, \nonumber\\
H^{(2)}\left(v_{5}^{(2)}\right) &= -g_2^2\left(v_3^{(2)}+v_4^{(2)}\right) + 2\left(g_1^2+g_2^2\right)v_5^{(2)} - g_1^2\,v_6^{(2)}, \nonumber\\
H^{(2)}\left(v_{6}^{(2)}\right) &= -g_1^2\,v_5^{(2)} + 2\left(g_1^2+g_2^2\right)v_6^{(2)} - g_2^2\,v_9^{(2)}, \nonumber\\
H^{(2)}\left(v_{7}^{(2)}\right) &= -\frac{g_2^2\left(3g_1^2+g_2^2\right)}{g_1^2+g_2^2}\,v_1^{(2)} - g_1^2\,v_4^{(2)} + \left(g_1^2+g_2^2\right)v_7^{(2)} + \frac{1}{2}\left(g_1^2-g_2^2\right)v_8^{(2)} \nonumber\\
&\quad + \frac{g_1g_2\left(g_1^2-g_2^2\right)}{g_1^2+g_2^2}\,v_{10}^{(2)} - \frac{g_2\left(g_1^2-g_2^2\right)}{2g_1}\,v_{12}^{(2)}, \nonumber\\
H^{(2)}\left(v_{8}^{(2)}\right) &= -g_2^2\left(v_1^{(2)}+v_6^{(2)}\right) - 2g_1^2\,v_4^{(2)} - \frac{1}{2}\left(g_1^2-g_2^2\right)\left(v_7^{(2)}+v_9^{(2)}\right) + \left(g_1^2+g_2^2\right)v_8^{(2)} \nonumber\\
&\quad - \frac{g_1\left(g_1^2-g_2^2\right)}{2g_2}\left(v_{11}^{(2)}-v_{13}^{(2)}\right), \nonumber\\
H^{(2)}\left(v_{9}^{(2)}\right) &= -g_1^2\,v_4^{(2)} - \frac{g_2^2\left(3g_1^2+g_2^2\right)}{g_1^2+g_2^2}\,v_6^{(2)} + \frac{1}{2}\left(g_1^2-g_2^2\right)v_8^{(2)} + \left(g_1^2+g_2^2\right)v_9^{(2)} \nonumber\\
&\quad + \frac{g_2\left(g_1^2-g_2^2\right)}{2g_1}\,v_{12}^{(2)} - \frac{g_1g_2\left(g_1^2-g_2^2\right)}{g_1^2+g_2^2}\,v_{14}^{(2)}, \nonumber\\
H^{(2)}\left(v_{10}^{(2)}\right) &= -g_1g_2\,v_2^{(2)} + \frac{g_2\left(g_1^2-g_2^2\right)}{2g_1}\,v_7^{(2)} + \left(g_1^2+g_2^2\right)v_{10}^{(2)} - \frac{1}{2}\left(g_1^2+g_2^2\right)v_{11}^{(2)}, \nonumber\\
H^{(2)}\left(v_{11}^{(2)}\right) &= \frac{g_2^3}{g_1}\,v_1^{(2)} - g_1g_2\,v_4^{(2)} + \frac{g_2\left(g_1^2+g_2^2\right)}{2g_1}\,v_8^{(2)} - g_2^2\,v_{10}^{(2)} \nonumber\\
&\quad + \left(g_1^2+g_2^2\right)v_{11}^{(2)} - \frac{g_2^2\left(g_1^2+g_2^2\right)}{2g_1^2}\,v_{12}^{(2)}, \nonumber\\
H^{(2)}\left(v_{12}^{(2)}\right) &= g_1g_2\left(v_1^{(2)}-v_6^{(2)}\right) - \frac{g_1\left(g_1^2+g_2^2\right)}{2g_2}\left(v_7^{(2)}-v_9^{(2)}\right) \nonumber\\
&\quad - \frac{g_1^2\left(g_1^2+g_2^2\right)}{2g_2^2}\left(v_{11}^{(2)}+v_{13}^{(2)}\right) + \left(g_1^2+g_2^2\right)v_{12}^{(2)}, \nonumber\\
H^{(2)}\left(v_{13}^{(2)}\right) &= g_1g_2\,v_4^{(2)} - \frac{g_2^3}{g_1}\,v_6^{(2)} - \frac{g_2\left(g_1^2+g_2^2\right)}{2g_1}\,v_8^{(2)} - \frac{g_2^2\left(g_1^2+g_2^2\right)}{2g_1^2}\,v_{12}^{(2)} \nonumber\\
&\quad + \left(g_1^2+g_2^2\right)v_{13}^{(2)} - g_2^2\,v_{14}^{(2)}, \nonumber\\
H^{(2)}\left(v_{14}^{(2)}\right) &= g_1g_2\,v_5^{(2)} - \frac{g_2\left(g_1^2-g_2^2\right)}{2g_1}\,v_9^{(2)} - \frac{1}{2}\left(g_1^2+g_2^2\right)v_{13}^{(2)} + \left(g_1^2+g_2^2\right)v_{14}^{(2)}.
\end{align}
Viewed this way, however, the picture is somewhat misleading. It suggests that all states mix, which is not quite the case: for even $M$ and even $L$, the system has a parity symmetry (see Section~\ref{sec:diluteXZ}). For instance, the characteristic polynomial
of the $M=2$, $L=6$ Hamiltonian factorises into two polynomials, of degree eight and six respectively. This becomes explicit when the Hamiltonian is written in the parity-manifest basis of Section~\ref{subsec:parity}, where it
takes block-diagonal form. The first block is $8\times 8$ and given by
\begin{align}
H^{(2)}\left(u_{1}^{(2,+)}\right) &= 2\left(g_1^2+g_2^2\right)u_{1}^{(2,+)} - g_1^2\,u_{2}^{(2,+)} - g_2^2\,u_{5}^{(2,+)}, \nonumber\\
H^{(2)}\left(u_{2}^{(2,+)}\right) &= -g_1^2\,u_{1}^{(2,+)} + 2\left(g_1^2+g_2^2\right)u_{2}^{(2,+)} - 2g_2^2\left(u_{3}^{(2,+)}+u_{4}^{(2,+)}\right), \nonumber\\
H^{(2)}\left(u_{3}^{(2,+)}\right) &= -g_2^2\,u_{2}^{(2,+)} + 2\left(g_1^2+g_2^2\right)u_{3}^{(2,+)}, \nonumber\\
H^{(2)}\left(u_{4}^{(2,+)}\right) &= -g_2^2\,u_{2}^{(2,+)} + 2\left(g_1^2+g_2^2\right)u_{4}^{(2,+)} - g_1^2\,u_{6}^{(2,+)}, \nonumber\\
H^{(2)}\left(u_{5}^{(2,+)}\right) &= -\frac{g_2^2\left(3g_1^2+g_2^2\right)}{g_1^2+g_2^2}\,u_{1}^{(2,+)} - 2g_1^2\,u_{4}^{(2,+)} + \left(g_1^2+g_2^2\right)u_{5}^{(2,+)} + \left(g_1^2-g_2^2\right)u_{6}^{(2,+)} \nonumber\\
&\quad + \frac{g_1g_2\left(g_1^2-g_2^2\right)}{g_1^2+g_2^2}\,u_{7}^{(2,+)}, \nonumber\\
H^{(2)}\left(u_{6}^{(2,+)}\right) &= -g_2^2\,u_{1}^{(2,+)} - 2g_1^2\,u_{4}^{(2,+)} - \frac{1}{2}\left(g_1^2-g_2^2\right)u_{5}^{(2,+)} \nonumber\\
&\quad + \left(g_1^2+g_2^2\right)u_{6}^{(2,+)} - \frac{g_1\left(g_1^2-g_2^2\right)}{2g_2}\,u_{8}^{(2,+)}, \nonumber\\
H^{(2)}\left(u_{7}^{(2,+)}\right) &= -g_1g_2\,u_{2}^{(2,+)} + \frac{g_2\left(g_1^2-g_2^2\right)}{2g_1}\,u_{5}^{(2,+)} + \left(g_1^2+g_2^2\right)u_{7}^{(2,+)} - \frac{1}{2}\left(g_1^2+g_2^2\right)u_{8}^{(2,+)}, \nonumber\\
H^{(2)}\left(u_{8}^{(2,+)}\right) &= \frac{g_2^3}{g_1}\,u_{1}^{(2,+)} - 2g_1g_2\,u_{4}^{(2,+)} + \frac{g_2\left(g_1^2+g_2^2\right)}{g_1}\,u_{6}^{(2,+)} - g_2^2\,u_{7}^{(2,+)} + \left(g_1^2+g_2^2\right)u_{8}^{(2,+)}\,, 
\end{align}
which is even under parity, while the other one is $6\times 6$, odd under parity and can be written as
\begin{align}
H^{(2)}\left(u_{1}^{(2,-)}\right) &= 2\left(g_1^2+g_2^2\right)u_{1}^{(2,-)} - g_1^2\,u_{2}^{(2,-)} - g_2^2\,u_{3}^{(2,-)}, \nonumber\\
H^{(2)}\left(u_{2}^{(2,-)}\right) &= -g_1^2\,u_{1}^{(2,-)} + 2\left(g_1^2+g_2^2\right)u_{2}^{(2,-)}, \nonumber\\
H^{(2)}\left(u_{3}^{(2,-)}\right) &= -\frac{g_2^2\left(3g_1^2+g_2^2\right)}{g_1^2+g_2^2}\,u_{1}^{(2,-)} + \left(g_1^2+g_2^2\right)u_{3}^{(2,-)} \nonumber\\
&\quad + \frac{g_1g_2\left(g_1^2-g_2^2\right)}{g_1^2+g_2^2}\,u_{4}^{(2,-)} - \frac{g_2\left(g_1^2-g_2^2\right)}{g_1}\,u_{6}^{(2,-)}, \nonumber\\
H^{(2)}\left(u_{4}^{(2,-)}\right) &= -g_1g_2\,u_{2}^{(2,-)} + \frac{g_2\left(g_1^2-g_2^2\right)}{2g_1}\,u_{3}^{(2,-)} + \left(g_1^2+g_2^2\right)u_{4}^{(2,-)} - \frac{1}{2}\left(g_1^2+g_2^2\right)u_{5}^{(2,-)}, \nonumber\\
H^{(2)}\left(u_{5}^{(2,-)}\right) &= \frac{g_2^3}{g_1}\,u_{1}^{(2,-)} - g_2^2\,u_{4}^{(2,-)} + \left(g_1^2+g_2^2\right)u_{5}^{(2,-)} - \frac{g_2^2\left(g_1^2+g_2^2\right)}{g_1^2}\,u_{6}^{(2,-)}, \nonumber\\
H^{(2)}\left(u_{6}^{(2,-)}\right) &= g_1g_2\,u_{1}^{(2,-)} - \frac{g_1\left(g_1^2+g_2^2\right)}{2g_2}\,u_{3}^{(2,-)} - \frac{g_1^2\left(g_1^2+g_2^2\right)}{2g_2^2}\,u_{5}^{(2,-)} + \left(g_1^2+g_2^2\right)u_{6}^{(2,-)}.
\end{align}
In this form, it is clear that these two blocks do not mix and we show them in two separate plots in Figures~\ref{fig:m2XZeven} and~\ref{fig:m2XZodd}.

\paragraph{Three-Magnon ($\boldsymbol{M=3}$) sector.}
There is no parity symmetry in this  Hamiltonian and the characteristic polynomial does not factorise. The Hamiltonian acts on this 19-dimensional basis as follows\begin{align}
H^{(3)}\left(v_{1}^{(3)}\right) &= 3\left(g_1^2+g_2^2\right)v_{1}^{(3)} - g_2^2\,v_{2}^{(3)} - g_1^2\,v_{3}^{(3)}, \nonumber\\
H^{(3)}\left(v_{2}^{(3)}\right) &= -\frac{4g_1^2g_2^2}{g_1^2+g_2^2}\,v_{1}^{(3)} + 2\left(g_1^2+g_2^2\right)v_{2}^{(3)} - g_1^2\,v_{4}^{(3)} \nonumber\\
&\quad + \frac{g_1g_2\left(g_1^2-g_2^2\right)}{g_1^2+g_2^2}\left(v_{8}^{(3)}-v_{11}^{(3)}\right), \nonumber\\
H^{(3)}\left(v_{3}^{(3)}\right) &= -\frac{4g_1^2g_2^2}{g_1^2+g_2^2}\,v_{1}^{(3)} + 2\left(g_1^2+g_2^2\right)v_{3}^{(3)} - g_2^2\,v_{4}^{(3)} \nonumber\\
&\quad - \frac{g_1g_2\left(g_1^2-g_2^2\right)}{g_1^2+g_2^2}\left(v_{6}^{(3)}-v_{8}^{(3)}\right), \nonumber\\
H^{(3)}\left(v_{4}^{(3)}\right) &= -\frac{1}{2}\left(g_1^2+g_2^2\right)\left(v_{2}^{(3)}+v_{3}^{(3)}\right) + 2\left(g_1^2+g_2^2\right)v_{4}^{(3)} - \frac{2g_1^2g_2^2}{g_1^2+g_2^2}\,v_{5}^{(3)} \nonumber\\
&\quad + \frac{g_2\left(g_1^2-g_2^2\right)}{2g_1}\,v_{7}^{(3)} + \frac{g_1\left(g_1^2-g_2^2\right)}{2g_2}\,v_{10}^{(3)} - \frac{g_1g_2\left(g_1^2-g_2^2\right)}{g_1^2+g_2^2}\,v_{13}^{(3)}, \nonumber\\
H^{(3)}\left(v_{5}^{(3)}\right) &= -\frac{2\left(g_1^4+g_2^4\right)}{g_1^2+g_2^2}\,v_{1}^{(3)} - \frac{1}{2}\left(g_1^2-g_2^2\right)\left(v_{2}^{(3)}-v_{3}^{(3)}\right) - \left(g_1^2+g_2^2\right)\left(v_{4}^{(3)}-v_{5}^{(3)}\right) \nonumber\\
&\quad + \frac{g_1g_2\left(g_1^2-g_2^2\right)}{g_1^2+g_2^2}\left(v_{6}^{(3)}+v_{11}^{(3)}\right) - \frac{g_2\left(g_1^2-g_2^2\right)^2}{2g_1\left(g_1^2+g_2^2\right)}\,v_{7}^{(3)} \nonumber\\
&\quad + \frac{g_1\left(g_1^2-g_2^2\right)^2}{2g_2\left(g_1^2+g_2^2\right)}\,v_{10}^{(3)} - \frac{g_1^2\left(g_1^2-g_2^2\right)}{g_1^2+g_2^2}\,v_{16}^{(3)} + \frac{g_2^2\left(g_1^2-g_2^2\right)}{g_1^2+g_2^2}\,v_{18}^{(3)}, \nonumber\\
H^{(3)}\left(v_{6}^{(3)}\right) &= -\frac{g_1\left(g_1^2-g_2^2\right)}{2g_2}\,v_{3}^{(3)} + 2\left(g_1^2+g_2^2\right)v_{6}^{(3)} - \frac{1}{2}\left(g_1^2+g_2^2\right)v_{7}^{(3)} - g_2^2\,v_{12}^{(3)}, \nonumber\\
H^{(3)}\left(v_{7}^{(3)}\right) &= \frac{g_1g_2\left(g_1^2-g_2^2\right)}{g_1^2+g_2^2}\,v_{4}^{(3)} - g_1^2\left(v_{6}^{(3)}+v_{8}^{(3)}\right) + 2\left(g_1^2+g_2^2\right)v_{7}^{(3)} - \frac{2g_1^2g_2^2}{g_1^2+g_2^2}\,v_{9}^{(3)}, \nonumber\\
H^{(3)}\left(v_{8}^{(3)}\right) &= \frac{g_2\left(g_1^2-g_2^2\right)}{2g_1}\,v_{2}^{(3)} + \frac{g_1\left(g_1^2-g_2^2\right)}{2g_2}\,v_{3}^{(3)} \nonumber\\
&\quad - \frac{1}{2}\left(g_1^2+g_2^2\right)\left(v_{7}^{(3)}+v_{10}^{(3)}\right) + 2\left(g_1^2+g_2^2\right)v_{8}^{(3)}, \nonumber\\
H^{(3)}\left(v_{9}^{(3)}\right) &= -\frac{g_1g_2\left(g_1^2-g_2^2\right)}{g_1^2+g_2^2}\,v_{5}^{(3)} - g_2^2\,v_{7}^{(3)} + 2\left(g_1^2+g_2^2\right)v_{9}^{(3)} - g_1^2\,v_{10}^{(3)} - \frac{g_1^4+g_2^4}{g_1^2+g_2^2}\,v_{13}^{(3)}, \nonumber\\
H^{(3)}\left(v_{10}^{(3)}\right) &= \frac{g_1g_2\left(g_1^2-g_2^2\right)}{g_1^2+g_2^2}\,v_{4}^{(3)} - g_2^2\left(v_{8}^{(3)}+v_{11}^{(3)}\right) - \frac{2g_1^2g_2^2}{g_1^2+g_2^2}\,v_{9}^{(3)} + 2\left(g_1^2+g_2^2\right)v_{10}^{(3)}, \nonumber\\
H^{(3)}\left(v_{11}^{(3)}\right) &= -\frac{g_2\left(g_1^2-g_2^2\right)}{2g_1}\,v_{2}^{(3)} - \frac{1}{2}\left(g_1^2+g_2^2\right)v_{10}^{(3)} + 2\left(g_1^2+g_2^2\right)v_{11}^{(3)} - g_1^2\,v_{14}^{(3)}, \nonumber\\
H^{(3)}\left(v_{12}^{(3)}\right) &= -\frac{4g_1^2g_2^2}{g_1^2+g_2^2}\,v_{6}^{(3)} - g_1^2\,v_{9}^{(3)} + \left(g_1^2+g_2^2\right)v_{12}^{(3)} \nonumber\\
&\quad + \frac{g_1g_2\left(g_1^2-g_2^2\right)}{g_1^2+g_2^2}\left(v_{15}^{(3)}-v_{17}^{(3)}\right), \nonumber\\
H^{(3)}\left(v_{13}^{(3)}\right) &= \frac{2g_1g_2\left(g_1^2-g_2^2\right)}{g_1^2+g_2^2}\,v_{1}^{(3)} - \frac{2g_1^2g_2^2}{g_1^2+g_2^2}\left(v_{6}^{(3)}+v_{11}^{(3)}\right) + \frac{g_2^2\left(g_1^2-g_2^2\right)}{g_1^2+g_2^2}\,v_{7}^{(3)} \nonumber\\
&\quad - \left(g_1^2+g_2^2\right)v_{9}^{(3)} - \frac{g_1^2\left(g_1^2-g_2^2\right)}{g_1^2+g_2^2}\,v_{10}^{(3)} - \frac{1}{2}\left(g_1^2-g_2^2\right)\left(v_{12}^{(3)}-v_{14}^{(3)}\right) \nonumber\\
&\quad + \left(g_1^2+g_2^2\right)v_{13}^{(3)} - \frac{g_1\left(g_1^2-g_2^2\right)^2}{2g_2\left(g_1^2+g_2^2\right)}\,v_{16}^{(3)} + \frac{g_2\left(g_1^2-g_2^2\right)^2}{2g_1\left(g_1^2+g_2^2\right)}\,v_{18}^{(3)}, \nonumber\\
H^{(3)}\left(v_{14}^{(3)}\right) &= -g_2^2\,v_{9}^{(3)} - \frac{4g_1^2g_2^2}{g_1^2+g_2^2}\,v_{11}^{(3)} + \left(g_1^2+g_2^2\right)v_{14}^{(3)} \nonumber\\
&\quad - \frac{g_1g_2\left(g_1^2-g_2^2\right)}{g_1^2+g_2^2}\left(v_{17}^{(3)}-v_{19}^{(3)}\right), \nonumber\\
H^{(3)}\left(v_{15}^{(3)}\right) &= -g_1^2\,v_{3}^{(3)} + \frac{g_2\left(g_1^2-g_2^2\right)}{2g_1}\,v_{12}^{(3)} + \left(g_1^2+g_2^2\right)v_{15}^{(3)} - \frac{1}{2}\left(g_1^2+g_2^2\right)v_{16}^{(3)}, \nonumber\\
H^{(3)}\left(v_{16}^{(3)}\right) &= -\frac{2g_1^2g_2^2}{g_1^2+g_2^2}\,v_{4}^{(3)} - \frac{g_1g_2\left(g_1^2-g_2^2\right)}{g_1^2+g_2^2}\,v_{9}^{(3)} \nonumber\\
&\quad - g_2^2\left(v_{15}^{(3)}+v_{17}^{(3)}\right) + \left(g_1^2+g_2^2\right)v_{16}^{(3)}, \nonumber\\
H^{(3)}\left(v_{17}^{(3)}\right) &= -\frac{g_2\left(g_1^2-g_2^2\right)}{2g_1}\,v_{12}^{(3)} - \frac{g_1\left(g_1^2-g_2^2\right)}{2g_2}\,v_{14}^{(3)} \nonumber\\
&\quad - \frac{1}{2}\left(g_1^2+g_2^2\right)\left(v_{16}^{(3)}+v_{18}^{(3)}\right) + \left(g_1^2+g_2^2\right)v_{17}^{(3)}, \nonumber\\
H^{(3)}\left(v_{18}^{(3)}\right) &= -\frac{2g_1^2g_2^2}{g_1^2+g_2^2}\,v_{4}^{(3)} - \frac{g_1g_2\left(g_1^2-g_2^2\right)}{g_1^2+g_2^2}\,v_{9}^{(3)} \nonumber\\
&\quad - g_1^2\left(v_{17}^{(3)}+v_{19}^{(3)}\right) + \left(g_1^2+g_2^2\right)v_{18}^{(3)}, \nonumber\\
H^{(3)}\left(v_{19}^{(3)}\right) &= -g_2^2\,v_{2}^{(3)} + \frac{g_1\left(g_1^2-g_2^2\right)}{2g_2}\,v_{14}^{(3)} - \frac{1}{2}\left(g_1^2+g_2^2\right)v_{18}^{(3)} + \left(g_1^2+g_2^2\right)v_{19}^{(3)}.
\end{align}

\section{Dynamical Temperley-Lieb structure from Felder's elliptic $R$-matrix}\label{app:Felder}
In this appendix we show that the bulk Hamiltonian of the spin chain built from Felder's elliptic dynamical $R$-matrix \cite{Felder:1994be} has a dynamical Temperley-Lieb structure, with a loop fugacity which genuinely depends on the dynamical variable. Let $\theta(u)=\theta_1(\pi u|\tau)/(\pi \theta_1'(0|\tau))$ be the odd Jacobi theta function for some modular parameter $\tau \in \mathbb{C}$ with $\mathrm{Im}(\tau)>0$, normalised such that $\theta'(0) = 1$. Define the standard combinations
\be
\rho(u)=\frac{\theta'(u)}{\theta(u)}\,,\qquad \phi(u,v)=\frac{\theta(u+v)}{\theta(u)\,\theta(v)}\,,
\ee
which are a linearly shifted Weierstrass $\zeta$-function and the Kronecker elliptic function, respectively. Felder's $R$-matrix acts on $\mathbb{C}^2\otimes\mathbb{C}^2$ as
\begin{align} \label{eq:RFelder}
R(u,\lambda)&=E_{11}\otimes E_{11}+E_{22}\otimes E_{22}+\alpha(u,\lambda)\,E_{11}\otimes E_{22}+\alpha(u,-\lambda)\,E_{22}\otimes E_{11}\nonumber\\
&\quad+\beta(u,\lambda)\,E_{12}\otimes E_{21}+\beta(u,-\lambda)\,E_{21}\otimes E_{12}\,,
\end{align}
with
\be
\alpha(u,\lambda)=\frac{\theta(u)\,\theta(\lambda+\eta)}{\theta(u+\eta)\,\theta(\lambda)}\,,\qquad \beta(u,\lambda)=\frac{\theta(u+\lambda)\,\theta(\eta)}{\theta(u+\eta)\,\theta(\lambda)}\,,
\ee
where $\lambda$ is the dynamical parameter. The $R$-matrix \eqref{eq:RFelder} originates in the Andrews-Baxter-Forrester face model \cite{Andrews:1984af}, and ultimately in Baxter's 6-vertex reformulation of the 8-vertex model \cite{Baxter:1972wg,Baxter:1972wf}. 

We consider a chain of $L$ spins, with $\ket{\uparrow}$ and $\ket{\downarrow}$ of weight $+1$ and $-1$, and in analogy with the colours of Section~\ref{sec:twist} we let $h_k$ be the operator measuring the weight at site $k$ and $h^{(i)}=h_1+\dots+h_i$. The dynamical variable $\lambda$ is the analogue of the leftmost colour, while
\be
\mu_i=\lambda+\eta\,h^{(i-1)}
\ee
plays the role of the colour of the region to the left of site $i$. In this notation,
\be
\check{R}_{i,i+1}(u,\lambda)=P_{i,i+1}\,R_{i,i+1}\big(u,\mu_i\big)
\ee
satisfies the braided dynamical Yang-Baxter equation \eqref{eq:dYBE} and $\check{R}_{i,i+1}(0,\lambda)=1$. In the basis $\big(\ket{\uparrow\uparrow},\ket{\uparrow\downarrow},\ket{\downarrow\uparrow},\ket{\downarrow\downarrow}\big)$ of sites $i,i+1$, the Hamiltonian density $\mathcal{H}_i=\check{R}_{i,i+1}(-u,\lambda)\,\allowbreak\check{R}_{i,i+1}'(u,\lambda)\big|_{u=0}$ with $\check{R}_{i,i+1}'(u,\lambda)=\frac{d}{du}\check{R}_{i,i+1}(u,\lambda)$ is
\be
\mathcal{H}_i=\begin{pmatrix}
0&0&0&0\\
0&-\rho(\mu_i)-\rho(\eta)&\phi(-\mu_i,\eta)&0\\
0&\phi(\mu_i,\eta)&\rho(\mu_i)-\rho(\eta)&0\\
0&0&0&0
\end{pmatrix}\,,
\ee
which follows, up to conventions, from evaluating the decomposition (C.2) in \cite{klabbersDeformedInozemtsevSpin2024} at zero. 
This can be split as
\be\label{eq:Felder-H}
\mathcal{H}_i=\frac{1}{\theta(\eta)}\,E_i(\lambda)+D_i(\lambda)\,,
\ee
where, in the same basis,
\be
E_i(\lambda)=\frac{1}{\theta(\mu_i)}\begin{pmatrix}
0&0&0&0\\
0&-\theta(\mu_i+\eta)&\theta(\mu_i-\eta)&0\\
0&\theta(\mu_i+\eta)&-\theta(\mu_i-\eta)&0\\
0&0&0&0
\end{pmatrix},\qquad
D_i(\lambda)=\begin{pmatrix}
0&0&0&0\\
0&F(\mu_i,\eta)&0&0\\
0&0&-F(\mu_i,-\eta)&0\\
0&0&0&0
\end{pmatrix},
\ee
and
\be
F(u,v)=\phi(u,v)-\rho(u)-\rho(v)
\ee
is regular at both $u=0$ and $v=0$ and vanishes in the trigonometric limit $\tau \to \I \infty$. 
The operators $E_i$ satisfy the relations \eqref{eq:dTL-relations} of the dynamical Temperley-Lieb algebra of \cite{Ren:2023mtn},
\be
E_i\,E_i=\kappa(\mu_i)\,E_i\,,\qquad E_i\,E_{i\pm1}\,E_i=E_i\,,\qquad E_i\,E_j=E_j\,E_i\,,\quad |i-j|>1\,,
\ee
with the loop fugacity
\be
\kappa(\mu)=-\frac{\theta(\mu+\eta)+\theta(\mu-\eta)}{\theta(\mu)}\,,
\ee
which depends on the dynamical variable. Since $\kappa$ is even in $\mu$, setting $\mu=\eta a$ gives $\kappa=-([a+1]+[a-1])/[a]$ with $[a]=\theta(\eta a)$. Up to the conventional sign $E_i\mapsto-E_i$ (see Section~\ref{sec:XXZ}), which preserves the cubic relation, this is the elliptic example $\kappa(a)=([a+1]+[a-1])/[a]$ of \cite[Proposition 2.30]{Ren:2023mtn}. 

Felder's $R$-matrix thus provides a non-trivial example of a dynamical Temperley-Lieb algebra corresponding to an integrable dynamical spin chain. It would be very interesting to understand the further extended algebra generated by the $E_i$ and $D_i$.

\providecommand{\href}[2]{#2}\begingroup\raggedright\endgroup

\end{document}